\documentclass[aps,preprint,preprintnumbers,amsmath,amssymb,floatfix,showpacs]{revtex4}
\usepackage[cp1251]{inputenc}
\usepackage[russian]{babel}
\usepackage{graphicx}
\usepackage{epsfig}
\usepackage{slashbox}

\begin{document}
\title{Современные методы расчета фотоионизации и ионизации электронным ударом
двухэлектронных атомов и молекул\\
Modern methods for calculations of photoionization and electron impact
ionization of two-electron atoms and molecules}
\author{В.В. Серов, В.Л. Дербов, Т.А. Сергеева}
\affiliation{Кафедра теоретической физики Саратовского
государственного университета им. Н. Г. Чернышевского, Саратов}
\author{С.И. Виницкий}
\affiliation{Лаборатория теоретической физики Объединенного
Института Ядерных Исследований, Дубна}

\author{V.V. Serov, V.L. Derbov, T.A. Sergeeva}
\affiliation{Department of Theoretical Physics, Saratov State University,}
\author{S.I. Vinitsky}
\affiliation{Bogolyubov Laboratory of Theoretical Physics, Joint Institute for Nuclear Research}

\begin{abstract}
Представлен обзор некоторых недавно разработанных методов расчета многократных
 дифференциальных сечений фотоионизации и ионизации электронным ударом атомов
 и молекул с двумя активными электронами. Методы основаны на оригинальных подходах
 к вычислению трехчастичных кулоновских волновых функций.
 Рассмотрены внешний комплексный скейлинг и формализм уравнения Шредингера с источником в правой части. Продемонстрирована эффективность временных подходов к задаче рассеяния, таких как параксиальное приближение и метод сопутствующих координат.
   Сформулирован оригинальный численный метод, разработанный авторами для решения
шестимерного уравнения Шредингера для атома с двумя активными электронами на основе преобразования Чанга-Фано и представления дискретной переменной.
    На основе численных экспериментов проанализировано пороговое поведение угловых
распределений двухэлектронной фотоионизации отрицательного иона водорода и атома гелия, а также многократные дифференциальные сечения ионизации электронным ударом молекул водорода и азота. Продемонстрирована некорректность закона Ванье для углового
   распределения двойной ионизации даже при очень малых энергиях.

A review of some recently developed methods of calculating multiple
differential cross-sections of photoionization and electron
impactionization of atoms and molecules having two active electrons is
presented. The methods imply original approaches to calculating
three-particle Coulomb wave functions.  The external complex scaling
method and the formalism of the Schr\"odinger equation with a source in the
right-hand side are considered. Efficiency of the time-dependent
approaches to the scattering problem, such as the paraxial approximation
and the time-dependent scaling, is demonstrated. An
original numerical method elaborated by the authors for solving the 6D
Schroedinger equation for an atom with two active electrons, based on the
Chang-Fano transformation and the discrete variable representation, is
formulated. Basing on numerical simulations, the threshold behavior of angular
distributions of two-electron photoionization of the negative hydrogen ion
and helium atom, and multiple differential cross-sections of electron
impact ionization of hydrogen and nitrogen molecules are analyzed. It is demonstrated that the Vanier law for the angular distribution of double ionisation is not correct even at very small energies.

\end{abstract}


\maketitle

\tableofcontents

\section{Введение}
Знание волновой функции сплошного спектра  трех заряженных частиц
требуется для расчета таких важных процессов в атомной и
молекулярной физике, как ионизация атомов и молекул медленным
электроном, двойная фотоионизация и т.д. Все эти процессы
характеризуются дифференциальным сечением,
расчет которого и является целью теории. Поскольку квантовомеханическая
задача трех тел, так же как и классическая, не имеет аналитического
решения, для расчетов приходится либо использовать приближенные функции
\cite{PopovReview2010}, корректные лишь при больших энергиях или
вовсе анзацные, либо применять численные методы \cite{Kalitkin1978}. При численном
решении задачи трех тел имеются две основные проблемы. Во-первых, это
высокая размерность задачи: после отделения движения центра масс она
описывается шестимерным уравнением Шредингера. Во-вторых, это
необходимость обеспечить физически корректную асимптотику волновой
функции, имеющую в данном случае весьма сложный вид \cite{PZV2011}.
На сегодняшний день было предложено достаточно много различных
численных схем, каждая из которых имеет свои преимущества и
недостатки. Их обзору и посвящена настоящая работа. С физической
точки зрения назначение включенных в обзор методов заключается в
обеспечении корректного теоретического описания новых экспериментов
с использованием техники совпадений, позволяющих одновременно
регистрировать энергии и импульсы всех вылетающих фрагментов.
Поэтому рассматриваемые расчетные методы иллюстрируются примерами
расчетов многократных дифференциальных сечений фото- и ударной
ионизации атомов и молекул с двумя активными электронами.
Значительная часть представленных методов разработана авторами
обзора
\cite{Serov2009,Serov2010,Serov2008,Serov2007,Serov2001,Serov2005,Serov2011,SerovSergeeva2010,Serov2012}.

Эксперименты по однократной и двойной ионизации электронным
ударом с одновременной регистрацией рассеянного электрона и
электронов, испущенных в результате ионизации,  являются полными в том смысле, что в них
измеряются  энергии и векторы импульса
всех вылетающих частиц. Такие эксперименты предоставляют уникальную возможность проверки различных
теоретических моделей и подходов. Исследование поведения
многократного дифференциального сечение таких процессов дает
ответы на вопросы, возникающие в различных областях знания, таких
как астрофизика, радиационное повреждение живой материи вторичными
электронами и физика плазмы.

Ионизация H$_2$, являющегося самым распространенным газом во
Вселенной, особенно интересна по ряду причин. Она реализуется
легче, чем ионизация атомного водорода, результаты можно
сравнивать с гелием, двойная ионизация диссоциативна и может
использоваться как источник протонов. Молекула водорода имеет
изотопы, поведение которых в подобных экспериментах дает
интересную информацию о колебательных эффектах \cite{jphysb2012}.

После пионерских работ по теории столкновений электронов с
молекулами \cite{Iijima,Bottcher,Fano,Shugard,McCarthy} и ранних
экспериментов по (e,2e) ионизации H$_2$
\cite{Ehrhard,Weigold,Crowe,Lahmam-Bennani1989} интерес к проблеме
ионизации двухатомных систем возродился в последние годы в связи с
разработкой экспериментальной техники одновременной регистрации
фрагментов, образующихся при ионизации двухатомных молекул
электронами \cite{Duguet,Dorn} или протонами
\cite{Dorner,Muramatsu,Takahashi,Reddish,Weber2004,Gisselbrecht}.
В настоящее время для двухатомных мишеней имеется много
теоретических подходов  от простейшей линейной комбинации атомных
матриц перехода  \cite{Hanssen,Weck1999} до более сложных моделей
и методов, например, с описанием двухцентрового кулоновского
континуума на основе подхода типа \cite{Pluvinage} (работа
\cite{Joulak,Chuluunbaatar2008}). Эта модель позже была использована для построения
двухэлектронных двухцентровых коррелированных произведений
\cite{Briggs} при исследовании двойной фотоионизации ($\gamma$,2e)
молекулы H$_2$, а также в работе  \cite{Weck2002} для простого
процесса (e,2e). Параллельно с указанными двухцентровыми подходами
используются одноцентровые модели, требующие большого числа
базисных функций. Такие модели использовались при попытках учета
эффектов второго порядка \cite{DalCappello2004,DalCappello2006}, а
также при применения сходящегося метода сильной связи каналов
(CCC) к двойной фотоионизации  H$_2$ \cite{Kheifets2005}. Следует
упомянуть также метод  парциальных волн с использованием
сепарабельности уравнения Шредингера в вытянутых сфероидальных
координатах \cite{Serov2005}.

Недавно были выполнены расчеты \textit{ab initio} двойной
фотоионизации молекулярного водорода
\cite{McCurdy2004,McCurdy2005,McCurdy2006}. Помимо своей очевидной
эффективности, применяемый в этих работах подход дает удачную
возможность адекватного описания состояний непрерывного спектра
двух медленных вылетающих электронов, которое представляет одну из
основных трудностей в задачах двойной ионизации. Одной из задач, рассматриваемых в настоящем обзоре,
является распространение указанного подхода на двойную ионизацию
электронным ударом с использованием вытянутой сфероидальной
системы координат, обладающей естественной симметрией двухатомных
систем и позволяющей разделить переменные в двухцентровом
одноэлектронном уравнении Шредингера.

Далее в тексте обзора, если не оговорено обратное, при записи формул используются атомные единицы,
в которых заряд  и масса  электрона, а также постоянная Планка  равны единице,
$e=m_{e}=\hbar=1$.

\section{Сведение задачи рассеяния к временному уравнению Шредингера с помощью приближения
параксиальной оптики}\label{sectionPA}

    В теории распространения световых пучков хорошо известно так называемое уравнение параксиальной оптики, которое вытекает из волнового уравнения при условии медленности продольного изменения огибающей пучка. По виду уравнение параксиальной оптики совпадает с временным уравнением Шредингера, но роль времени в нем играет продольная координата. В работе \cite{Serov2001} аналогичный подход был предложен для задачи рассеяния быстрых (нерелятивистких) электронов на атомах и молекулах. В работе \cite{Serov2007} этот подход впервые применен к двухкратной ударной ионизации, а в  работе \cite{Serov2011} впервые реализован без дополнительных приближений.

Стационарное уравнение Шредингера, описывающее частицу с массой
$\mu$, зарядом $q$ и начальным импульсом $k_i$, налетающую на
мишень (атом или молекулу), первоначально находящуюся в стационарном
состоянии с энергией $\epsilon_i$, имеет вид
\begin{eqnarray}
\left[-\frac{1}{2\mu}\nabla_{\mathbf{r}_0}^2+\hat{H}(\mathbf{r}_1)+V(\mathbf{r}_1,\mathbf{r}_0)\right]\Phi(\mathbf{r}_0,\mathbf{r}_1)
=\left(\frac{k_i^2}{2\mu}+\epsilon_i\right)\Phi(\mathbf{r}_0,\mathbf{r}_1),\label{StSchr}
\end{eqnarray}
где $\mathbf{r}_0$ --- радиус-вектор налетающей частицы,
$\mathbf{r}_1$ --- радиус-вектор электрона мишени (для простоты
полагаем, что в мишени имеется всего один активный электрон),
$\hat{H}$ - эффективный гамильтониан мишени, а $V$ --- эффективный
потенциал взаимодействия налетающей частицы с мишенью. При $r_0\to
\infty$ волновая функция должна иметь асимптотический вид в виде
суммы падающей плоской волны и расходящихся сферических волн
\begin{eqnarray}
\Phi(\mathbf{r}_0,\mathbf{r}_1)=\varphi_i(\mathbf{r}_1)\exp\left(ik_iz_0\right)+\sum_nf_n(\mathbf{k}_n)\varphi_n(\mathbf{r}_1)\frac{\exp\left(ik_nr_0\right)}{r_0},\label{Phi_z_minus_infty}
\end{eqnarray}
где $\varphi_i(\mathbf{r}_1)$ --- начальное состояние мишени, $\varphi_n(\mathbf{r}_1)$ --- возможные конечные состояния мишени с энергией $\epsilon_n$, $n$ --- набор квантовых чисел, однозначно определяющих состояние мишени, $f_n(\mathbf{k}_n)$ --- амплитуда рассеяния быстрого электрона, если мишень после рассеяния оказалась в состоянии $n$.

Представим волновую функцию в виде
\begin{eqnarray}
\Phi(\mathbf{r}_0,\mathbf{r}_1)=\tilde{\Psi}\left(\mathbf{r}_\perp,z_0,\mathbf{r}_1\right)\exp\left(ik_iz_0\right),
\end{eqnarray}
где $\mathbf{r}_\perp=(x_0,y_0)$ --- двухкомпонентный вектор, составленный
из координат  $\mathbf{r}_0$, перпендикулярных $\mathbf{k}_i$, а $\tilde{\Psi}\left(\mathbf{r}_\perp,z_0,\mathbf{r}_1\right)$ --- огибающая. Тогда уравнение (\ref{StSchr}) примет вид
\begin{eqnarray}
&&i\frac{k_i}{\mu}\frac{\partial\tilde{\Psi}\left(\mathbf{r}_\perp,z_0,\mathbf{r}_1\right)}{\partial z_0}+\frac{1}{2\mu}\frac{\partial^2\tilde{\Psi}\left(\mathbf{r}_\perp,z_0,\mathbf{r}_1\right)}{\partial z_0^2}\nonumber
\\&&=\left[-\frac{1}{2\mu}\nabla_\perp^2+\hat{H}(
\mathbf{r}_1)+V(\mathbf{r}_1,\mathbf{r}_0)-\epsilon_i\right]\tilde{\Psi}\left(\mathbf{r}_\perp,z_0,\mathbf{r}_1\right).
\end{eqnarray}
 Если $k_i\gg 1$, то первое слагаемое в левой части этого уравнения будет много больше второго  и, следовательно, второй производной от $\tilde{\Psi}\left(\mathbf{r}_\perp,z_0,\mathbf{r}_1\right)$ по $z_0$ можно пренебречь. Это и является параксиальным приближением. Если ввести параметр $t=z_0\mu/k_i$ и сделать замену $\Psi(\mathbf{r}_1,\mathbf{r}_\perp,t)=
\tilde{\Psi}\left(\mathbf{r}_\perp,k_it/\mu,\mathbf{r}_1\right)\exp(i\epsilon_it)$,
мы получим  пятимерное временное уравнение Шредингера
\begin{eqnarray}
i\frac{\partial\Psi(\mathbf{r}_1,\mathbf{r}_\perp,t)}{\partial t}=\left[-\frac{1}{2\mu}\nabla_\perp^2+\hat{H}(\mathbf{r}_1)+V\left(\mathbf{r}_1,\mathbf{r}_\perp,\frac{k_i}{\mu}t\right)\right]\Psi(\mathbf{r}_1,\mathbf{r}_\perp,t).\label{paraxialeq}
\end{eqnarray}
Вследствие (\ref{Phi_z_minus_infty}), начальное условие для уравнения (\ref{paraxialeq}) имеет вид
\begin{eqnarray}
\Psi(\mathbf{r}_1,\mathbf{r}_\perp,t_0)=\varphi_i(\mathbf{r}_1)\exp(-i\epsilon_it_0),
\end{eqnarray}
при $t_0\to -\infty$.

Отличие параксиального приближения от широко известного эйконального приближения состоит в том, что вторые производные по поперечным координатам быстрой
частицы не считаются пренебрежимо малыми.

\subsection{Извлечение амплитуды рассеяния из огибающей}
После получения волновой функции $\Psi(\mathbf{r}_1,\mathbf{r}_\perp,t)$ необходимо
извлечь из нее информацию о наблюдаемых величинах, то есть амплитуды рассеяния.
Выполним преобразование Фурье по поперечным координатам рассеянной частицы
\begin{eqnarray}
\psi_{\mathbf{k}_{\perp}}(\mathbf{r}_1,t)=\frac{1}{2\pi}e^{i\frac{k_{\perp}^2}{2\mu}t}\int\exp(-i\mathbf{k}_{\perp}\mathbf{r}_\perp)\Psi(\mathbf{r}_1,\mathbf{r}_\perp,t)\,d\mathbf{r}_\perp. \label{FouriePsiPA}
\end{eqnarray}
Возьмем проекцию $\psi_{\mathbf{k}_{\perp}}(\mathbf{r}_1,t)$ на
некоторое состояние мишени $\varphi_n(\mathbf{r}_1)$
\begin{eqnarray}
c_n(\mathbf{k}_{\perp},t)=\int\varphi_n^*(\mathbf{r}_1)\exp(i\epsilon_nt)\psi_{\mathbf{k}_{\perp}}(\mathbf{r}_1,t)\label{c_n_PA_def}
\end{eqnarray}
и выполним обратное преобразование Фурье
\begin{eqnarray}
\Psi_n(\mathbf{r}_\perp,t)=\frac{1}{2\pi}\int \exp\left(i\mathbf{k}_{\perp}\mathbf{r}_\perp-i\frac{k_{\perp}^2}{2\mu}t\right)c_n(\mathbf{k}_{\perp},t)d\mathbf{k}_{\perp}.
\end{eqnarray}
При больших $t$ и $r_\perp$ подынтегральное выражение быстро
осциллирует при всех $\mathbf{k}_{\perp}$, за исключением значений
вблизи стационарной точки $\mathbf{k}_{\perp 0}$, где по
определению его фаза не имеет линейной
зависимости от $\mathbf{k}_{\perp}$
\begin{eqnarray}
\left.\frac{\partial}{\partial\mathbf{k}_{\perp}}\left(\mathbf{k}_{\perp}\mathbf{r}_\perp-\frac{k_{\perp}^2}{2\mu}t\right)\right|_{\mathbf{k}_{\perp}=\mathbf{k}_{\perp 0}}=0,
\end{eqnarray}
то есть $\mathbf{k}_{\perp 0}=\mu\mathbf{r}_\perp/t=\mu\mathbf{v}_\perp$. Поэтому
\begin{eqnarray}
\Psi_n(\mathbf{r}_\perp,t)\simeq\frac{1}{2\pi}c_n(\mathbf{k}_{\perp 0},t)\int \exp\left(i\mathbf{k}_{\perp}\mathbf{r}_\perp-i\frac{k_{\perp}^2}{2\mu}t\right)d\mathbf{k}_{\perp}=\frac{\mu}{it}c_n(\mathbf{k}_{\perp 0},t)\exp\left(i\frac{k_{\perp 0}^2}{2\mu}t\right).\label{Psi_n_asymp_PA}
\end{eqnarray}

Из сравнения с уравнениями (\ref{Phi_z_minus_infty}) и (\ref{c_n_PA_def}) очевидно,
что если $n\neq i$, т. е. если состояние мишени не совпадает с начальным, должно быть
\begin{eqnarray}
\Psi_n(\mathbf{r}_\perp,t\to\infty)=f_n(\mathbf{k}_n)\frac{\exp\left[i(k_nr_0-k_iz_0+(\epsilon_n-\epsilon_i)t)\right]}{r_0}.\label{Psi_n_asymp}
\end{eqnarray}
Поскольку $r_0/t=k_i\sqrt{1+(r_\perp/z_0)^2}/\mu=k_i\sqrt{1+(k_{\perp 0}/k_i)^2}/\mu=k_i/\mu+k_{\perp 0}^2/2\mu k_i+O(k_{\perp 0}^2)$, а $k_n=\sqrt{k_i^2-2\mu\Delta\epsilon}=k_i-\mu\Delta\epsilon/k_i+O[(\Delta\epsilon)^2]$ (здесь мы ввели $\Delta\epsilon=\epsilon_n-\epsilon_i$ --- изменение энергии мишени), фаза выражения (\ref{Psi_n_asymp})
\begin{eqnarray}
k_nr_0-k_iz_0+(\epsilon_n-\epsilon_i)t=[k_nr_0/t-k_i^2/\mu+\Delta\epsilon]t\simeq\frac{k_{\perp 0}^2}{2\mu}t.\label{phase_PA}
\end{eqnarray}
Таким образом, (\ref{Psi_n_asymp_PA}) совпадает с (\ref{Psi_n_asymp}), если положить
\begin{eqnarray}
f_n(\mathbf{k}_n)=-i\sqrt{k_i^2+k_{\perp}^2}c_n(\mathbf{k}_{\perp},t\to\infty).
\end{eqnarray}
Очевидно, что здесь $\mathbf{k}_{\perp}$  --- поперечные компоненты импульса
рассеянной частицы $\mathbf{k}_n$ и $k_{\perp}=k_n\sin\theta_s$,
где $\theta_s$ --- угол рассеяния. Также можно заметить,
что $\mathbf{k}_{\perp}=-\mathbf{K}_{\perp}$,
$\mathbf{K}=\mathbf{k}_n-\mathbf{k}_i$ --- переданный мишени импульс.

Отношение членов, которыми мы пренебрегли в (\ref{phase_PA}), к
главному члену в фазе волны $k_iz_0$ дает оценку точности
параксиального приближения и, соответственно, условие его
применимости
\begin{eqnarray}
\frac{(\Delta\epsilon+E_\perp)^2}{8E_i^2}\ll 1,
\end{eqnarray}
где $E_\perp=K_\perp^2/2\mu$ --- энергия поперечного движения рассеянного электрона, $E_i$ --- энергия налетающего электрона.

\subsection{Параксиальное приближение с первым борновским приближением}
В наших работах \cite{Serov2001,Serov2007} предложен метод PA1B, основанный на комбинации параксиального приближения (PA) с первым борновским
приближением (1B). Метод PA1B основан на непосредственном вычислении фурье-образа $\psi_{\mathbf{k}_{\perp}}(\mathbf{r}_1,t)$, определяемого выражением (\ref{FouriePsiPA}), путем решения приближенного уравнения для него.
Преимуществом PA1B является то, что он требует численного решения уравнения с числом размерностей, равным числу степеней свободы мишени, в то время как точный PA приводит к уравнению с числом размерностей, большим на два.

Фурье-образ $\psi_{\mathbf{k}_{\perp}}(\mathbf{r}_1,t)$ удовлетворяет уравнению
\begin{eqnarray}
i\frac{\partial\psi_{\mathbf{k}_{\perp}}(\mathbf{r}_1,t)}{\partial t}&=& \hat{H}(\mathbf{r}_1)\psi_{\mathbf{k}_{\perp}}(\mathbf{r}_1,t)
 +\int\exp\left[i\frac{k_{\perp}^2-k_{\perp}'^2}{2\mu}t\right] V_{\mathbf{k}_{\perp}-\mathbf{k}_{\perp}'}(\mathbf{r}_1,t)\psi_{\mathbf{k}_{\perp}'}(\mathbf{r}_1,t)d\mathbf {k}_{\perp}', \label{BasicEqPrep}
\end{eqnarray}
где
\begin{equation}
V_{\mathbf{k}_{\perp}}(\mathbf{r}_1,t)=
\frac{1}{(2\pi)^2}\int\exp\left(-i\mathbf{k}_{\perp}\mathbf{r}_{\perp}\right)
V\left(\mathbf{r}_1,\mathbf{r}_\perp,\frac{k_i}{\mu}t\right)d\mathbf{r}_{\perp}.
\end{equation}
Предположим, что амплитуда рассеянной волны гораздо меньше, чем падающей.
Тогда в интеграле в правой части можно положить
\begin{eqnarray}
\psi_{\mathbf{k}_{\perp}'}(\mathbf{r}_1,t)\simeq 2\pi\delta(\mathbf{k}_{\perp}')\varphi_i(\mathbf{r}_1)\exp(-i\epsilon_it).
\end{eqnarray}
Такое приближение эквивалентно применению первого борновского приближения.
В результате уравнение (\ref{BasicEqPrep}) превращается в неоднородное
временное уравнение
\begin{eqnarray}
i\frac{\partial \psi_{\mathbf{k}_{\perp}}(\mathbf{r},t)}{\partial t}&=&
\hat{H}(\mathbf{r})\psi_{\mathbf{k}_{\perp}}(\mathbf{r},t)+F_{\mathbf{k}_{\perp}}(\mathbf{r},t), \label{PA1Beq}
\end{eqnarray}
с начальным условием
$\psi_{\mathbf{k}_{\perp}}(\mathbf{r},-\infty)=0$. Член-источник в этом уравнении
имеет вид
\begin{eqnarray}
F_{\mathbf{k}_{\perp}}(\mathbf{r},t)=2\pi
\exp\left[i\left(\frac{k_{\perp}^2}{2\mu}-\epsilon_i\right)t\right]V_{\mathbf{k}_{\perp}}(\mathbf{r},t)\varphi_i(\mathbf{r})
\label{PA1Bsource0}
\end{eqnarray}

В качестве примера применения PA1B рассмотрим потенциал взаимодействия частицы с зарядом $q$
с водородоподобным ионом с зарядом ядра $Z$
\begin{equation}
V(\mathbf{r}_1,\mathbf{r}_0)=-\frac{Zq}{r_0}+\frac{q}{|\mathbf{r}_1-\mathbf{r}_0|}.\label{Vhydr}
\end{equation}
Фурье-образ кулоновского потенциала по поперечным координатам можно получить
из известной формулы для кулоновского потенциала в импульсном представлении
\begin{eqnarray}
\int\exp(-i\mathbf{k}\mathbf{r})\frac{1}{r}d\mathbf{r}=\frac{4\pi}{k^2}=\frac{4\pi}{k_z^2+k_{\perp}^2}
\end{eqnarray}
путем обратного преобразования Фурье
\begin{eqnarray}
\int\exp(-i\mathbf{k}\mathbf{r}_{\perp})\frac{1}{r}d\mathbf{r}_{\perp}=
\frac{1}{2\pi}\int_{-\infty}^{\infty}\exp(ik_zz)\frac{4\pi}{k_z^2+k_{\perp}^2}=\frac{2\pi}{k_{\perp}}\exp(-k_{\perp}|z|).
\end{eqnarray}
Для потенциала (\ref{Vhydr}) это дает
\begin{eqnarray}
V_{\mathbf{k}_{\perp}}(\mathbf{r}_1,t)=\frac{q}{2\pi k_{\perp}}\left[\exp(-k_{\perp}|k_it/\mu-z_1|-i\mathbf{k}_{\perp}\mathbf{r}_{1\perp})-Z\exp(-k_{\perp}|k_it/\mu|)\right].
\label{MatrixEl}
\end{eqnarray}
В результате  член-источник (\ref{PA1Bsource0}) принимает вид
\begin{eqnarray}
F_{\mathbf{k}_{\perp}}(\mathbf{r},t)=\frac{q}{k_{\perp}}
e^{i\left(k_{\perp}^2/2\mu-\epsilon_i\right)t}
\left[e^{-k_{\perp}|k_i t/\mu-z|-i\mathbf{k}_{\perp}\cdot\mathbf{r}_{\perp}}-Ze^{-k_{\perp}|k_i t/\mu|}\right]\varphi_i(\mathbf{r}).
\label{PA1Bsource}
\end{eqnarray}
Поскольку он экспоненциально стремится к нулю при $|t|\to\infty$, можно положить $\psi_{\mathbf{k}_{\perp}}(\mathbf{r},t_0)=0$, где $t_0<0$, $|t_0|\gg \mu/(k_{\perp}k_i)$ и вести счет до $t\gg \mu/(k_{\perp}k_i)$.

\subsection{Представление амплитуды ионизации методом Фурье-разложения по времени}
Амплитуда ионизации может быть выражена из $\psi_{\mathbf{k}_{\perp}}(\mathbf{r},t)$,
полученной с помощью (\ref{FouriePsiPA}), как проекция на состояние континуума мишени
\begin{equation}
f(\Omega_s,E_e,\Omega_e)=-i k_i \lim_{t\to\infty} \langle
\mathbf{k}_e|\psi_{\mathbf{k}_{\perp}}(\mathbf{r},t)\rangle
e^{iE_et}.\label{f_via_psiKperp}
\end{equation}
Здесь ${\mathbf k}_e$ --- импульс испущенного электрона,
$E_e=k_e^2/2$ --- его энергия,
$|\mathbf{k}_e\rangle\equiv\varphi_{\mathbf{k}_e}^{(-)}(\mathbf{r})$ --- волновая функция континуума мишени.

Но мы использовали для вычисления амплитуды ионизации метод, предложенный в работе \cite{Selin1999},
который не требует знания точной волновой функции мишени.
Этот подход основан на Фурье-разложении по времени потока вероятности сквозь
некоторую замкнутую поверхность
\begin{eqnarray}
f=-i k_i \int_{t_0}^{T}dt\oint_{S}\mathbf{n}_SdS\cdot
 \mathbf{j}
 \left[\psi_{\mathbf{k}_{\perp}}(\mathbf{r},t),\chi_{\mathbf{k}_e}^{(-)*}(\mathbf{r}) e^{iE_et}\right]. \label{IonizationAmplitude}
\end{eqnarray}
Здесь вектор потока вероятности, введенный в \cite{Selin1999}
\begin{eqnarray}
\mathbf{j}[\psi,\varphi]\equiv\frac{i}{2}[\psi\nabla\varphi-\varphi\nabla\psi],
\end{eqnarray}
$T$ --- время, до которого симулировалась эволюция,
 $S$ --- замкнутая поверхность, окружающая систему (обычно сфера радиуса $r_S$),
 $\vec{n}_S$ --- нормальный вектор поверхности,
 $\chi_{\mathbf{k}_e}^{(-)}(\mathbf{r})$ --- функция,
 стремящаяся к $\varphi_{\mathbf{k}_e}^{(-)}(\mathbf{r})$ при $r\to\infty$.
 В работе \cite{Serov2011} в качестве $\chi_{\mathbf{k}_e}^{(-)}(\mathbf{r})$
 использовались квазиклассические функции, которые отличаются
 от точных $\varphi_{\mathbf{k}_e}^{(-)}(\mathbf{r})$ на $O(1/r^2)$.

Однако при использовании данного метода возникает серьёзная проблема: волновая функция
$\psi_{\mathbf{K}_{\perp}}(\mathbf{r},t)$ обычно не обращается в
нуль на $S$ даже при очень большом $T$. Это следствие того, что в
$\psi_{\mathbf{K}_{\perp}}(\mathbf{r},t)$ вносят заметный вклад
высоковозбужденные связанные состояния мишени и медленные
испущенные электроны. Вследствие этого, ур. (\ref{IonizationAmplitude}) дает
результат, сильно осциллирующий с изменением $T$. Мы преодолели этот
артефакт, предположив, что
\begin{eqnarray}
\psi_{\mathbf{k}_{\perp}}(\mathbf{r}_S,t>T)\simeq
\psi_{\mathbf{k}_{\perp}}(\mathbf{r}_S,T)\exp[-iE_{eff}(\mathbf{r}_S,T)(t-T)],
\end{eqnarray}
где $E_{eff}$ --- некоторая комплексная эффективная энергия. В результате
интеграл в области $t\in (T,\infty)$ может быть вычислен аналитически.
Тогда выражение (\ref{IonizationAmplitude}) превращается в
\begin{eqnarray}
f=-i k_i \oint_{S}\vec{n}_S\cdot\left\{\mathbf{j}\left[
\int_{0}^{T}e^{iE_et}\psi_{\mathbf{k}_{\perp}}(\mathbf{r},t)dt-\frac{e^{iE_eT}}{i(E_e-E_{eff})}\psi_{\mathbf{k}_{\perp}}(\mathbf{r},t),
 \chi_{\mathbf{k}_e}^{(-)*}(\mathbf{r}) \right]\right\}dS.
 \label{IonizationAmplitudeCorr}
\end{eqnarray}
В работе \cite{Serov2011} эффективная энергия вычислялась с помощью выражения
\begin{eqnarray} E_{eff}(\mathbf{r}_S,T)=\frac{i}{\psi_{\mathbf{k}_{\perp}}(\mathbf{r}_S,t)}\frac{\partial \psi_{\mathbf{k}_{\perp}}(\mathbf{r}_S,t)}{\partial t}.
\end{eqnarray}
Условие применимости этого приближения имеет вид $\left|dE_{eff}/dT\right|/E_e^2\ll 1$. При $T\to\infty$
оно переходит в $[U(r_S)/E_e]^2\sim 1/r_S^2\ll 1$, что по порядку сходимости совпадает с точностью
квазиклассической $\chi_{\mathbf{k}_e}^{(-)}(\mathbf{r})$.

 Для получения достаточно точных результатов с помощью этого метода необходимо выполнить эволюцию на большом промежутке времени $T$, много большем, чем характерное время взаимодействия с налетающим электроном (которое можно оценить как $\mu/(k_{\perp}k_i)$, как показано в предыдущем разделе).
  Нет никакого смысла выполнять затратное решение пятимерного уравнения (\ref{paraxialeq}) на всем промежутке времени $T$. Поэтому в работе \cite{Serov2011} решение (\ref{paraxialeq}) выполнялось только до $T_{\mathrm{PA}}\gg \mu/(k_{\perp}k_i)$, после чего из пятимерной функции $\Psi(\mathbf{r}_1,\mathbf{r}_\perp,T_{\mathrm{PA}})$
  с помощью (\ref{FouriePsiPA}) извлекалось $\psi_{\mathbf{K}_{\perp}}(\mathbf{r},T_{\mathrm{PA}})$
  для интересующего угла рассеяния. Затем $\psi_{\mathbf{K}_{\perp}}(\mathbf{r},T_{\mathrm{PA}})$
  использовалось в качестве начального условия для трехмерного временного уравнения (\ref{PA1Beq}),  которое  решалось уже до $t=T$.

\subsection{Приближение одного активного электрона}
 Чтобы расчет ударной ионизации многоэлектронных мишеней ударом быстрого электрона в рамках параксиального приближения стал практически возможным, для мишени необходимо применить приближение одного активного электрона. В работе \cite{Serov2011} такое приближение строится на основе приближения Хартри-Фока и приближения ``замороженных'' оболочек.

Начнем с временного уравнения Хартри-Фока для системы во внешнем поле
\begin{eqnarray}
i\frac{\partial\psi_i(\mathbf{r},t)}{\partial t}=\hat{F}\left[\{\psi_j(\mathbf{r}',t)\}_{j=1}^{N_o}\right]\psi_i(\mathbf{r},t)+v(\mathbf{r},t)\psi_i(\mathbf{r},t).
\end{eqnarray}
Здесь $N_o=N_e/2$ --- число заполненных оболочек, $N_e$ --- число электронов в системе, $v(\mathbf{r},t)$ --- внешнее поле. Оператор Фока для произвольного набора ортогональных волновых функций $\{\varphi_j\}_{j=1}^{N_o}$ имеет вид
\[
\hat{F}\left[\{\varphi_j\}_{j=1}^{N_o}\right] =
\hat{h}+\sum_{j=1}^{N_o}\left(2\hat{J}[\varphi_j]-\hat{K}[\varphi_j]\right).
\]
Он содержит  одноэлектронный гамильтониан
\[
\hat{h}=-\frac{1}{2}\nabla^2+u(\mathbf{r}),
\]
где $u(\mathbf{r})$ --- поле ядер, а также кулоновский оператор
\[\hat{J}[\varphi]\psi(\mathbf{r})=\int\frac{\left|\varphi(\mathbf{r}')\right|^2}{|\mathbf{r}-\mathbf{r}'|}\,d\mathbf{r}'\,\psi(\mathbf{r})\]
и обменный оператор
\[\hat{K}[\varphi]\psi(\mathbf{r})=\varphi(\mathbf{r})\int{\frac{\varphi^*(\mathbf{r}')\psi(\mathbf{r}')}{|\mathbf{r}-\mathbf{r}'|}d\mathbf{r}'}.\]

Предположим, что все орбитальные функции за исключением $i-$й не меняются в течении процесса
\[
\psi_j(\mathbf{r},t)=\left\{
\begin{array}{ll}
\psi(\mathbf{r},t),& j=i;\\
\varphi_j(\mathbf{r})\exp(-i\epsilon_j t),& j\neq i;
\end{array}
\right.
\]
где $\{\varphi_j\}_{j=1}^{N_o}$ --- решения стационарного уравнения Хартри-Фока.
\begin{equation}
\hat{F}\left[\{\varphi_j\}_{j=1}^{N_o}\right]\varphi_i(\mathbf{r})=\epsilon_i\varphi_i(\mathbf{r}).\label{eigenFock}
\end{equation}

Мы можем ввести эффективный потенциал остаточного иона после вылета $i$-го электрона
\[w_i(\mathbf{r})=2\sum_{j=1}^{N_o}\hat{J}[\varphi_j]-\hat{J}[\varphi_i]=\sum_{j=1}^{N_o}(2-\delta_{ij})\int\frac{\left|\varphi_j(\mathbf{r}_2)\right|^2}{r_{12}}\,d\mathbf{r}_2.\]
Остаточный оператор с формальной точки зрения имеет вид
\[\hat{X}_i=\hat{F}\left[\{\psi_j\}_{j=1}^{N_o}\right]-\left[\hat{h}+w_i(\mathbf{r})\right]
=2\hat{J}[\psi]-\hat{J}[\varphi_i]-\sum_{j=1}^{N_o}\hat{K}[\psi_j],\]
то есть
\[\hat{X}_i\psi=\left\{\hat{J}[\psi]-\hat{J}[\varphi_i]-\sum_{j\neq i}\hat{K}[\varphi_j]\right\}\psi,\]
но поскольку $\hat{J}[\psi]$ описывает электрон на той же
оболочке с противоположным спином, состояние которого в рамках
приближения ``замороженных'' оболочек мы так же можем считать
неизменным в ходе процесса, то приходим к обменному оператору
\[\hat{X}_i=-\left\{\sum_{j\neq i}\hat{K}[\varphi_j]\right\}.\]
Корректное введение этого оператора приводит к интегральному
уравнению. Поскольку обмен существенен только в том случае, когда
электрон находится вблизи молекулы, мы можем ввести приближенный
оператор обмена
\[\hat{X}_i^{N}=-\hat{I}^{N}\left\{\sum_{j\neq i}\hat{K}[\varphi_j]\right\}\hat{I}^{N},\]
где проекционный оператор на подпространстве решений ур. (\ref{eigenFock})
\[\hat{I}^{N}\psi(\mathbf{r})=\sum_{k=1}^{N}\varphi_k(\mathbf{r})\int\varphi_k^*(\mathbf{r}')\psi(\mathbf{r}')d\mathbf{r}'.\]
Тогда эффективный гамильтониан мишени принимает вид
\begin{eqnarray}
\hat{H}=\hat{h}+w_i(\mathbf{r})+\hat{X}_i^N. \label{effHam}
\end{eqnarray}
Если $N\geq N_o$, то приближенный оператор будет обеспечивать корректные орбитальные
энергии для основного состояния. В работе \cite{Serov2011} авторы полагали $N=N_o$,
тем самым фактически пренебрегая обменом для соcтояний континуума и возбужденных состояний.
Эффективный потенциал иона, очевидно, будет иметь вид
\begin{eqnarray}
U_i(\mathbf{r})=u(\mathbf{r})+w_i(\mathbf{r}).
\end{eqnarray}

Потенциал взаимодействия налетающего электрона с молекулой в уравнении (\ref{paraxialeq})
складывается из эффективного потенциала остаточного иона и потенциала взаимодействия
налетающего электрона с активным электроном молекулы
\begin{eqnarray}
V(\mathbf{r}_1,\mathbf{r}_0)=U_i(\mathbf{r}_0)+\frac{1}{|\mathbf{r}_1-\mathbf{r}_0|}.
\end{eqnarray}

Детали численной схемы для решения пятимерного временного уравнения (\ref{paraxialeq})
изложены в Приложении \ref{sect5Dsolution}.

\subsection{Сравнение с другими методами}

\begin{figure}[t]
\includegraphics[angle=-90,width=0.5\textwidth]{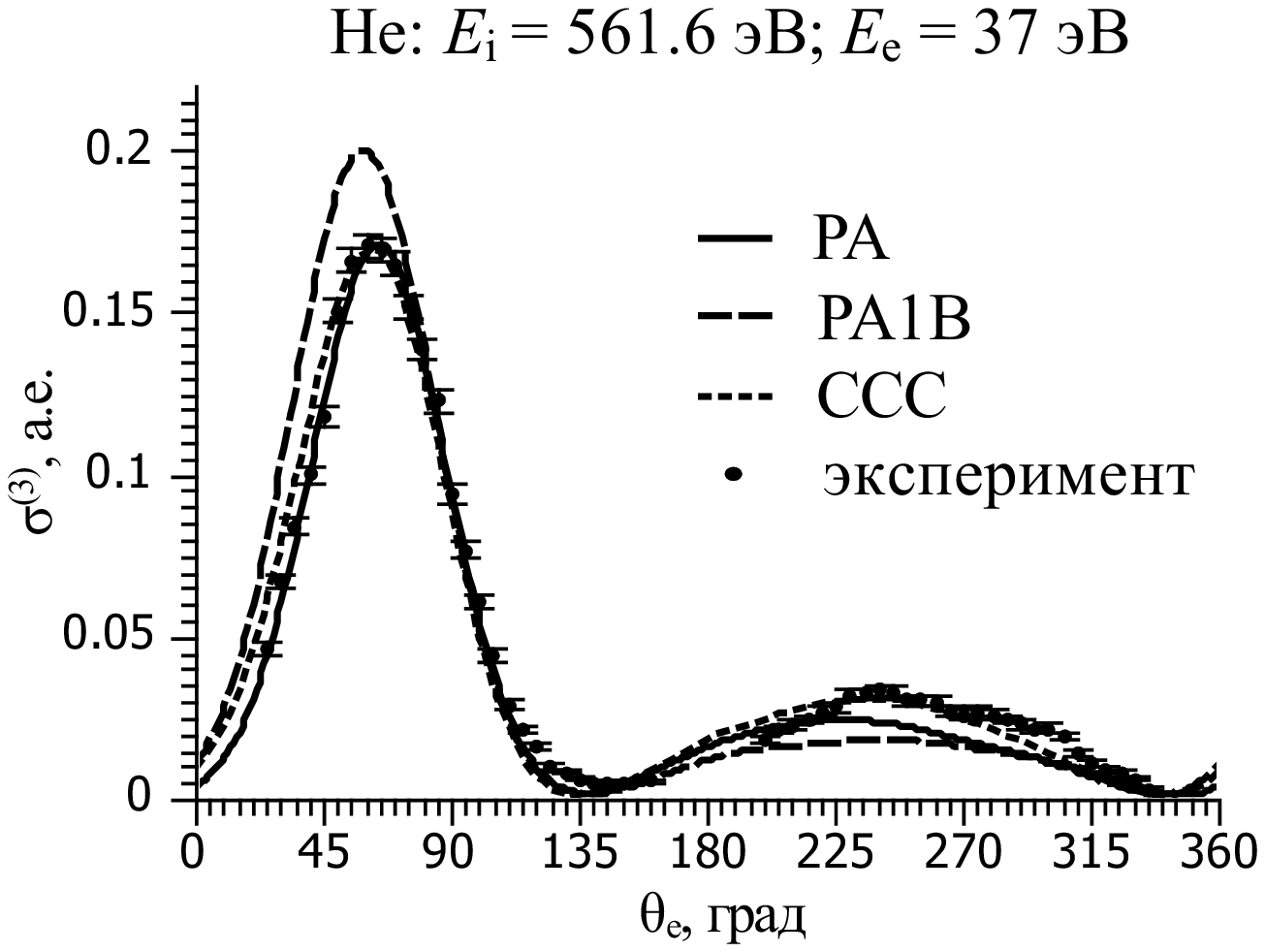}\\
\textit{а}\\
\includegraphics[angle=-90,width=0.5\textwidth]{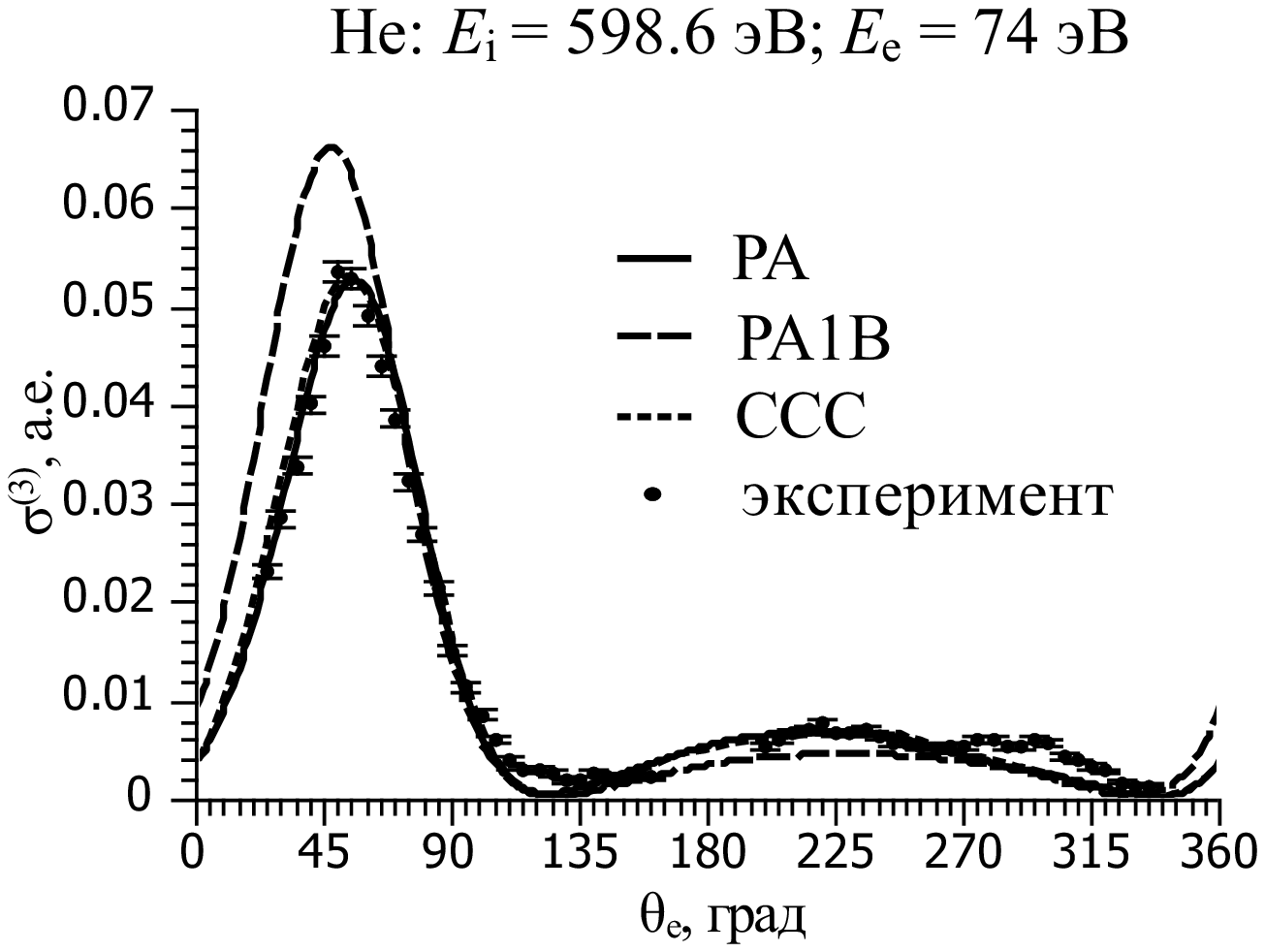}\\
\textit{б}\\
\caption{3ДС для процесса He$(e,2e)$ как функция угла вылета
$\theta_{e}$ для энергии вылетевшего электрона (\textit{а}) $E_e$=37 эВ и (\textit{б})
$E_e$=74 эВ: результаты PA (сплошная кривая),
PA1B  (штриховая кривая), CCC \cite{LahmamBennani2008}
(пунктирная кривая) и экспериментальные данные
\cite{LahmamBennani2008} (кружки).}\label{FIGsigmaHe}
\end{figure}

Для тестирования метода в работе \cite{Serov2011} был проведен расчет
3ДС однократной ионизации гелия ударом быстрого электрона  при параметрах эксперимента
\cite{LahmamBennani2008}: энергия налетающего электрона $E_i=500$
эВ, энергия испущенного электрона $E_e=37$ эВ и $E_e=74$ эВ, в
третьем наборе экспериментальных данных \cite{LahmamBennani2008}
слишком большая энергия испущенного электрона $E_e=205$ эВ для
расчета с помощью PA. На рис. \ref{FIGsigmaHe} демонстрируются
результаты PA, PA1B, экспериментальные данные
\cite{LahmamBennani2008} и  результаты CCC из
\cite{LahmamBennani2008} (нормированные на наилучшее совпадение
двойных пиков с результатами PA, поскольку в \cite{LahmamBennani2008}
они даны в произвольных единицах). Видно, что наши результаты PA
очень хорошо совпадают как с
экспериментом, так и с CCC данными, и хотя результаты CCC несколько
лучше воспроизводят пик отдачи при $E_e=37$ эВ, при $E_e=74$ эВ
результаты PA неотличимы от CCC результатов.

\begin{figure}[t]
\includegraphics[angle=-90,width=0.5\textwidth]{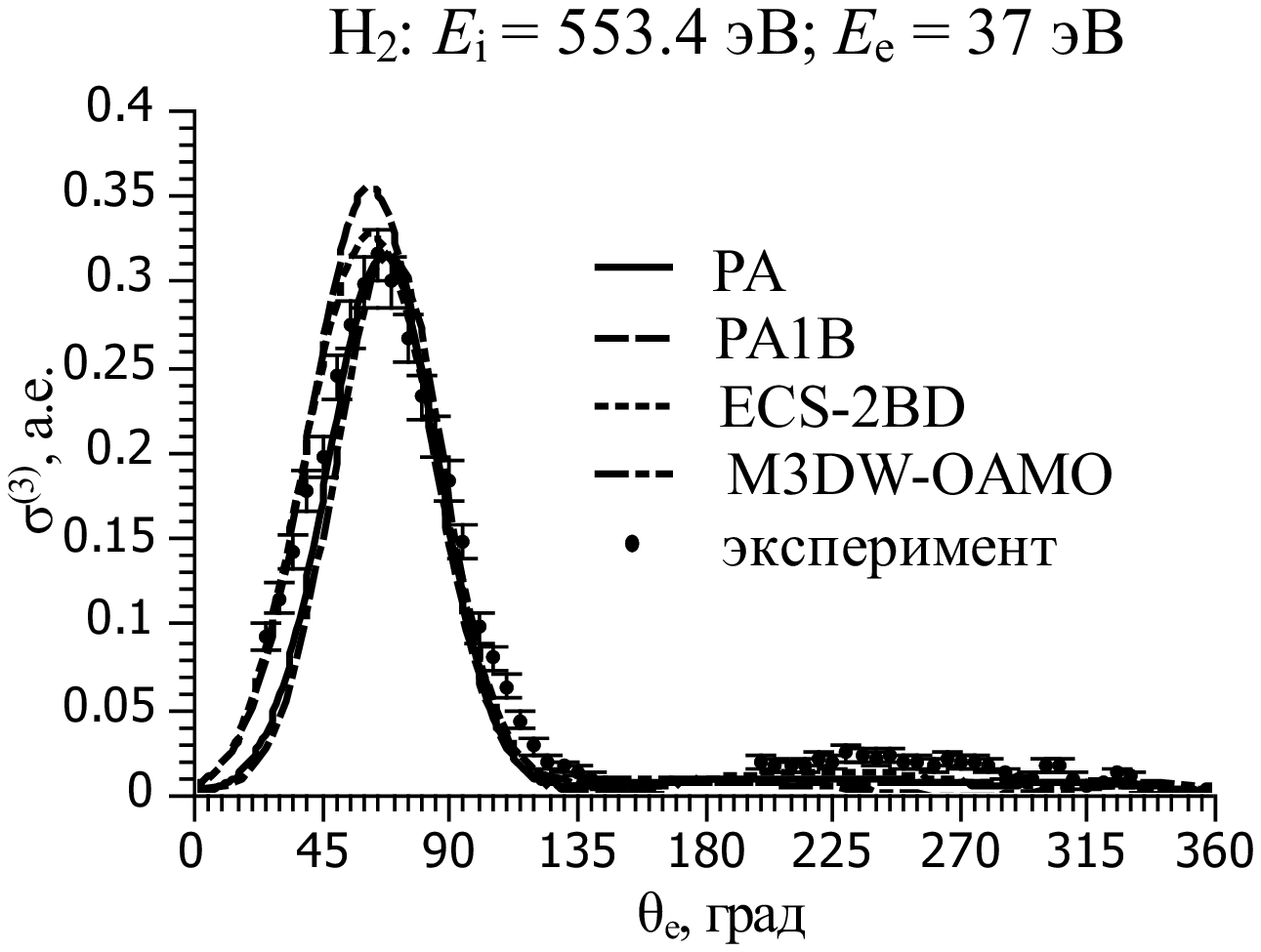}\\
\textit{а}\\
\includegraphics[angle=-90,width=0.5\textwidth]{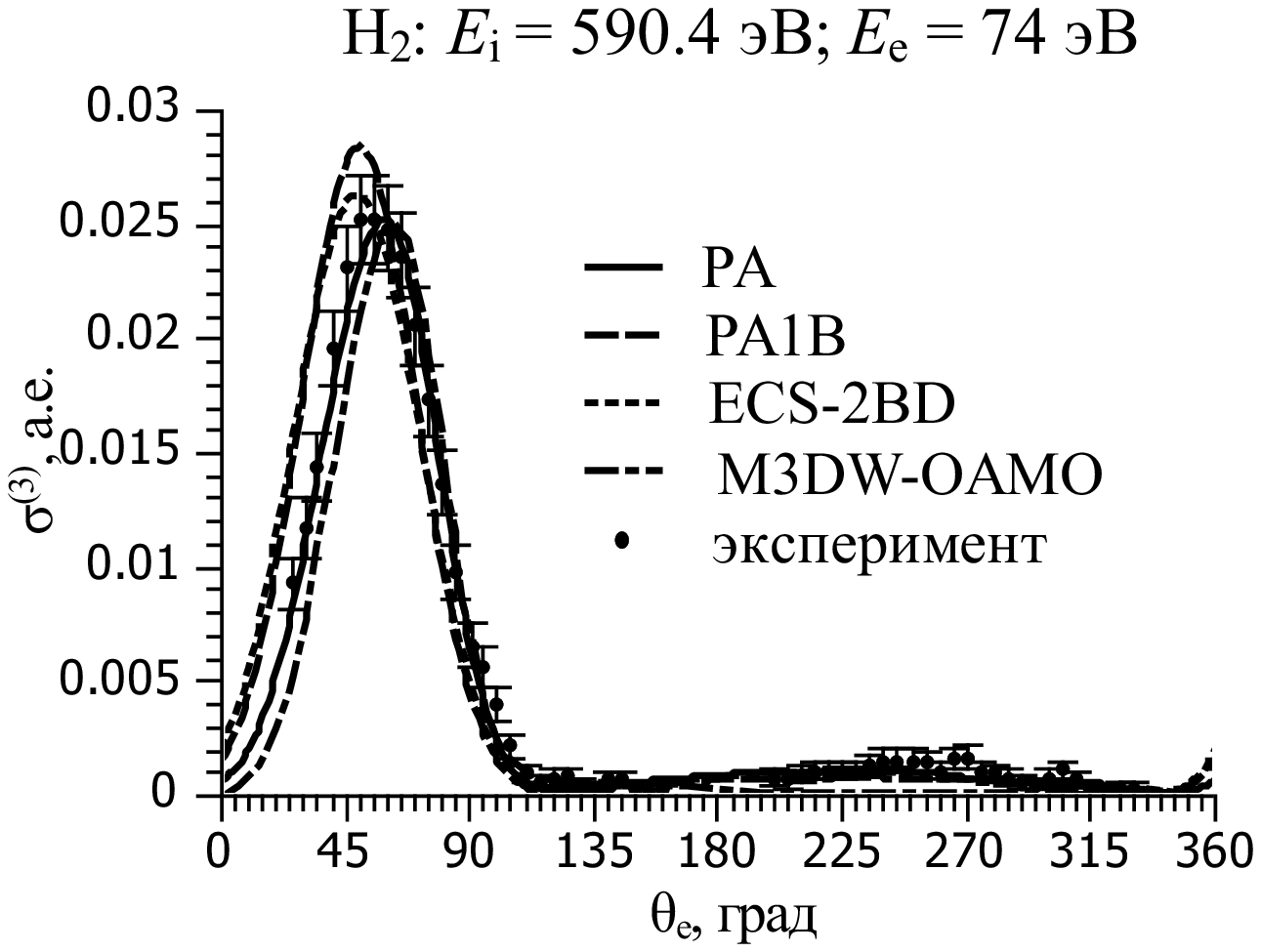}\\
\textit{б}\\
\caption{3ДС для процесса H$_2(e,2e)$  как функция угла вылета
$\theta_{e}$ для энергий вылетающего электрона (\textit{а}) $E_e$=37 эВ и (\textit{б})
$E_e$=74 эВ: результаты PA  (сплошная кривая),
PA1B  (штриховая кривая), ECS-2BD \cite{Serov2010} (пунктирная кривая),
M3DW-OAMO \cite{LahmamBennani2008} (штрих-пунктирная кривая) и экспериментальные данные \cite{LahmamBennani2008}
(кружки).}\label{FIGsigmaH2}
\end{figure}

Также для параметров эксперимента \cite{LahmamBennani2008}
в работе \cite{Serov2011} был проведен расчет 3ДС однократной ионизации неориентированной
молекулы H$_2$ ударом быстрого электрона.
3ДС рассчитывалось для набора фиксированных ориентаций молекулы,
а затем усреднялось по направлению молекулярной оси для получения
3ДС для неориентированной молекулы. На рис. \ref{FIGsigmaH2},
кроме результатов PA, PA1B и экспериментальных данных
\cite{LahmamBennani2008}, показаны результаты расчета методом
внешнего комплексного скейлинга с учетом второго борновского члена
в дипольном приближении (ECS-2BD) (см. раздел \ref{sectECS}), а так же
результаты M3DW-OAMO (Molecular 3-body Distorted Wave approximation with
Orientation Averaged Molecular Orbital, приближение трехтельной
молекулярной искажённой волны совместно с приближением усредненной по направлению молекулярной орбитали) \cite{LahmamBennani2008}. Экспериментальные
данные и результаты M3DW-OAMO нормированы на наилучшее совпадение
двойных пиков с результатами PA.
Совпадение наших результатов PA с экспериментальными лучше, чем у M3DW-OAMO как по
положению основных пиков, так и по величине пиков отдачи, хотя при
$E_e=37$ эВ наши результаты, как и для гелия, несколько недооценивают величину
пика отдачи. ECS-2BD результаты хорошо совпадают по величине с
результатами PA, но дают сильно заниженную величину углового
сдвига относительно направления вектора передачи импульса
$\mathbf{K}$. Поскольку в ECS-2BD учитывалась только бидипольная
компонента второго борновского члена, то есть линейный по
радиус-векторам обоих электронов мишени вклад в разложение второго
борновского члена, а в PA --- только взаимодействие
одного из электронов мишени, можно сделать вывод, что основной
вклад в угловое смещение вносят компоненты второго борновского
члена, зависящие от координат только одного из электронов. Это
также объясняет неспособность ECS-2BD воспроизвести форму углового
распределения двойной ионизации H$_2$ \cite{Serov2010}.

\begin{figure}[t]
\begin{center}
\parbox{0.48\textwidth}{\includegraphics[angle=-90,width=0.5\textwidth]{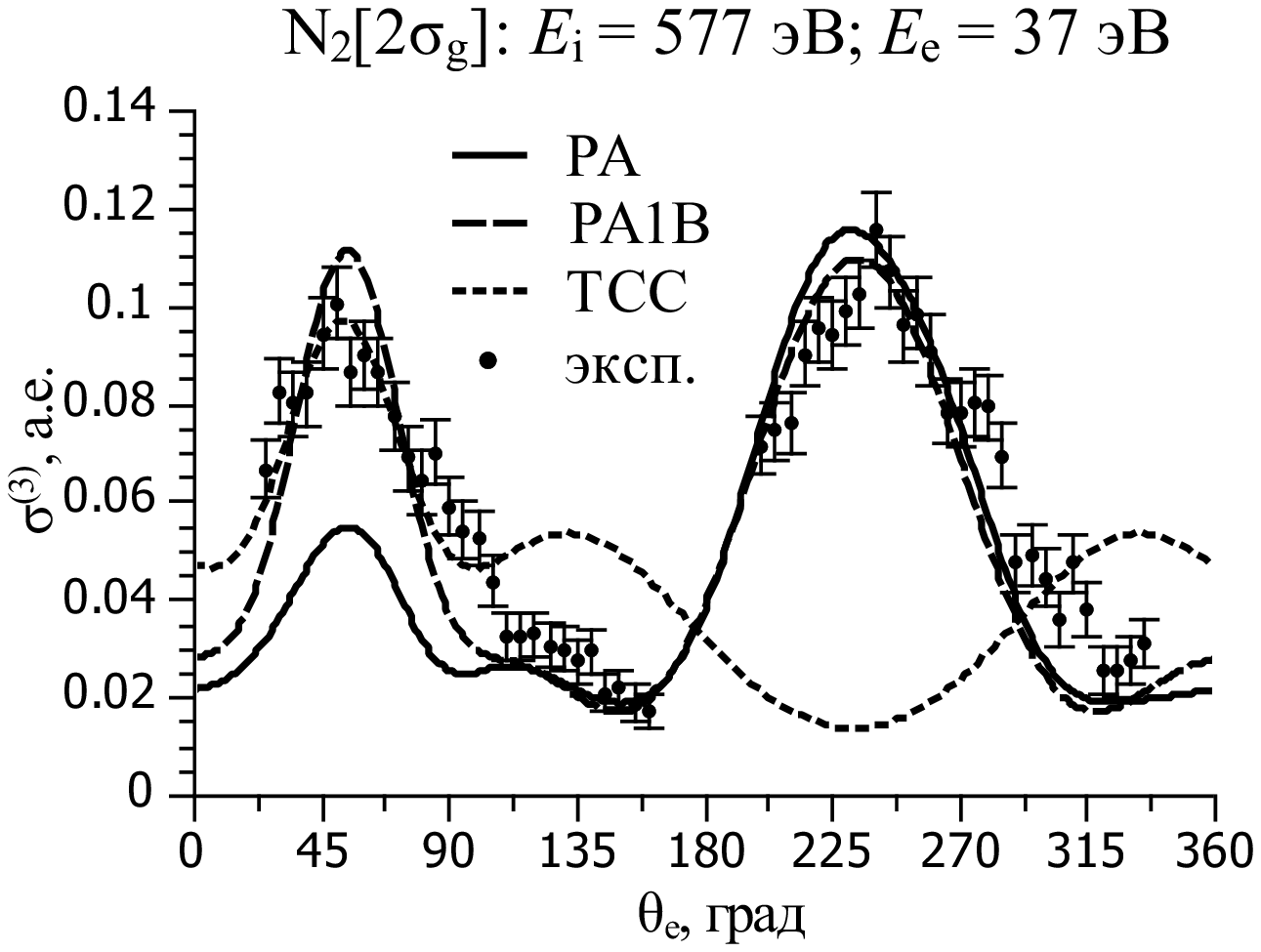}\\ \textit{а}}
\parbox{0.48\textwidth}{\includegraphics[angle=-90,width=0.5\textwidth]{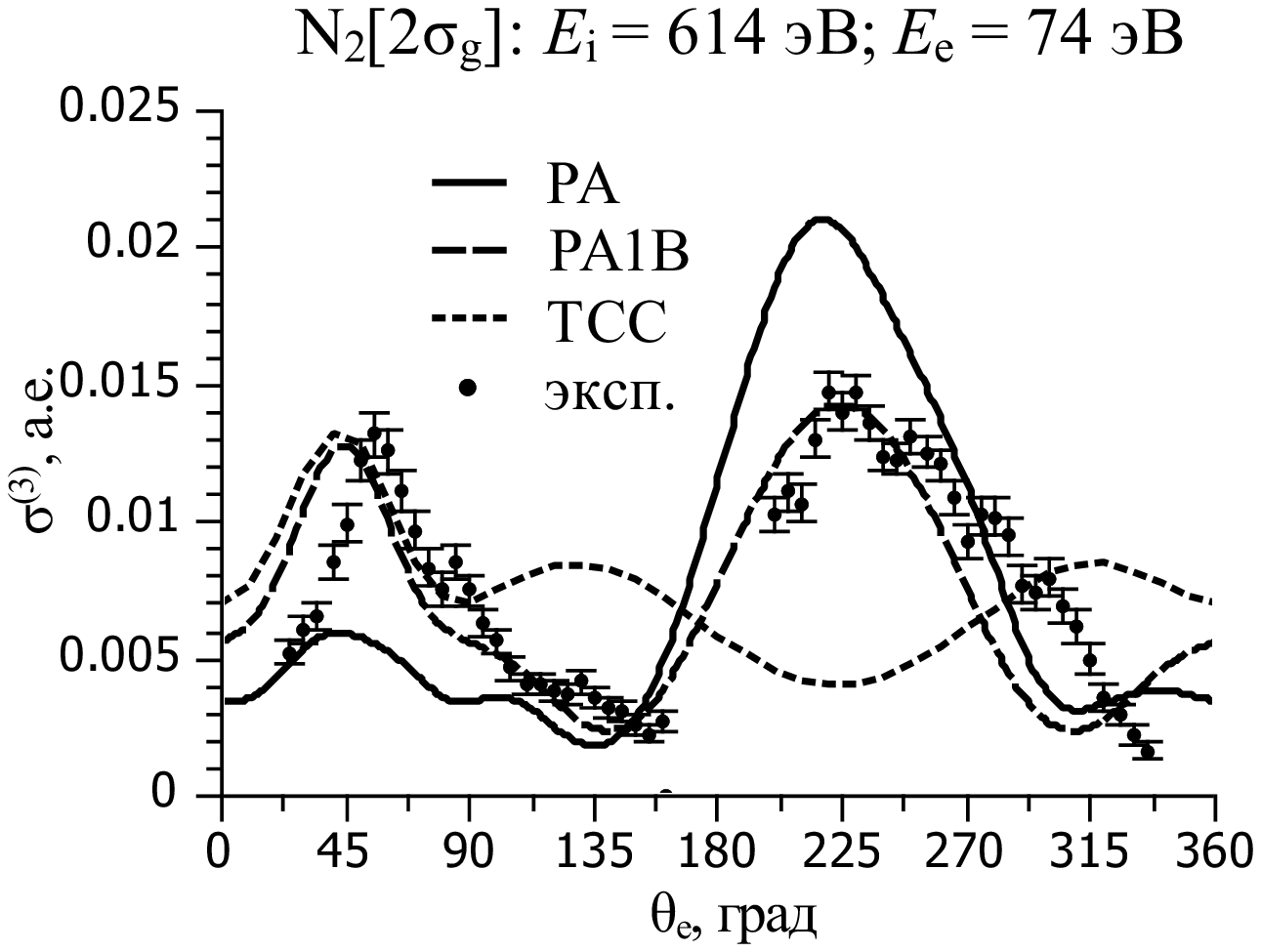}\\ \textit{б}}\\
\vspace{1cm}
\parbox{0.48\textwidth}{\includegraphics[angle=-90,width=0.5\textwidth]{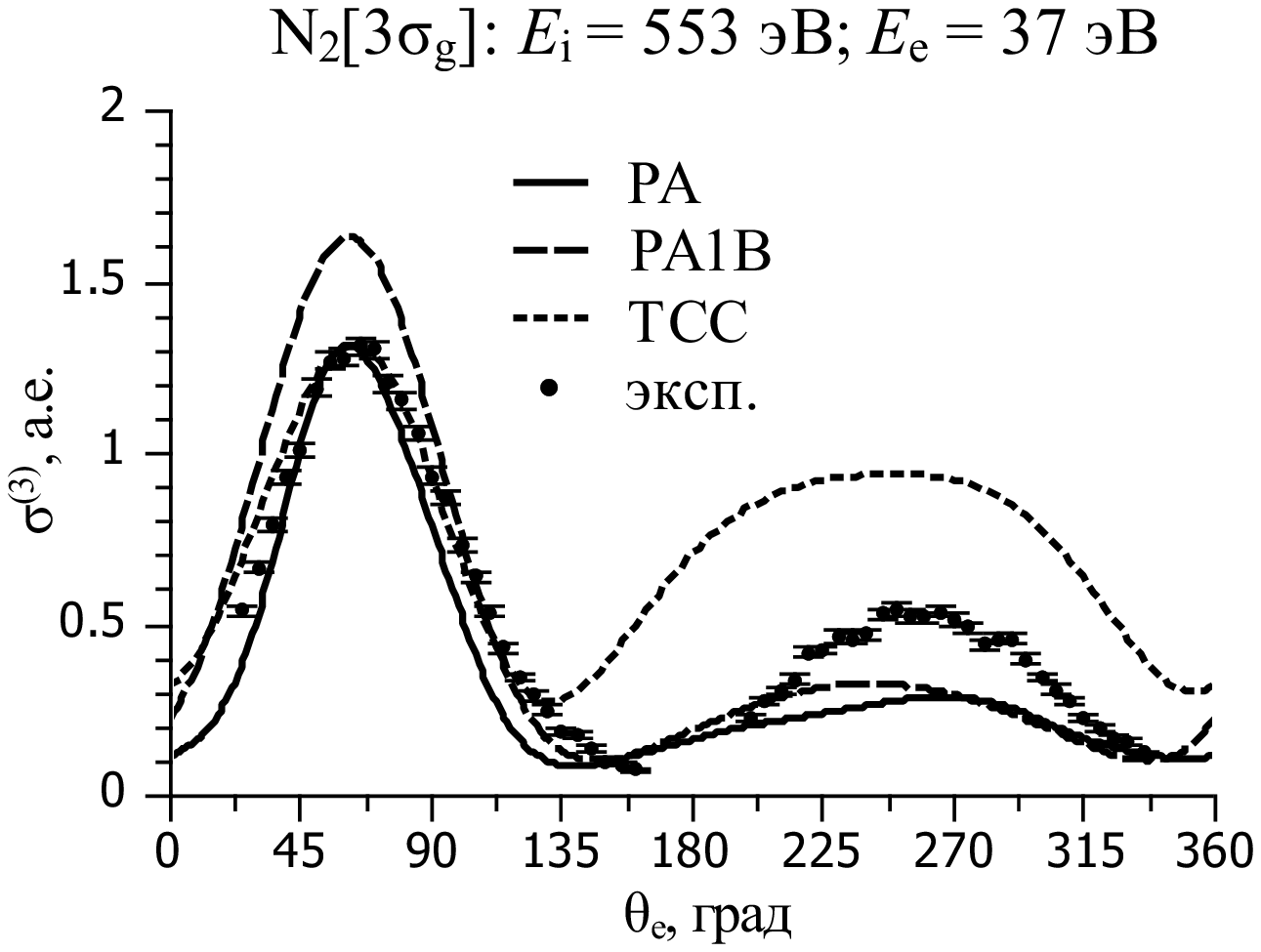}\\ \textit{в}}
\parbox{0.48\textwidth}{\includegraphics[angle=-90,width=0.5\textwidth]{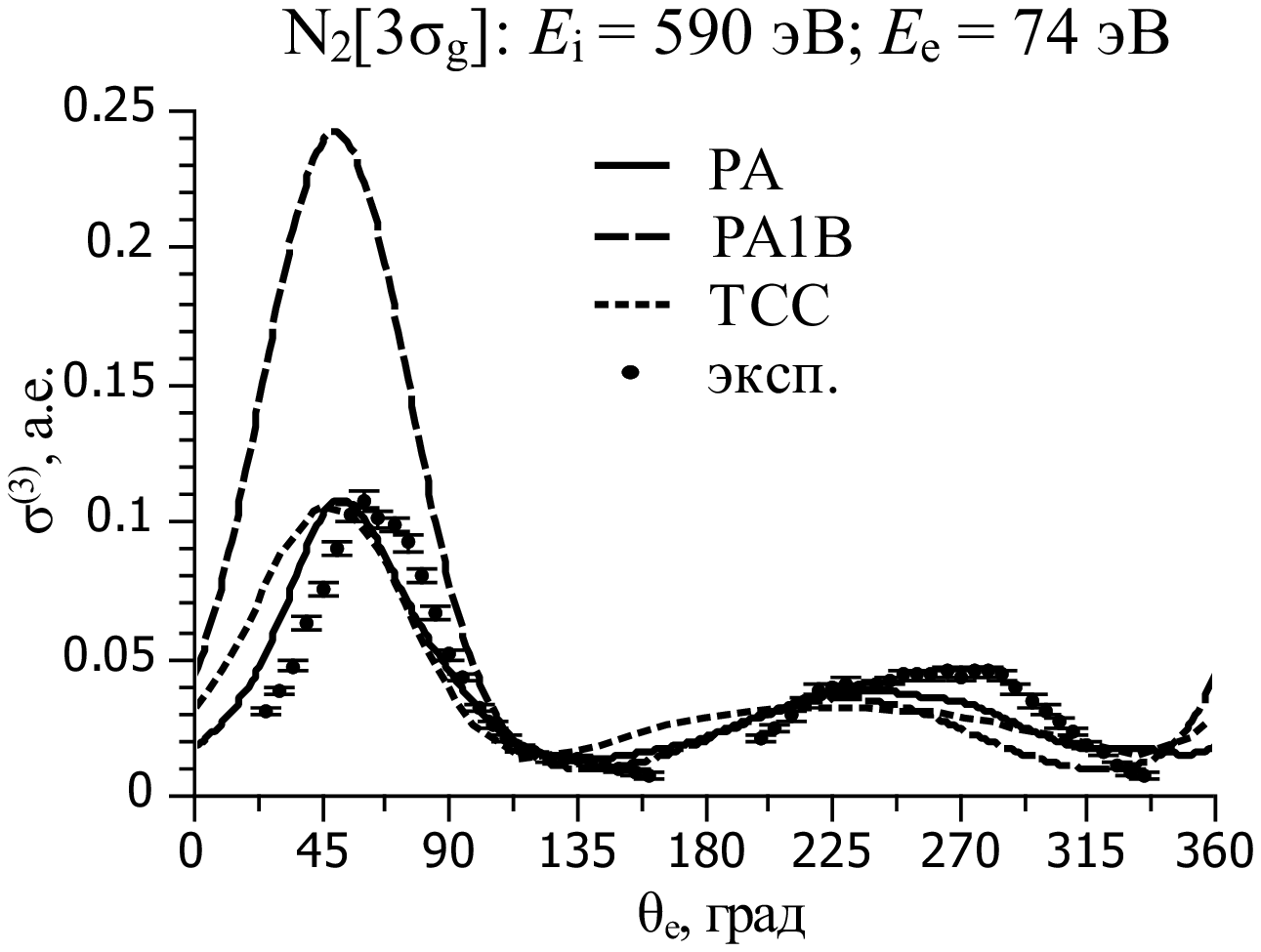}\\ \textit{г}}
\end{center}
\caption{3ДС для N$_2(e,2e)$ процесса на $2\sigma_g$ орбитали (\textit{а},\textit{б}) и внешних орбиталей (\textit{в},\textit{г}) как функция угла испускания $\theta_{e}$ для энергии рассеянного электрона $E_s=500$ эВ и угла рассеяния $\theta_s=-6^\circ$: PA (сплошная кривая) и PA1B (штриховая кривая). Также показаны результаты TCC-1B (пунктирная кривая) и экспериментальные данные (кружки).}\label{FIGsigma2sg}
\end{figure}

В качестве еще одного примера рассмотрим выполненный в
\cite{Serov2011} расчет 3ДС для неориентированной молекулы N$_2$ при
параметрах экспериментов \cite{LahmamBennani2007,LahmamBennani2009}.
В качестве функций начальных состояний орбиталей N$_2$
использовались вариационные функции \cite{Scherr1955}. На рис.
\ref{FIGsigma2sg}, \textit{а},\textit{б} показаны результаты PA,
PA1B и экспериментальные данные \cite{LahmamBennani2009}, а также
результаты TCC-1B \cite{LahmamBennani2008} для ионизации N$_2$ с
вырыванием электрона с внутренней $2\sigma_g$-оболочки. Поскольку
PA1B гораздо ближе к экспериментальным данным, чем PA,
экспериментальные данные и результаты TCC-1B нормированы на PA1B.
Можно предположить, что столь обескураживающий провал PA связан с
использованием приближения одного активного электрона и,
соответственно, с пренебрежением как изменением состояния остальных
электронов мишени при взаимодействии с налетающим электроном, так и
электрон-электронной корреляцией в мишени. Поскольку
последняя приводит к увеличению среднего
расстояния между электронами, она, очевидно, должна приводить к
уменьшению воздействия внешних оболочек на проходящий сквозь них
выбитый внутренний электрон (по сравнению с приближением
фиксированных внешних оболочек). На первый борновский член, по-видимому, этот эффект не оказывает значительного влияния, а для
высших борновских членов его неучет приводит к сильному завышению их
вклада, и поэтому PA1B оказался гораздо ближе к экспериментальным
данным, чем PA.

На рис. \ref{FIGsigma2sg}, \textit{в},\textit{г} показано 3ДС для ионизации N$_2$
с вырыванием электрона с внешних оболочек. Мы
рассчитали вклады в 3ДС от ионизации с $3\sigma_g$, $1\pi_u$ и
$2\sigma_u$--оболочек N$_2$ и просуммировали их с коэффициентами
1, 0.78 и 0.32, следуя \cite{LahmamBennani2009}. Здесь
экспериментальные данные и TCC-1B ремасштабированы на величину
двойного пика в PA. Наши результаты PA в данном случае очевидным
образом ближе к экспериментальным точкам, чем результаты PA1B и
TCC-1B, хотя PA заметно недооценивает величину пика отдачи для
$E_e=37$ эВ и угловой сдвиг основного пика при $E_e=74$ эВ. Вероятно, эти расхождения связаны с динамикой внешних оболочек
мишени.

\begin{figure}[t]
\begin{center}
\parbox{0.48\textwidth}{\includegraphics[width=0.5\textwidth]{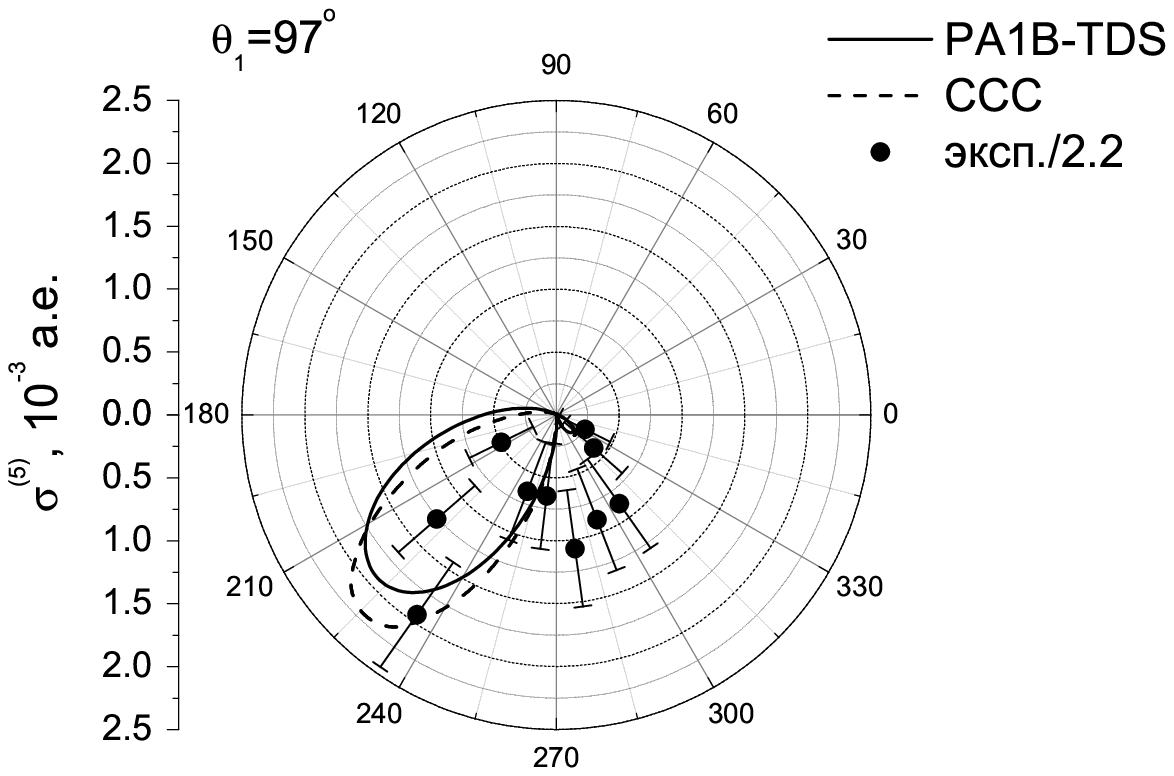}\\ \textit{а}}
\parbox{0.48\textwidth}{\includegraphics[width=0.5\textwidth]{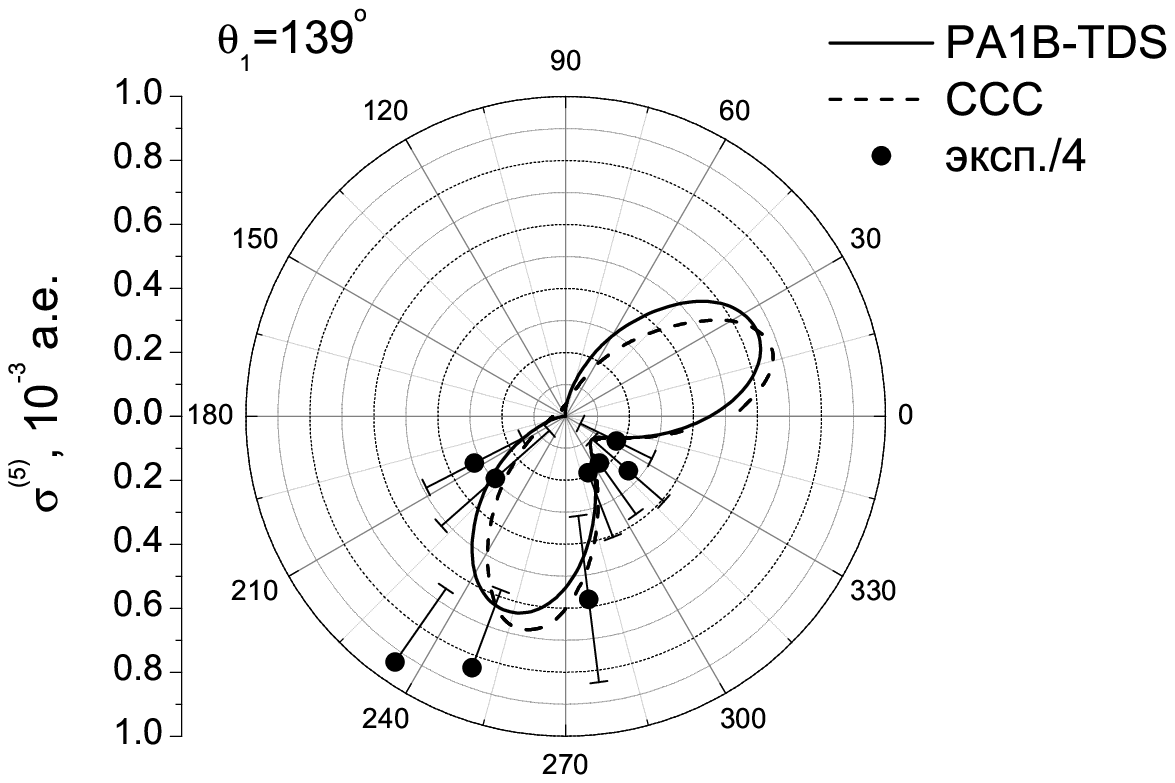}\\ \textit{б}}
\parbox{0.48\textwidth}{\includegraphics[width=0.5\textwidth]{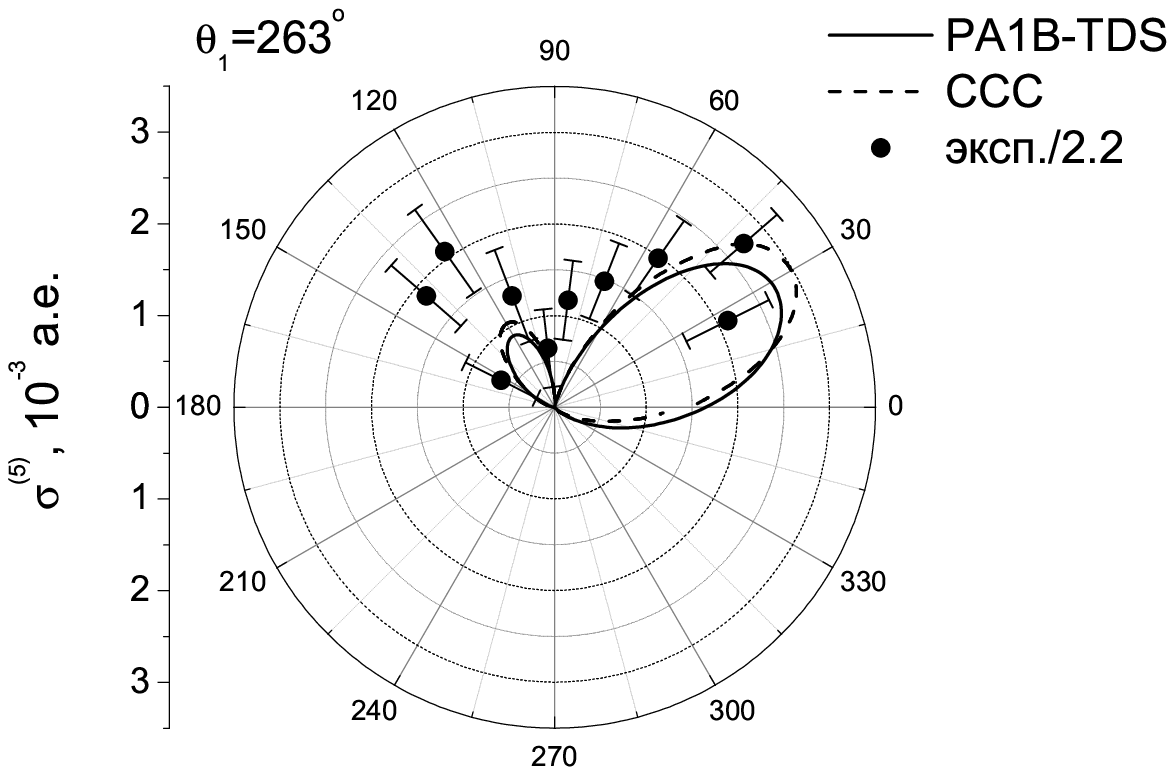}\\ \textit{в}}
\end{center}
 \caption{5ДС двойной ионизации атома He ударом быстрого электрона в зависимости
от $\theta_2$ при энергии налетающего электрона $E_i=5600$ эВ, угле рассеяния $\theta_s=0.45$, энергии вылетающих электронов $E_1=E_2$=10 эВ и углах вылета первого электрона а)$\theta_1=97^o$; б)$\theta_1=139^\circ$; в)$\theta_1=263^\circ$.
Результаты PA1B-TDS (сплошная кривая), CCC (штриховая кривая)
\cite{Kheifets2004} и экспериментальные данные
\cite{Lahmam-Bennani1999e3e}.}\label{e3e}
\end{figure}

Поскольку PA1B приводит к уравнению с числом размерностей, равным
числу степеней свободы мишени, для мишеней с малым числом электронов
оно может эффективно применяться и без одноэлектронного приближения
для мишени. В работе \cite{Serov2007} метод PA1B в комбинации с
методом сопутствующих координат (TDS, time-dependent scaling, см.
раздел \ref{expgridTDSE}) был использован для расчета двойной ионизации
атома гелия ударом быстрого электрона. Стимулом к этой работе явилось
наличие серьезных расхождений между экспериментальными данными
\cite{Lahmam-Bennani1999e3e} и сходящимся методом сильной связи
(CCC, Convergent close coupling) \cite{Kheifets2004}. Было высказано
предположение \cite{ShablovPopov2002}, что расхождение между
ССС-расчетами и экспериментом обусловлено неправильным
асимптотическим поведением функций двухэлектронного континуума,
даваемых ССС. На рис. \ref{e3e} показано многократное
дифференциальное сечение двойной ионизации гелия как функция угла
вылета одного электрона при фиксированном угле вылета другого
электрона. Наши результаты близки к результатам расчетом по методу
ССС \cite{Kheifets2004}  и в несколько раз меньше по величине, чем
экспериментальные данные \cite{Lahmam-Bennani1999e3e}. Поскольку
подход PA1B-TDS \cite{Serov2007} радикально отличается от CCC,
гипотеза о том, что расхождение между ССС-расчетами и экспериментами
обусловлено неправильным асимптотическим поведением ССС-функций
двухэлектронного континуума, не подтвердилась.

\section{Решение задачи рассеяния на основе формализма
 стационарного уравнения Шредингера с источником и комплексного скейлинга}\label{sectECS}

В работах \cite{Serov2009,Serov2010} описана вычислительная
процедура на основе формализма уравнения Шредингера с источником в
правой части (driven Schrodinger equation,
см. \cite{ReviewMcCurdy2004} и ссылки в этой работе) и внешнего
комплексного скейлинга (ВКС) \cite{Simon1979} в применении к
двойной ионизации фотонами \cite{ReviewMcCurdy2004} и электронами.

Комплексный скейлинг, то есть поворот радиальной координаты в
уравнении Шредингера в комплексную плоскость, первоначально был
предложен для исследования аналитических свойств S-матрицы
\cite{Regge1962}. Было показано, что при преобразовании
\[ r\to r e^{i\eta}\]
энергии состояний континуума поворачиваются в энергетической
плоскости на угол $2\eta$, энергии квазистационарных состояний
приобретают их истинную мнимую часть, а энергии стационарных
состояний не меняются. Все это связано с тем, что расходящаяся
волна после такого преобразования приобретает экспоненциально
падающую с радиусом амплитуду. Б. Саймон \cite{Simon1979}
предложил метод внешнего комплексного скейлинга (ECS) (exterior complex scaling),
который заключается в том, что поворот производится не во всей области
определения $r$, а только начиная с некоторой точки $r_{s}$
\[
r\to\left\{
 \begin{array}{ll}
 r,& r<r_{s}; \\
 r_{s}+e^{i\eta}(r-r_{s}),& r>r_{s},
 \end{array} \right. \eqno (I.5) \label{ECSdef}
\]
и показал, что при таком преобразовании спектр меняется так же,
как и при обычном комплексном скейлинге. Преимуществом ECS
является то, что при $r<r_{s}$ функция неизменна. Если при
решении ищется рассеянная волна, то, воспользовавшись ECS, можно
решать уравнение Шредингера с граничными условиями Дирихле. Это
снимает необходимость знания конкретного асимптотического вида
волновой функции.

Переход к формализму уравнения Шредингера с источником начнем с общего выражения
для амплитуды ионизации
\begin{eqnarray}
f(\mathbf{k})=\langle
\varphi_{\mathbf{k}}^{(-)}|\hat\mu|\varphi_{0}\rangle\label{trans_ampl},
\end{eqnarray}
где $\varphi_{0}$ представляет начальное состояние мишени, $\hat\mu$
--- оператор возмущения для данного процесса, ${\mathbf{k}}$ --- набор
импульсов вылетевших электронов и $\varphi_{\mathbf{k}}^{(-)}$ ---
волновая функция континуума мишени, построенная как сумма падающей волны и
входящей сферической волны, удовлетворяющая стационарному уравнению
Шредингера
\begin{eqnarray}
(\hat{H}-E)\varphi_{\mathbf{k}}^{(-)} = 0.
\end{eqnarray}
Гамильтониан мишени можно представить как $\hat{H}=\hat{H}_0+V$. Здесь
$V(\mathbf{r}_1,...,\mathbf{r}_{N_e})=\sum_{i\neq\,j}^{N_e}\frac{1}{|\mathbf{r}_i-\mathbf{r}_j|}$ ---
потенциал межэлектронного взаимодействия,  $N_e$ --- число
электронов мишени. Оператор
$\hat{H}_0=\sum_{\alpha=1}^{N_e}[-\frac{1}{2}\nabla_\alpha^2+U_\alpha(\mathbf{r}_\alpha)]$
представляет гамильтониан электронов в поле ядер, описываемом
потенциалом $U_\alpha(\mathbf{r}_\alpha)$, $\mathbf{r}_\alpha$ --- координаты электронов мишени. Пусть пробная функция
$\chi_{\mathbf{k}}^{(-)}$ удовлетворяет уравнению
\begin{eqnarray}
(\hat{H}_0-E)\chi_{\mathbf{k}}^{(-)}=0.\label{chi_eq}
\end{eqnarray}
Пользуясь уравнением Липпмана-Швингера
\begin{eqnarray*}
\chi_{\mathbf{k}}^{(-)}&=&\varphi_{\mathbf{k}}^{(-)}+\frac{1}{E-\hat{H}-i\epsilon}(-V)\chi_{\mathbf{k}}^{(-)}
\end{eqnarray*}
и заменяя $\varphi_{\mathbf{k}}^{(-)}$, находим
\begin{eqnarray*}
f(\mathbf{k})=\langle
[1+(E-\hat{H}-i\epsilon)^{-1}V]\chi_{\mathbf{k}}^{(-)}|\hat\mu|\varphi_{0}\rangle=\langle
\chi_{\mathbf{k}}^{(-)}|[1+V(E-\hat{H}+i\epsilon)^{-1}]\hat\mu|\varphi_{0}\rangle.
\end{eqnarray*}
Используя тождество
 $
 1+V(E-\hat{H}+i\epsilon)^{-1}\equiv(E-\hat{H}_0)(E-\hat{H}+i\epsilon)^{-1},
 $
получаем следующее выражение для амплитуды перехода
\begin{eqnarray}
f(\mathbf{k})=\langle\chi_{\mathbf{k}}^{(-)}|E-\hat{H}_0|\psi^{(+)}\rangle,
\label{temp_f}
\end{eqnarray}
где  ``волновая функция первого порядка'' $\psi^{(+)}$
удовлетворяет уравнению Шредингера с источником в правой части
\begin{eqnarray}
(\hat{H}-E)\psi^{(+)} = -\hat\mu\varphi_{0}\label{drivenSchr}
\end{eqnarray}
с граничными условиями уходящей волны.

Теперь рассмотрим область $\mathcal{V}\in\Re^{3n_e}$ в
конфигурационном пространстве, размерность которого определяется числом испущенных электронов  $n_e$. Гамильтониан всего пространства можно
расщепить на два эрмитовых оператора --- внутренний
\begin{eqnarray*}
 \hat{H}_{in}=\left\{
 \begin{array}{ll}
 \hat{H}_0+\hat{L}_S, & \mathbf{r}\in\mathcal{V};\\
 0,& \mathbf{r}\not\in\mathcal{V},
 \end{array}
 \right.
\end{eqnarray*}
  и внешний
\begin{eqnarray*}
 \hat{H}_{out}=\left\{
 \begin{array}{ll}
 0, & \mathbf{r}\in\mathcal{V};\\
 \hat{H}_0-\hat{L}_S,& \mathbf{r}\not\in\mathcal{V}.
 \end{array}
 \right.
\end{eqnarray*}
Здесь $\mathbf{r}=\{\mathbf{r}_\alpha\}_{\alpha=1}^{n_e}$ --- набор
радиус-векторов электронов. Оператор Блоха $\hat{L}_S$
удовлетворяет формуле Грина
\begin{eqnarray*}
\int_{\Re^{3n_e}}\varphi \hat{L}_S\psi
\,\mathrm{d}V=\frac{1}{2}\oint_\mathcal{S}
\varphi(\mathbf{n}_S\nabla)\psi \,\mathrm{d}\mathcal{S},
\end{eqnarray*}
где $\mathcal{S}$ представляет поверхность, ограничивающую объем
$\mathcal{V}$. Используя такое разделение, перепишем амплитуду
перехода (\ref{temp_f}) в виде
\begin{eqnarray*}
f(\mathbf{k})&=&
\langle\chi_{\mathbf{k}}^{(-)}|E-\hat{H}_0-\hat{L}_S|\psi^{(+)}\rangle_{\mathbf{r}\in\mathcal{V}}
+\langle\chi_{\mathbf{k}}^{(-)}|E-\hat{H}_0+\hat{L}_S|\psi^{(+)}\rangle_{\mathbf{r}\not\in\mathcal{V}}.
\end{eqnarray*}

Используя свойство эрмитовости и (\ref{chi_eq}),  получаем для
внутреннего интеграла
\begin{eqnarray*}
 \langle\chi_{\mathbf{k}}^{(-)}|E-\hat{H}_0-\hat{L}_S|\psi^{(+)} \rangle_{\mathbf{r}\in\mathcal{V}}= \langle\psi^{(+)}|E-\hat{H}_0-\hat{L}_S|\chi_{\mathbf{k}}^{(-)}\rangle_{\mathbf{r}\in\mathcal{V}}^*=
 -\langle\psi^{(+)}|\hat{L}_S|\chi_{\mathbf{k}}^{(-)}\rangle^*
\end{eqnarray*}
и
\begin{eqnarray*}
 \langle\chi_{\mathbf{k}}^{(-)}|E-\hat{H}+V+\hat{L}_S|\psi^{(+)} \rangle_{\mathbf{r}\not\in\mathcal{V}}=
 \langle\chi_{\mathbf{k}}^{(-)}|\hat{L}_S|\psi^{(+)}\rangle
 +\langle\chi_{\mathbf{k}}^{(-)}|V|\psi^{(+)} \rangle_{\mathbf{r}\not\in\mathcal{V}}
\end{eqnarray*}
для внешнего. Это позволяет записать формулу для амплитуды в виде
\begin{eqnarray}
f(\mathbf{k})&=&-\langle\psi^{(+)}|\hat{L}_S|\chi_{\mathbf{k}}^{(-)}\rangle^*
 + \langle\chi_{\mathbf{k}}^{(-)}|\hat{L}_S|\psi^{(+)}\rangle+\langle \chi_{\mathbf{k}}^{(-)}|V|\psi^{(+)}\rangle_{\mathbf{r}\not\in\mathcal{V}}.\label{f_via_L}
\end{eqnarray}

Теперь, предполагая, что объем  $\mathcal{V}$ выбран таким
образом, что $V\rightarrow 0$ для $\mathbf{r}\not\in\mathcal{V}$,
 можем записать
\begin{eqnarray}
f(\mathbf{k})&=&i\oint_\mathcal{S} \left(\mathbf{n}_S\cdot
\mathbf{j}[\psi^{(+)},\chi_{\mathbf{k}}^{(-)}]\right)
\mathrm{d}\mathcal{S}, \label{f_via_flux}
\end{eqnarray}
где
\begin{eqnarray*}
\mathbf{j}[\psi,\varphi]=\frac{i}{2}\left[\psi\nabla\varphi^*-\varphi^*\nabla\psi\right]
\end{eqnarray*}
представляет поток вероятности \cite{Selin1999}.

В случае обычного кулоновского взаимодействия между электронами
$V=1/|\mathbf{r}_1-\mathbf{r}_2|$ третий член в (\ref{f_via_L})
расходится, однако выражение (\ref{f_via_flux}) и в этом случае
 работает \cite{McCurdy2003,McCurdy2001} и дает амплитуды, отличающиеся от точных только фазовым множителем. Этот множитель зависит от объема
$\mathcal{V}$, но не влияет на какие-либо физические наблюдаемые.

Рассмотрим двухатомную молекулу с фиксированным межъядерным
вектором $\mathbf{R}=R\mathbf{n}_R$. Потенциал притяжения ядер в
случае H$_2^+$ и H$_2$ дается выражением
\[
U(\mathbf{r})=-\frac{1}{|\mathbf{r}-\frac{\mathbf{R}}{2}|}-\frac{1}{|\mathbf{r}+\frac{\mathbf{R}}{2}|}
.
\]
Для описания электронов в такой молекуле логичнее всего использовать
конфокальные эллиптические (вытянутые сфероидальные) координаты
\[
\xi=\frac{|\mathbf{r}-\frac{\mathbf{R}}{2}|+|\mathbf{r}+\frac{\mathbf{R}}{2}|}{R}
\in [1,\infty );\,\,\,
\eta=\frac{|\mathbf{r}-\frac{\mathbf{R}}{2}|-|\mathbf{r}+\frac{\mathbf{R}}{2}|}{R}
\in [-1,1];\,\,\, \phi \in [0,2\pi).
\]
Стационарное уравнение Шредингера для одноэлектронных систем с
двумя кулоновскими центрами, таких как H$_2^+$, в этой системе
координат допускает разделение переменных, что является следствием
его изначальной двухцентровой симметрии. Поэтому, в
отличие от \cite{McCurdy2004,McCurdy2005,McCurdy2006}, где для молекулы H$_2$
использовались сферические координаты, в работах \cite{Serov2009,Serov2010,Serov2012}
применялись сфероидальные. Преимущество такого выбора в том, что сингулярные точки
двухцентрового потенциала расположены на границе области определения $\xi=1$,
что снимает проблему наличия разрыва первой производной волновой функции
в точках сингулярности потенциала. Чтобы подчеркнуть отличие от метода ECS
\cite{McCurdy2004,McCurdy2005,McCurdy2006}, состоящее в
выборе координатной системы,  будем называть метод, использованный в \cite{Serov2009,Serov2010,Serov2012},
 внешним комплексным скейлингом в вытянутых сфероидальных координатах
 (prolate spheroidal exterior complex scaling, PSECS).
 Детали численной схемы для решения шестимерного стационарного уравнения
 в сфероидальных координатах изложены в Приложении \ref{sectH2solution}.

\subsection{Пробные функции для вычисления амплитуд ионизации}
Для использования формулы (\ref{f_via_flux}) необходима пробная функция,
асимптотический вид которой близок к асимптотическому виду волновой функции
в соответствующем канале реакции. Пробная функция для простой ионизации
с образованием $H_2^+$ в состоянии $(nlm)$ дается выражением
\begin{eqnarray}
\chi_{\mathbf{k}nlm}^{(-)}(\mathbf{r}_1,\mathbf{r}_2)=\chi_{\mathbf{k}}^{(-)}(\mathbf{r}_1)\varphi_{nlm}(\mathbf{r}_2),\label{SItestFunction}
\end{eqnarray}
где $\varphi_{nlm}(\mathbf{r})$ представляет волновую функцию
связанного электрона в ионе $H_2^+$, а
$\chi_{\mathbf{k}}^{(-)}(\mathbf{r})$ --- двухцентровую кулоновскую волновую
функцию континуума для импульса $\mathbf{k}$ и экранированного
заряда $Z_+=1$. Естественным выбором поверхности $\mathcal{S}$ в
(\ref{f_via_flux}) является сфероид, определяемый уравнением
$\xi=\xi_S$. Компонента плотности потока вероятности вдоль
нормального вектора поверхности $\mathrm{d}\mathcal{S}$ приобретает вид
\begin{eqnarray*}
j_{\xi}\mathrm{d}\mathcal{S}=\frac{R}{4}(\xi_S^2-1)\left.\left[ \chi_{\mathbf{k}}^{(-)*}\frac{\partial\psi_{\mathbf{k}nlm}}{\partial\xi_1}-\psi_{\mathbf{k}nlm}\frac{\partial\chi_{\mathbf{k}}^{(-)*}}{\partial\xi_1}
 \right]\right|_{\xi_1=\xi_S}\,\mathrm{d}\eta_1\,\mathrm{d}\phi_1,
\end{eqnarray*}
где
$\psi_{\mathbf{k}nlm}(\mathbf{r}_1)=\langle\varphi_{nlm}(\mathbf{r}_2)|\psi^{(+)}(\mathbf{r}_1,\mathbf{r}_2)\rangle$.

Двухцентровая кулоновская волновая функция континуума может быть выражена через сумму
сфероидальных парциальных волн
\begin{eqnarray}
\chi_{\mathbf{k}}^{(-)}(\mathbf{r})=(2\pi)^{3/2}4\pi\sum_{lm}\Upsilon_{klm}^*(\cos\theta_k,\phi_k)
i^{l}e^{-i\delta_{klm}}T_{ml}\left(c,\xi\right)\Upsilon_{klm}(\eta,\phi),
\end{eqnarray}
где $c=kR/2$, $T_{ml}(c,\xi)$ --- радиальные кулоновские сфероидальные
функции \cite{Komarov1976}, $\delta_{klm}$ --- фазовые сдвиги, а  ``сфероидальные гармоники''
\begin{equation}
\Upsilon_{klm}(\eta,\varphi)=S_{ml}\left(c,\eta\right)
\frac{\exp(im\varphi)}{\sqrt{2\pi}};\qquad\Upsilon_{0lm}(\cos\theta,\varphi)=Y_{lm}(\theta,\varphi),\label{Upsdef}
\end{equation}
определяются через угловые сфероидальные функции
$S_{ml}\left(c,\eta\right)$ \cite{Komarov1976}.
В свою очередь, сфероидальные функции $T_{ml}(c,\xi)$ и $S_{ml}(c,\eta)$ получаются численным
решением разделенных уравнений
\begin{eqnarray*}
 \left[ \frac{d}{d\xi }(\xi^{2}-1)\frac{d}{d\xi }+RZ_{+}\xi
-\frac{m^{2}}{\xi^{2}-1}+c^{2}\xi^{2}+A_{ml}(c)\right] T_{ml}(c,\xi)&=&0, \\
 \left[ \frac{d}{d\eta }(1-\eta^{2})\frac{d}{d\eta }
-\frac{m^{2}}{1-\eta^{2}}-c^{2}\eta ^{2}-A_{ml}(c)\right]
S_{ml}(c,\eta)&=&0
\end{eqnarray*}
с использованием b-сплайнов \cite{deBoor2001} и разложения по полиномам Лежандра,
соответственно, с той же сеткой и размером базиса, что и в
(\ref{psi_expansion})(здесь $A_{ml}$ - константа разделения).

При расчете двойной ионизации в качестве поверхности $\mathcal{S}$
использовался гиперсфероид, удовлетворяющий уравнению
$(\xi_1-1)^2+(\xi_2-1)^2=(\xi_S-1)^2$. В качестве пробной функции
для двойной ионизации проще всего взять произведение функций
одноэлектронного континуума
\begin{eqnarray}
\chi_{\mathbf{k}_1\mathbf{k}_2}^{(-)}(\mathbf{r}_1,\mathbf{r}_2)=\chi_{\mathbf{k}_1}^{(-)}(\mathbf{r}_1)\chi_{\mathbf{k}_2}^{(-)}(\mathbf{r}_2).\label{DItestFunction}
\end{eqnarray}
Если пробная функция двойной ионизации не ортогональна функции, описывающей однократную ионизацию,
то в расчитанной с помощью формулы (\ref{f_via_flux}) амплитуде двойной ионизации появляется
паразитный вклад от однократной ионизации, крайне медленно сходящийся к нулю с увеличением радиуса
поверхности, на которой вычисляются амплитуды. Чтобы этого избежать,
обе функции континуума $\chi_{\mathbf{k}_{1,2}}^{(-)}(\mathbf{r})$ в (\ref{DItestFunction})
должны быть ортогональны всем волновым функциям $\varphi_{nlm}(\mathbf{r})$
связанных состояний электрона однократно-ионизированного иона.
Для этого, согласно \cite{ReviewMcCurdy2004},  функция $\chi_{\mathbf{k}}^{(-)}(\mathbf{r})$
должна быть собственной функцией того же гамильтониана, собственной функцией которого является
волновая функция $\varphi_{nlm}(\mathbf{r})$, то есть, в случае молекулы водорода,
нужно использовать кулоновскую сфероидальную функцию $\chi_{\mathbf{k}}^{(-)}(\mathbf{r})$
при $Z_+=2$.

\subsection{Оператор возмущения для взаимодействия с быстрым налетающим электроном}
Для начала выпишем оператор возмущения для однофотонной ионизации под
действием периодического внешнего поля бесконечной длительности в координатной калибровке
\begin{eqnarray}
\hat\mu
=\sum_{\alpha=1}^{n_e}\mathbf{e}\cdot\mathbf{r}_{\alpha},\label{lgauge}
\end{eqnarray}
здесь $\mathbf{e}$  --- вектор поляризации падающего излучения.

 Для взаимодействия двухэлектронной
двухцентровой молекулы с быстрым налетающим электроном борновский
член первого порядка оператора
возбуждения записывается в виде \cite{Serov2009}
\begin{eqnarray}
\hat{\mu}_{1B}&=&-\frac{1}{2\pi}\langle\mathbf{k}_s|V|\mathbf{k}_i\rangle=
-\frac{2}{K^2}\left[e^{i\mathbf{K}\cdot\mathbf{r}_1}+e^{i\mathbf{K}\cdot\mathbf{r}_2}-e^{i\mathbf{K}\cdot\mathbf{R}/2}-e^{-i\mathbf{K}\cdot\mathbf{R}/2}\right]
, \label{FBexp}
\end{eqnarray}
где ${\mathbf{k}_i}$ --- импульс налетающего электрона, ${\mathbf{k}_s}$
--- импульс рассеянного электрона, а ${\mathbf{K}}$ --- импульс,
передаваемый мишени от налетающего электрона с радиус-вектором ${\mathbf{r}_0}$,
${\mathbf{r}_{1,2}}$ --- радиус-векторы молекулярных электронов,
вектор $\mathbf{R}$ определяет межъядерное расстояние и ориентацию
молекулярной оси,
$|\mathbf{k}\rangle\equiv|\exp(i\mathbf{k}\cdot\mathbf{r}_0)\rangle$.

Первый борновский член описывает однократное взаимодействие налетающего электрона с мишенью. При этом двойная ионизация может произойти посредством двух возможных механизмов --- ``стряхивания'' и ``выбивания'' \cite{Schneideretal2002}.  Оба этих процесса начинаются с выбивания одного из электронов мишени налетающим электроном. Второй электрон после этого может вылететь вследствие резкого изменения действующего на него эффективного потенциала (``стряхивание'') или быть выбитым первым электроном (``выбивание''). Но, как было показано в \cite{Serov2009}, при энергиях налетающего электрона $E_i<1$ кэВ в двойную ионизацию вносит существенный вклад процесс последовательной двойной ионизации, который не принимается во внимание в первом борновском приближении. Этот процесс заключается в том, что налетающий электрон последовательно выбивает каждый из электронов мишени. Для его учёта в оператор возмущения должен быть включён и второй борновский член. Вопрос применимости второго борновского приближения к проблеме
ударной ионизации подробно рассмотрен в \cite{Shablov2010}.

Второй борновский член в амплитуде перехода дается выражением \cite{Serov2010}
\begin{eqnarray*}
f_{2B}=-\frac{1}{2\pi}\sum_n\int\frac{d\mathbf{k}}{(2\pi)^3}\frac{\langle\mathbf{k}_sf|V|\mathbf{k}n\rangle\langle\mathbf{k}n|V|\mathbf{k}_ii\rangle}{k_i^2/2+E_0-k^2/2-E_n+i\epsilon},
\end{eqnarray*}
где
$|\mathbf{k}_ii\rangle\equiv|e^{i\mathbf{k}_i\cdot\mathbf{r}_0}\psi_i(\mathbf{r}_1,\mathbf{r}_2)\rangle$,
$|\mathbf{k}_sf\rangle\equiv|e^{i\mathbf{k}_s\cdot\mathbf{r}_0}\psi_f(\mathbf{r}_1,\mathbf{r}_2)\rangle$
и
$|\mathbf{k}n\rangle\equiv|e^{i\mathbf{k}\cdot\mathbf{r}_0}\psi_n(\mathbf{r}_1,\mathbf{r}_2)\rangle$
представляют начальное, конечное и промежуточное состояния
системы. Здесь $E_0$ и $E_n$ --- энергии начального и
промежуточного состояний мишени.  Для упрощения вычислений в \cite{Serov2010}
использовано приближение замыкания для функции Грина (см.
\cite{Walters1984} и ссылки в этой работе), которое состоит в
замене  $E_{n}$ в знаменателе на некоторое $E_{t}$, одинаковое для
всех каналов. Это позволяет использовать соотношение полноты
$\sum_n\psi_n^*(\mathbf{r})\psi_n(\mathbf{r}')=\delta(\mathbf{r}-\mathbf{r}')$
и получить выражение
\begin{eqnarray*}
f_{2B}=-\frac{1}{2\pi}\langle\psi_f|\int\frac{d\mathbf{k}}{(2\pi)^3}\frac{\langle\mathbf{k}_s|V|\mathbf{k}
\rangle\langle\mathbf{k}|V|\mathbf{k}_i\rangle}{k_i^2/2+E_0-k^2/2-E_{t}+i\epsilon}|\psi_i\rangle.
\end{eqnarray*}

В результате сравнения с общим выражением для
амплитуды $f_{2B}=\langle\psi_f|\hat{\mu}_{2B}|\psi_i\rangle$
можно записать оператор возмущения для второго борновского члена как
\begin{eqnarray}
\hat{\mu}_{2B}=-\frac{1}{2\pi}\int\frac{d\mathbf{k}}{(2\pi)^3}\frac{\langle\mathbf{k}_s|V|\mathbf{k}\rangle\langle\mathbf{k}|V|\mathbf{k}_i\rangle}{k_i^2/2+E_0-k^2/2-E_{t}+i\epsilon}.
\label{mu2Bclosure}
\end{eqnarray}

В рамках приближения замыкания результат не должен быть
чувствителен к конкретному выбору $E_t$, если $E_t$ близко к энергии доминирующего
промежуточного канала. В работе \cite{Serov2010} это было проверено путем вычисления
многократного дифференциального сечения (МДС) для нескольких
значений $E_t\in (E_0,E_f)$, где $E_f$  ---  энергия конечного
состояния мишени. При этом было обнаружена независимость МДС
от $E_t$, что оправдывает использование приближения замыкания.

Для расчетов \cite{Serov2010} использовалось выражение $E_t=E_0+k_i(k_i-k_s)/2\simeq (E_0+E_f)/2$,
 предложенное в \cite{Kheifets2004}.При ударной ионизации быстрым электроном потенциальная энергия
взаимодействия между налетающим электроном и мишенью H$_2$ в
атомных единицах дается выражением
\begin{eqnarray*}
V(\mathbf{r}_0,\mathbf{r}_1,\mathbf{r}_2)&=&\frac{1}{|\mathbf{r}_1-\mathbf{r}_0|}
+\frac{1}{|\mathbf{r}_2-\mathbf{r}_0|}-\frac{1}{|\mathbf{R}/2-\mathbf{r}_0|}-\frac{1}{|\mathbf{R}/2+\mathbf{r}_0|}.
\end{eqnarray*}
С помощью аналитического выражения для матричного элемента $\langle\mathbf{k}|V|\mathbf{k}'\rangle$,
 оператор (\ref{mu2Bclosure}) можно выразить через
\begin{eqnarray}
\mathcal{W}(\mathbf{k}_i,\mathbf{K};\mathbf{r}_1,\mathbf{r}_2)
&=&-\frac{1}{2\pi}\int\frac{d\mathbf{k}}{(2\pi)^3}\frac{\frac{4\pi}
{|\mathbf{q}_2|^2}\exp(i\mathbf{q}_2\mathbf{r}_2)\frac{4\pi}{|\mathbf{q}|^2}
\exp(i\mathbf{q}\mathbf{r}_1)}{k_i^2/2+E_0-k^2/2-E_{t}+i\epsilon}\nonumber\\
&=&-\frac{\exp(i\mathbf{K}\mathbf{r}_2)}{\pi^2
k_i}\mathcal{I}(\mathbf{k}_i,\mathbf{K};\mathbf{r}_1-\mathbf{r}_2).\label{SBWr1r2}
\end{eqnarray}
Здесь введены векторы промежуточной передачи импульса
$\mathbf{q}=\mathbf{k}_i-\mathbf{k}$,
$\mathbf{q}_2=\mathbf{k}-\mathbf{k}_s=\mathbf{K}-\mathbf{q}$.
Интеграл $\mathcal{I}(\mathbf{k}_i,\mathbf{K};\mathbf{r})$ дается
формулой (\ref{2BIntegral}). Подробности процедуры
интегрирования приведены в Приложении \ref{appendixSB}.

Непосредственное применение в операторе возмущения соотношений (\ref{lgauge}), (\ref{FBexp})
 и  (\ref{SBWr1r2}) неудобно в случае неориентированных молекул, так
как их использование требует решения уравнения (\ref{drivenSchr}) для каждой ориентации $\mathbf{R}$.

Фотоионизационный оператор возмущения (\ref{lgauge}) можно выразить
через $\Sigma$- и $\Pi$-волны, соответствующие направлениям
поляризации падающего излучения вдоль и поперек молекулярной оси.
Решив (\ref{drivenSchr}) для каждой из этих ситуаций по отдельности
и получив амплитуды ионизации, через них можно выразить амплитуду
для произвольного направления молекулярной оси \cite{McCurdy2006}.

Первый борновский оператор (\ref{FBexp}) можно выразить через разложение плоской волны
\cite{Morse53} в вытянутых сфероидальных координатах
\begin{equation}
\exp\left(i\mathbf{Kr}\right)=4\pi
\sum_{M=-\infty}^{\infty}\,\sum_{L=|M|}^{\infty}\Upsilon_{KLM}^{*}(\cos\theta_{KR},\varphi_{KR})i^{L}\Upsilon_{KLM}(\eta,\varphi)\mathrm{je}_{ML}\left(c,\xi\right),
\label{PWE}
\end{equation}
где $\theta_{KR},\varphi_{KR}$ --- углы, задающие направление вектора передачи
импульса $\mathbf{K}$ в системе координат, привязанной к направлению молекулярной оси,
 $\mathrm{je}_{ML}\left(c,\xi\right)$ --- эллиптическая функция Бесселя.
Уравнение (\ref{drivenSchr}) решается для каждого из членов парциального разложения (\ref{PWE})
 по отдельности
\begin{eqnarray}
\!\!(\hat{H}-E)\psi^{(+)}_{LM}(\mathbf{r}_1',\mathbf{r}_2')\!=\!
\frac{8\pi i^{L}}{K^2}
\left[\Upsilon_{KLM}(\eta_1,\varphi_1)\mathrm{je}_{ML}\!+\!\Upsilon_{KLM}(\eta_2,\varphi_2)\mathrm{je}_{ML}\left(c,\xi_2\right)\right]\varphi_0(\mathbf{r}_1',\mathbf{r}_2')
,
\end{eqnarray}
в результате чего получается парциальная волновая функция
$\psi^{(+)}_{LM}(\mathbf{r}_1',\mathbf{r}_2')$, где
$\mathbf{r}_{1,2}'$  --- векторы координат в системе, привязанной к
молекуле, $Oz'||\mathbf{R}$. Из этих парциальных функций с помощью
выражения (\ref{f_via_flux}) можно рассчитать парциальные амплитуды
двойной ионизации $f_{LM}$. А из парциальных амплитуд можно получить
амплитуду первого борновского процесса для произвольной ориентации
молекулярной оси
\begin{equation}
f_{1B}=\sum_{LM}f_{LM}\Upsilon_{KLM}^{*}(\cos\theta_{KR},\varphi_{KR}).
\end{equation}

Для второго борновского члена (\ref{SBWr1r2}) аналитическое разложение
по парциальным волнам отсутствует. Поэтому для него приходится использовать
мультипольное разложение. В работе \cite{Serov2010} учитывался только бидипольный член
\begin{eqnarray}
\hat{\mu}_{2BD}(\mathbf{r}_1,\mathbf{r}_2)=\sum_{M_1,M_2=-1}^{1}\mathcal{M}^{M_1M_2}(x_{1M_1}+x_{2M_1})(x_{1M_2}+x_{2M_2}),\label{dipole2B}
\end{eqnarray}
который, предположительно, должен вносить ведущий вклад в двойную ионизацию.
Здесь $x_{\alpha,\pm 1}=\frac{1}{\sqrt{2}}(\mp
x_{\alpha}-iy_{\alpha})$, $x_{\alpha 0}=z_{\alpha}$, а тензор
второго ранга
\begin{eqnarray}
\mathcal{M}^{M_1M_2}=\left.\frac{\partial^2\mathcal{W}(\mathbf{r}_1,\mathbf{r}_2)}{\partial
x_{1M_1}\partial
x_{2M_2}}\right|_{\mathbf{r}_1=0,\,\mathbf{r}_2=0}.\label{Dtensor}
\end{eqnarray}
Тогда с помощью (\ref{dipole2B}) можно представить
(\ref{drivenSchr}) в виде системы из девяти несвязанных уравнений
\begin{eqnarray}
(\hat{H}-E)\psi^{(+)}_{M_1M_2}(\mathbf{r}_1',\mathbf{r}_2') =
-(x_{1M_1}'+x_{2M_1}')(x_{1M_2}'+x_{2M_2}')\varphi_0(\mathbf{r}_1',\mathbf{r}_2').\label{drivenSchrD}
\end{eqnarray}
Из парциальных функций $\psi^{(+)}_{M_1M_2}$ с помощью выражения
(\ref{f_via_flux}) можно рассчитать парциальные амплитуды
$f_{M_1M_2}$. Окончательно, амплитуда второго борновского процесса
для произвольной ориентации молекулярной оси может быть представлена
как
\begin{eqnarray}
f_{2B}=\sum_{M_1,M_2=-1}^{1}\mathcal{M}'{}^{M_1M_2}(\mathbf{n}_R)f_{M_1M_2},
\label{f2Bassembling}
\end{eqnarray}
где $\mathcal{M}'$ соответствует тензору, определенному в
(\ref{Dtensor}), преобразованному в молекулярную систему координат
при фиксированной ориентации  $\mathbf{R}$ с помощью формулы для
контравариантных тензоров \cite{Varshalovich1989}
\begin{eqnarray*}
\mathcal{M}'{}^{M_1M_2}(\mathbf{n}_R)&=&\sum_{M_1',M_2'=-1}^{1}\mathcal{M}^{M_1'M_2'}D^{1*}_{M_1'M_1}(\varphi_{R},\theta_{R},0)D^{1*}_{M_2'M_2}(\varphi_{R},\theta_{R},0),
\end{eqnarray*}
где $D^{l}_{mm'}(\alpha,\beta,\gamma)$ --- D-функция Вигнера.
Заметим, что уравнение (\ref{drivenSchrD}) нужно решать только для
четырех комбинаций $(M_1,M_2)$: $(0,0)$, $(-1,1)$, $(1,1)$ и $(1,0)$.
Все другие $\psi^{(+)}_{M_1M_2}$ можно вывести из этих четырех
значений с помощью симметрии относительно перестановки электронов
$\psi^{(+)}_{M_2M_1}(\mathbf{r}_1,\mathbf{r}_2)=\psi^{(+)}_{M_1M_2}(\mathbf{r}_1,\mathbf{r}_2)$
и аксиальной симметрии  $\langle \ell_1m_1\ell_2m_2|
\psi^{(+)}_{-M_1,-M_2}\rangle=\langle \ell_1,-m_1,\ell_2,-m_2|
\psi^{(+)}_{M_1,M_2}\rangle$.

Как показано в работе \cite{Kheifets2004}, дипольное приближение
для второго борновского члена в обычном виде дает сильно
завышенную оценку, поскольку зависимость билинейного оператора (\ref{dipole2B})
от координат становится заметно больше полного второго борновского члена (\ref{SBWr1r2})
уже на расстоянии порядка размера атома. Метод коррекции, предложенный в \cite{Kheifets2004},
несовместим с используемой в \cite{Serov2010}
 процедурой. Поэтому дипольное приближение было скорректировано
альтернативным способом. Был введен тензор
\begin{eqnarray}
\mathcal{W}^{M_1M_2}(r_1,r_2)=\oint\oint
Y_{1M_1}^*(\Omega_1)Y_{1M_2}^*(\Omega_2)\mathcal{W}(\mathbf{r}_1,\mathbf{r}_2)d\Omega_1d\Omega_2.
\end{eqnarray}
Дипольное приближение (\ref{dipole2B}) эквивалентно
билинейному приближению
$\mathcal{W}^{M_1M_2}(r_1,r_2)\simeq\frac{4\pi}{3}\mathcal{M}^{M_1M_2}r_1r_2$.
Можно использовать билинейное приближение вида
$\frac{4\pi}{3}\widetilde{\mathcal{M}}^{M_1M_2}r_1r_2$ такое, что оно точно
совпадет с $\mathcal{W}^{M_1M_2}(r_1,r_2)$ для некоторого радиуса $r_{mol}$.
Из соотношения
\[ \mathcal{W}^{M_1M_2}(r_{mol},r_{mol})=\frac{4\pi}{3}\widetilde{\mathcal{M}}^{M_1M_2}r_{mol}^2
\]
следует
\begin{eqnarray}
\widetilde{\mathcal{M}}^{M_1M_2}=\frac{3}{4\pi}\frac{\mathcal{W}^{M_1M_2}(r_{mol},r_{mol})}{r_{mol}^2}.
\label{CDtensor}
\end{eqnarray}
При $r_{mol}\to 0$ выражение (\ref{CDtensor}) совпадает с (\ref{Dtensor}).

В работе \cite {Serov2010} радиус $r_{mol}$ был выбран равным
радиусу наивысшей электронной плотности в невозмущенной волновой
функции, которая равна правой части (п.ч.) (\ref{drivenSchrD}), то
есть максимуму выражения
$|\text{[п.ч.(\ref{drivenSchrD})]}|^2r_1^2r_2^2$. С использованием
некоррелированной начальной функции
$\psi_i(\mathbf{r}_1,\mathbf{r}_2)=\varphi_1(\mathbf{r}_1)\varphi_1(\mathbf{r}_2)$
после усреднения по углам в \cite{Serov2010} $r_{mol}$ было получено
из условия
\begin{eqnarray*}
\frac{\partial}{\partial
r}\left.\left[r^4\oint|\varphi_1(\mathbf{r})|^2d\Omega\right]\right|_{r_{mol}}=0.
\end{eqnarray*}
Для He с использованием одноэкспоненциальной функции
$\varphi_1(\mathbf{r})\sim\exp(-\zeta r)$, $\zeta=27/16$
было получено $r_{mol}=1.19$. Для H$_2$ была выбрана функция Коулсона
$\varphi_1(\mathbf{r})\sim\exp(-\zeta|\mathbf{r}-\mathbf{R}/2|)+\exp(-\zeta|\mathbf{r}+\mathbf{R}/2|)$, $\zeta=1.197$; при этом получилось $r_{mol}=1.76$. Подчеркнем, что некоррелированные функции использовались в \cite{Serov2010} только для оценки $r_{mol}$. При решении же
уравнения (\ref{drivenSchrD}) использовалась полностью
коррелированная начальная функция (см. подробности в
\cite{Serov2009}).

\subsection{Численные расчеты и сравнение с другими работами}

\begin{figure}[t]
\includegraphics[angle=-90,width=0.5\textwidth]{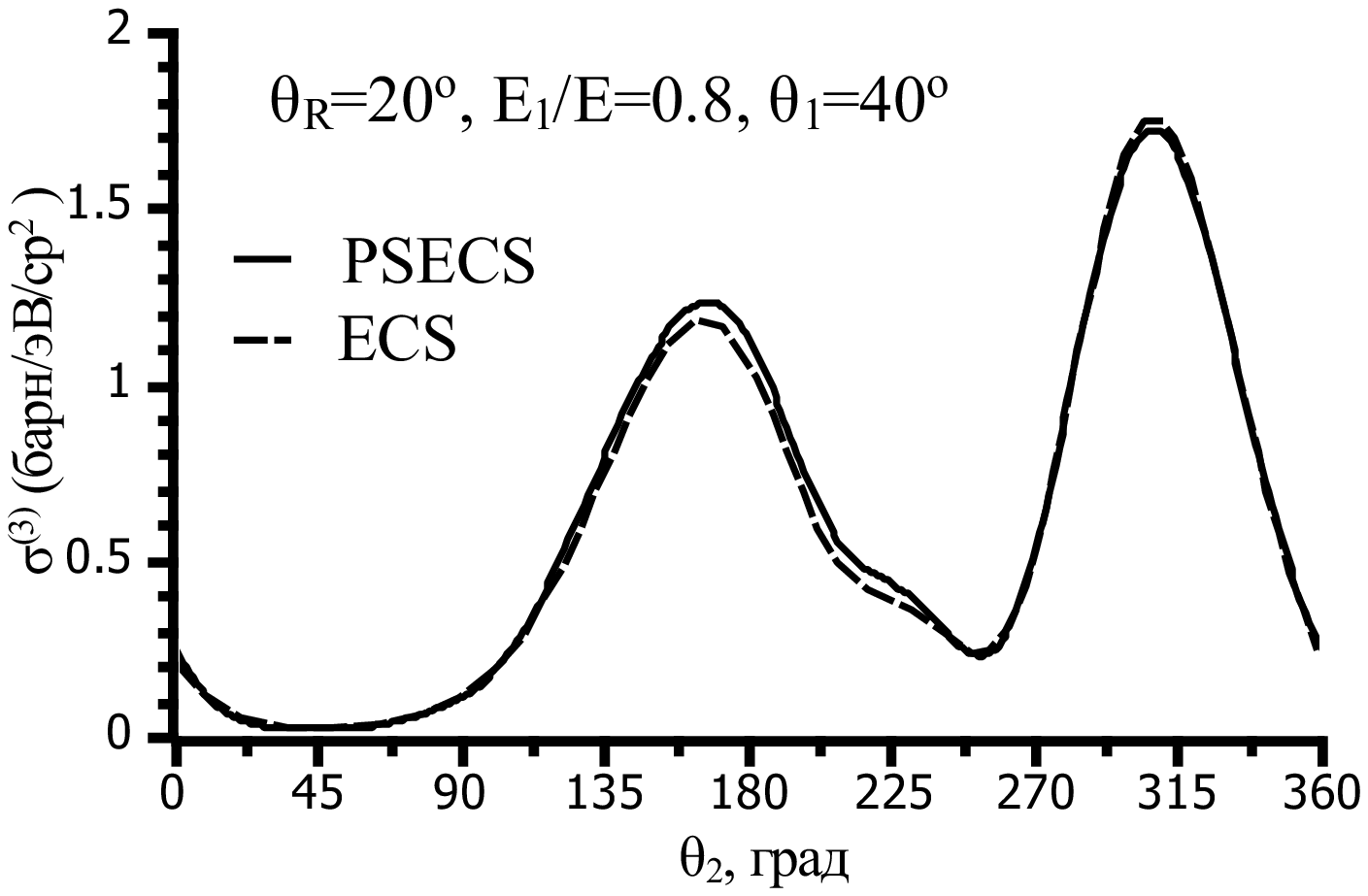}\\
 \textit{а}\\[-0.4cm]
\includegraphics[angle=-90,width=0.5\textwidth]{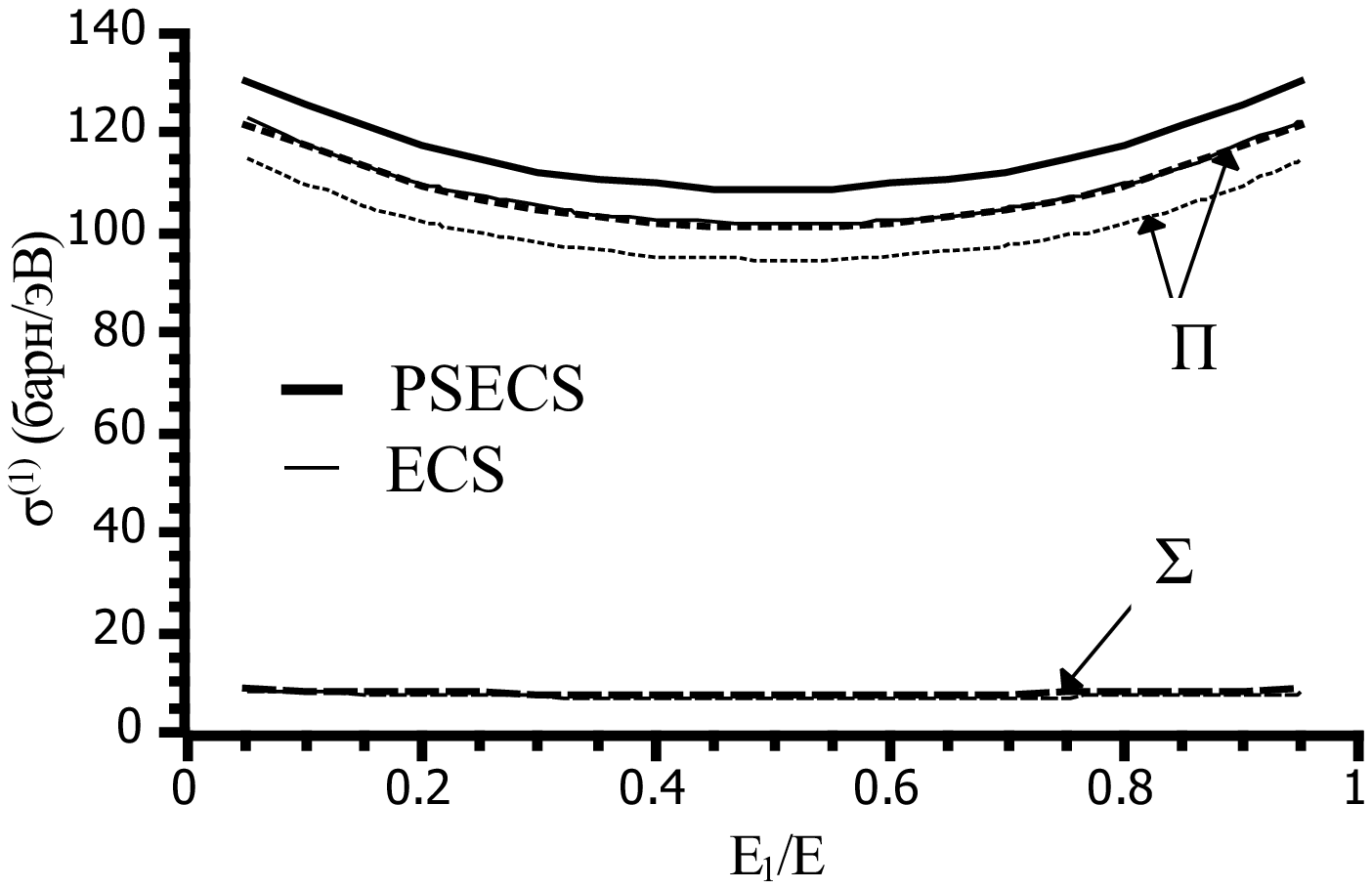}\\
 \textit{б}\\[-0.4cm]
\caption{Двухкратная однофотонная фотоионизация H$_2$ при энергии фотона $\omega=75$ эВ: \textit{а}) 3ДС как функция $\theta_1$ для $E_1=0.8E$, $\theta_2=40^{\circ}$,
углы между направлением поляризации и молекулярной осью
$\theta_R=20^{\circ}$, $\phi_R=0^{\circ}$; \textit{б}) ДС как функция
энергии вылета $E_1$ (сплошные кривые) и вклады $\Sigma_u$ (штриховые кривые) и
$\Pi_u$ (пунктирные кривые) компонент.
Толстые кривые --- результаты PSECS, тонкие кривые --- результаты ECS в сферических координатах \cite{McCurdy2006}.}\label{figMC}
\end{figure}

В \cite{Serov2009} с помощью внешнего комплексного скейлинга в
сфероидальных координатах (PSECS, prolate spheroidal external
complex scaling) был выполнен расчет дифференциальных сечений
двойной фотоионизации H$_2$. Это было сделано для того,  чтобы
сравнить PSECS с существующими теоретическими результатами
\cite{McCurdy2006} внешнего комплексного скейлинга в сферических
координатах (ECS), которые, в свою очередь, согласуются с
экспериментальными результатами \cite{Gisselbrecht,Weber2004}. Как
можно видеть из рис. \ref{figMC}, \textit{а}, две кривые,
соответствующие 3ДС для ориентированной молекулы H$_2$,
\[\sigma^{(3)}(\omega,E_1,\theta_1,\phi_1,\theta_2,\phi_2;\mathbf{R})=\frac{4\pi^2\omega}{c}k_1k_2|f(\mathbf{k}_1,\mathbf{k}_2;\mathbf{R})|^2,\]
в зависимости от одного из углов вылета, полученные с помощью ECS
и PSECS, близки друг к другу.

Но PSECS дает интегральное сечение $\sigma=2.77$ кбарн, а ECS ---
$\sigma=2.61$ кбарн \cite{McCurdy2006}. Различие между этими двумя
значениями намного больше, чем порядок численной ошибки в
PSECS-расчетах (см. Приложение \ref{sectH2solution}). Однократное
дифференциальное сечение (ДС)
$\frac{d\sigma}{dE_1}=\frac{1}{3}\left(\frac{d\sigma^{(\Sigma)}}{dE_1}+2\frac{d\sigma^{(\Pi)}}{dE_1}\right)$
и вклады в него $\Sigma_u$- и $\Pi_u$-компонент (для молекул, ориентированных
параллельно и перпендикулярно поляризации излучения, соответственно), показанные на рис.
\ref{figMC}, \textit{б}, также демонстрируют такое расхождение.
Метод PSECS дает заметно большие значения, чем ECS,
так что кривая $\Pi_u$, рассчитанная с помощью PSECS, фактически
совпадает с кривой полного ДС, рассчитанного с помощью ECS.

Вероятно, хорошее совпадение PSECS и ECS для углового
распределения объясняется тем, что в ECS расчетах в статье
\cite{McCurdy2006} использовался угловой базис с максимальным
угловым моментом $l_{max}=7$, а для интегрального сечения
--- $l_{max}=5$. Таким образом, результаты PSECS
\cite{Serov2009} точнее при меньшем угловом базисе (см. Приложение
\ref{sectH2solution}), что демонстрирует преимущество использования
сфероидальных координат. Впрочем, различие между результатами
\cite{Serov2009} и результатами \cite{McCurdy2006} намного меньше
ошибки существующих экспериментальных данных для интегрального
сечения двойной ионизации \cite{Dujardin1987,Kossmann1989}. Более
поздние расчеты \cite{LiangTao2010}, в которых также использовались
сфероидальные координаты, дали странный результат --- заметное
отклонение от результатов \cite{McCurdy2006}, но в противоположную
сторону по сравнению с \cite{Serov2009}. Еще позднее появилась
работа \cite{Xiaoxi2011}, где с помощью временного подхода в
сфероидальных координатах были получены результаты,  совпадающие с
\cite{Serov2009}. В работе \cite{Serov2012} метод PSECS использован для изучения
 проявления двухцентровой интерференции в однофотонной двухкратной
 ионизации молекулы водорода с неравновесным расстоянием между ядрами.

\begin{figure}[t]
\includegraphics[angle=-90,width=0.5\textwidth]{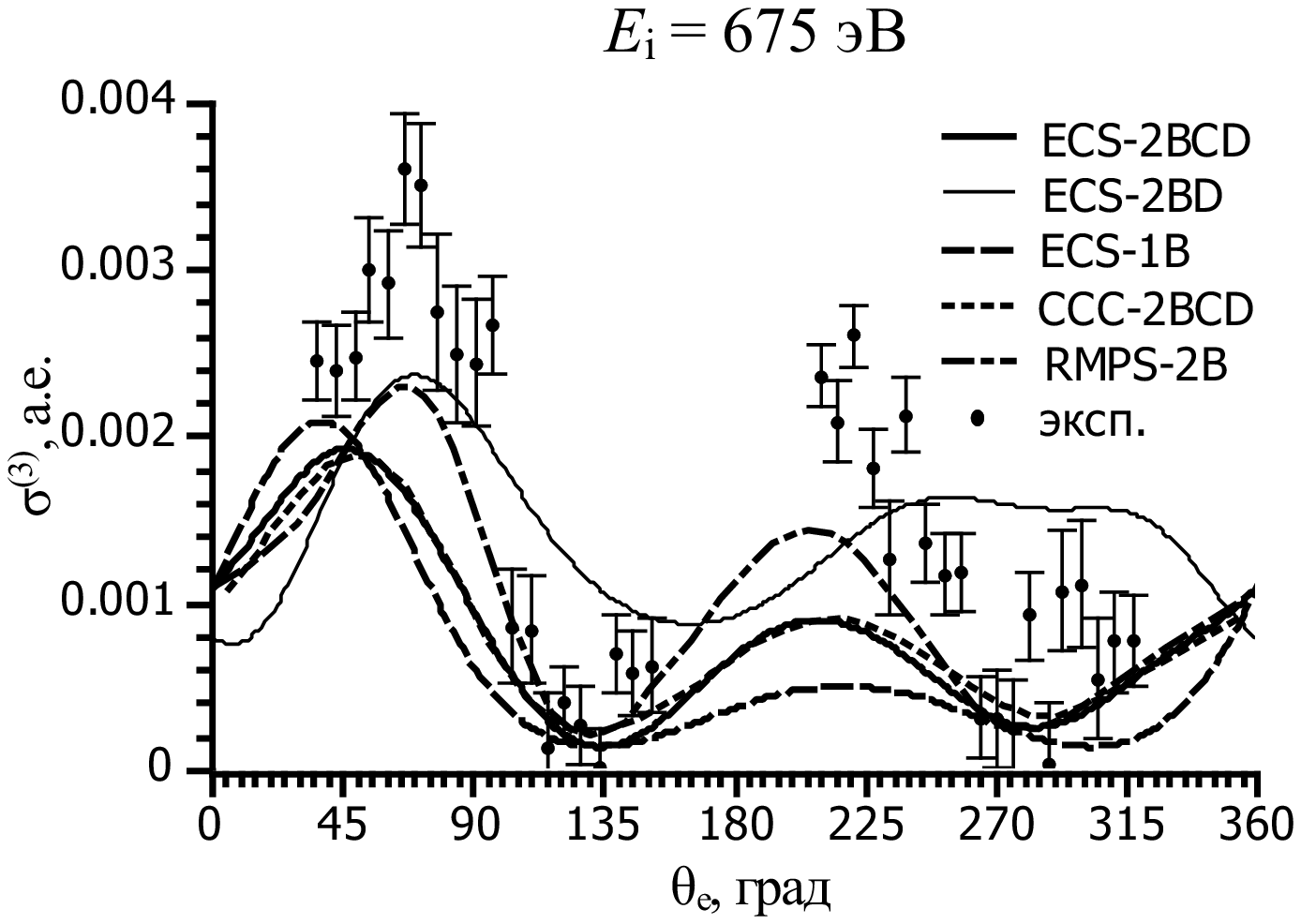}\\
 \textit{а}\\
\includegraphics[angle=-90,width=0.5\textwidth]{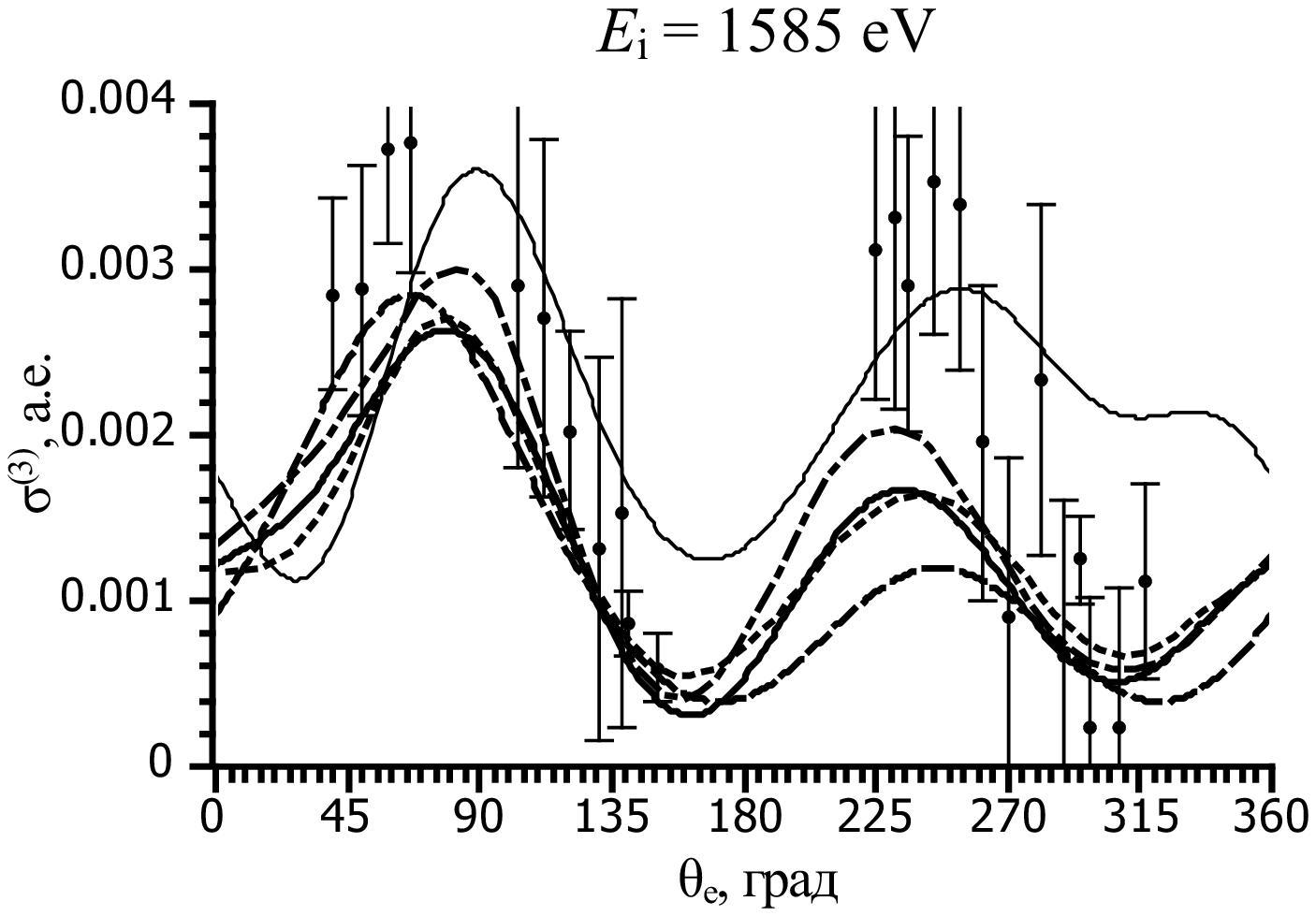}\\
 \textit{б}
\caption{3ДС ударной ионизации He с возбуждением остаточного иона в
зависимости от $\theta_e$: \textit{а}) энергия рассеянного электрона
$E_s=570$ эВ, угол рассеяния $\theta_s=-4^{\circ}$, энергия
вылетевшего электрона $E_e=40$ эВ; \textit{б})$E_s=1500$ эВ,
$\theta_s=-4^{\circ}$, $E_e=20$ эВ. Результаты ECS--2BD
(тонкая сплошная кривая), ECS--1B (штриховая кривая), CCC--2BCD \cite{Kheifets2004} (пунктирная кривая), $R$MPS--2B \cite{Fang2001} (штрих-пунктирная кривая),
экспериментальные данные \cite{Avaldi1998} (кружки).}
\label{fig_e2eHe}
\end{figure}

Для демонстрации различия корректированного и некорректированного бидипольного
второго борновского члена в \cite{Serov2010} проведен расчет 3ДС
ионизации-возбуждения атома гелия электронным ударом с
образованием остаточного иона в возбужденном состоянии $n=2$
при параметрах эксперимента \cite{Avaldi1998}.
В \cite{Serov2010} PSECS метод с оператором возмущения, содержащим первый
борновский член (\ref{FBexp}) и второй борновский член (\ref{dipole2B})
в оригинальном бидипольном приближении (то есть когда в соотношение (\ref{f2Bassembling})
входит тензор (\ref{Dtensor})) обозначается ECS--2BD, а когда в  (\ref{f2Bassembling})
 используется скорректированный тензор (\ref{CDtensor}) --- ECS--2BCD.
 Кривые без учета второго борновского члена обознаются ECS--1B.
На рис. \ref{fig_e2eHe} проводится сравнение результатов, полученных методами PSECS и
 ССС с учетом второго борновского члена в бидипольном
приближении (CCC--2BCD) из работы \cite{Kheifets2004}, а также методом
$R$-матрицы с псевдосостояниями с учетом полного второго борновского члена
($R$MPS--2B) \cite{Fang2001}. Видно, что наши результаты ECS--2BCD
очень близки к результатам  CCC--2BCD, несмотря на то, что мы
используем совершенно другой подход к коррекции бидипольного приближения.
Однако из сравнения с экспериментальными данными \cite{Avaldi1998} и с кривой
$R$MPS--2B ясно, что бидипольного приближения недостаточно,
чтобы воспроизвести положение максимумов 3ДС, и, следовательно,
может потребоваться учет других членов мультипольного разложения второго борновского члена.

В \cite{Serov2009,Serov2010} был проведен расчет четырехкратного дифференциального
сечения (4ДС)
\[\sigma^{(4)}(E_i,\theta_s,\phi_s,E_1,E_2,\theta_1,\phi_1)=\frac{d^4\sigma}{d\Omega_sdE_1dE_d\Omega_1}=\oint\sigma^{(5)}d\Omega_2\]
двойной ионизации H$_2$ ударом быстрого электрона для случайно ориентированных молекул.
Была выбрана та же динамическая ситуация, что и в эксперименте \cite{Lahmam-Bennani2002}:
энергия налетающего электрона полагалась $E_i=612$ эВ, энергия
рассеянного электрона $E_s=500$ эВ, угол рассеяния
$\theta_s=1.5^{\circ}$, регистрируется только один из вылетающих
электронов, имеющий энергию  $E_1=51$ эВ. Закон сохранения энергии
позволяет экспериментаторам вычислить энергию ненаблюдаемого вылетающего электрона
$E_2=10$ эВ.

\begin{figure}[t]
\includegraphics[angle=0,width=0.85\textwidth]{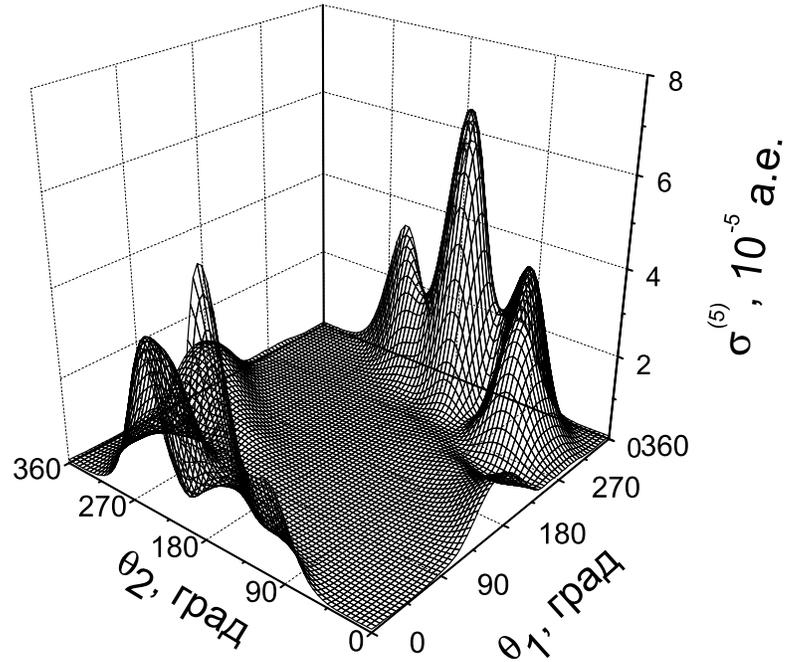}
\caption{5ДС для двухкратной ударной ионизации неориентированного H$_2$ как функция углов испускания $\theta_1$ и $\theta_2$ при $E_i=612$ эВ, $\theta_s=-1.5^{\circ}$, $E_s=500$ эВ, $E_1=51$ эВ, рассчитанное с помощью PSECS с учетом бидипольного второго борновского члена.}\label{figH2e3e1}
\end{figure}

Пятикратное дифференциальное сечение (5ДС) ионизации неориентированной молекулы (рис. \ref{figH2e3e1})
\[\sigma^{(5)}(E_i,\theta_s,\phi_s,E_1,\theta_1,\phi_1,E_2,\theta_2,\phi_2)=\frac{d^5\sigma}{d\Omega_sdE_1d\Omega_1dE_2d\Omega_2}=\frac{1}{4\pi}\oint\sigma^{(5)}(\mathbf{R})d\Omega_R\]
вычислялось путем интегрирования по направлению молекулярной оси 5ДС
ионизации ориентированной молекулы
\[\sigma^{(5)}(E_i,\theta_s,\phi_s,E_1,\theta_1,\phi_1,E_2,\theta_2,\phi_2;\mathbf{R})=\frac{k_1k_2k_s}{k_i}|f(\mathbf{k}_1,\mathbf{k}_2,\mathbf{K};\mathbf{R})|^2.\]

\begin{figure}[t]
\includegraphics[angle=-90,width=0.5\textwidth]{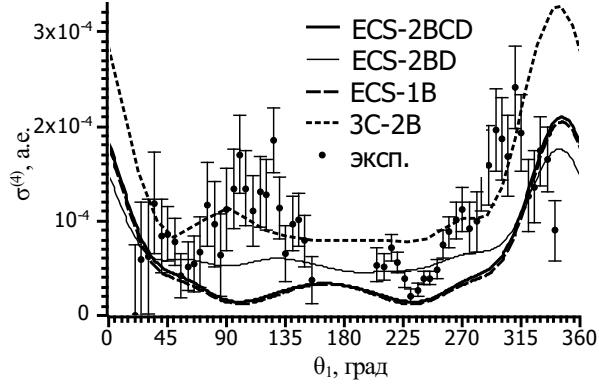}
\caption{4ДС для (e,3-1e)--процесса на неориентированном H$_2$ как функция
$\theta_1$ при $E_i=612$ эВ, $\theta_s=-1.5^{\circ}$, $E_s=500$ эВ,
$E_1=51$ эВ.
Результаты, полученные с помощью PSECS с учетом бидипольного второго борновского члена (тонкая сплошная кривая), с корректированным бидипольным вторым борновским членом  (толстая сплошная кривая), и только с первым борновским членом (штриховая кривая). Также показаны результаты 3C расчетов с учетом второго борновского члена \cite{DalCappello2004} (пунктирная кривая) и экспериментальные данные \cite{Lahmam-Bennani2002} (кружки).}\label{figH2e3e}
\end{figure}

Сравнение результатов PSECS с экспериментальными данными
\cite{Lahmam-Bennani2002} и результатами теоретических расчетов с
применением приближенной 3С-функции \cite{DalCappello2004} показано
на рис. \ref{figH2e3e}. Экспериментальные данные
\cite{Lahmam-Bennani2002} получены в произвольных единицах. Поэтому
в \cite{Serov2009,Serov2010} их нормировали так,
чтобы добиться наилучшего визуального совпадения
 с кривыми PSECS. Как показывает рис.
\ref{figH2e3e}, второй борновский член в бидипольном приближении не
вносит существенной коррекции по сравнению с первым борновским
приближением, в частности, совершенно не меняет положение
максимумов распределения. По-видимому, это связано с тем, что
импульс, передававаемый молекуле при каждом из двух взаимодействий с
налетающим электроном, при использовании бидипольного приближения
фактически полагается равным нулю.

\section{Метод сопутствующих координат для решения временного уравнения Шредингера}\label{expgridTDSE}

Начнем с временного уравнения Шредингера
\begin{equation}
i\frac{\partial
\psi(\mathbf{r},t)}{\partial t}=
\hat{H}_0(\mathbf{r})\psi(\mathbf{r},t).\label{TDSEorig}
\end{equation}
Здесь гамильтониан
\begin{equation}
\hat{H}_0(\mathbf{r})=-\frac{1}{2}\nabla^2+U(\mathbf{r})
\end{equation}
полагаем не зависящим от времени, т.е. рассматриваем ситуацию после прекращения
внешнего воздействия на систему.

Решение $\psi(\mathbf{r},t)$ уравнения (\ref{TDSEorig}) можно разложить
по собственным функциям системы
\begin{eqnarray}
\psi(\mathbf{r},t) =\int C(\mathbf{k}) \varphi_{\mathbf{k}}^{(-)}(\mathbf{r}) e^{-i Et} d\mathbf{k}
 +\sum_{nlm}C_{nlm}\varphi_{nlm}(\mathbf{r})e^{-i E_{nlm}
t}.\label{int}
\end{eqnarray}
Здесь $E=k^2/2$, $\varphi_{\mathbf{k}}(\mathbf{r})$ ---- волновые
функции сплошного спектра мишени, нормированные как
\[
\int\varphi_{\mathbf{k}'}^{(-)}(\mathbf{r})
\varphi_{\mathbf{k}}^{(-)}(\mathbf{r})d{\mathbf r}= \delta(\mathbf{k}'-\mathbf{k}),
\]
а $\varphi_{nlm}(\mathbf{r})$  --- функции связанных состояний мишени.

Согласно \cite{Mott65}, в случае если потенциал $U(\mathbf{r})$ --- короткодействующий,
асимптотическое решение есть
\begin{equation}
 \psi(\mathbf{r}\to \infty,t\to \infty) = \frac{1}{(i t)^{3N_e/2}}
\exp\left(\frac{i}{2}\frac{r^2}{t}\right)C(\mathbf{k}_{0}), \label{assol0}
\end{equation}
где
\[
\mathbf{k}_{0}= \frac{\mathbf{r}}{t}
\]
представляет стационарную точку интеграла в (\ref{int}),
 $N_e$ --- количество испущенных при ионизации электронов.

Если потенциал содержит дальнодействующий кулоновский потенциал, то есть
\[U(r\to\infty)=-\frac{Z}{r},\]
то асимптотическое решение принимает вид
\begin{equation}
 \psi(\mathbf{r}\to \infty,t\to \infty) = \frac{1}{(i t)^{3N_e/2}}
\exp\left(\frac{i}{2}\frac{r^2}{t}+i\frac{Z}{k_e}\ln 2k_e^2
t\right)C(\mathbf{k}_{0}), \label{assol}
\end{equation}
где стационарную точку интеграла $\mathbf{k}_{0}$ приближенно можно описать выражением
\[
\mathbf{k}_{0}= \mathbf{k}_{e}-Z\frac{\ln 2k_e^2 t}{k_e^2
t}\frac{\mathbf{k}_{e}}{k_e},
\]
где $\mathbf{k}_{e}=\mathbf{r}/t$  --- импульс электрона на больших
расстояниях от центра.

Численное представление $\psi(\mathbf{r},t)$ сталкивается с
проблемами, связанными с конечным размером пространственной сетки.
Когда волновой пакет распространяется в течение длительного
времени $t$, компонента волновой функции, относящаяся к сплошному
спектру, расширяется и отражается от границы сетки конечного
размера. Более того, пространственный градиент фазы возрастает со
временем, так что волновая функция становится сильно
осциллирующей. Чтобы избежать этих трудностей, Сидки \cite{Sidky2000}
предложил использовать зависящее от времени масштабное преобразование \cite{Vinitsky1985}
\begin{eqnarray}
\mathbf{r}&=&a(t){\boldsymbol{\xi}},\label{Rxi}
\end{eqnarray}
где $\boldsymbol{\xi}$  --- вектор сопутствующих координат, а $a =
a(t)$ --- масштабный множитель, который зависит только от времени.

Физически данный прием означает расширение координатной сетки вместе с возникшим
при ионизации волновым пакетом. Нужно отметить, что впервые он был предложен
в работе \cite{Vinitsky1985} для корректного адиабатического представления
трехчастичной кулоновской задачи. В рамках атомной физики он был затем распространен
на динамику ионизации при столкновениях атомов с другими атомами и ионами \cite{Ovchinnikov2004},
взаимодействия атомов и молекул с электромагнитными импульсами \cite{Sidky2000,Kaschiev2003} и,
 наконец, ударной однократной и двукратной ионизации гелия \cite{Serov2001,Serov2007}.
 Подобный подход использовался также в физике плазмы для вакуумного расширения
 классической одномерной бесстолкновительной двухкомпонентной плазмы,
 а также квантового электронного газа в плоской геометрии \cite{PlasmaTDS1,PlasmaTDS2}.
 Кроме того, он применялся в квантовой оптике для свободного расширения конденсата
 Бозе-Эйнштейна \cite{OpticsTDS1,OpticsTDS2} и даже в астрофизике для описания свойств подобия и законов масштабирования для излучающих жидкостей \cite{AstroTDS}.
  Адаптивные сетки применяются в нелинейной оптике для описания распространения
  световых пучков и импульсов с сильно изменяющимися в процессе эволюции пространственными
  и временными масштабами (см., например, \cite{Fibich2003}).
  Обзор ряда применений метода сопутствующих координат приведен в \cite{Piraux2011}.

Подстановка (\ref{Rxi}) во временное уравнение Шредингера (\ref{TDSEorig})
дает уравнение, содержащее первые производные по координатам и член с мнимой константой.
Чтобы привести его к виду, аналогичному исходному временному уравнению Шредингера,
необходимо сделать замену волновой функции
\begin{equation}
\psi(\mathbf{r},t)= \frac{1}{a^{3N_e/2}} \exp \left( \frac{i}{2}a
\dot a\xi^2 \right)\Psi(\boldsymbol{\xi},t), \label{psiphi}
\end{equation}
где $\dot a=\frac{da}{dt}$ и $\Psi(\boldsymbol{\xi},t)$  ---
``пилотная волновая функция'' решения $\psi({\mathbf r},t)$,
удовлетворяющая временному уравнению шредингеровского типа
\cite{Vinitsky1985}
\begin{eqnarray}
 i\,{\frac {\partial }{\partial t}}\,\Psi(\boldsymbol{\xi}, \,t)
&=& \left[ \hat H_0(a(t)\boldsymbol{\xi}) +\frac {1}{2} a(t) \ddot
a(t)\xi^{2}\right]\Psi (\boldsymbol{\xi},t). \label{PREq}
\end{eqnarray}
Заметим, что если $\ddot a>0$, то спектр оператора в квадратных скобках чисто
дискретный. Это снимает теоретические вопросы по поводу правомерности
приближения континуума конечным набором базисных функций.

Рассмотрим масштабный параметр, линейно растущий в асимптотической
области:
\begin{eqnarray}
a(t\to\infty)=\dot a_{\infty} t;\qquad \dot a_{\infty}>0.
\end{eqnarray}
Сравнивая (\ref{assol}) и (\ref{psiphi}), приходим к
асимтотическому виду пилотной функции при больших $t$
\cite{Kaschiev2003}
\begin{eqnarray}
\Psi(\boldsymbol{\xi},t)=(-i\dot a_{\infty})^{3N_e/2}C(\mathbf{k}_{0})
\exp\left(i\frac{Z}{k_e}\ln 2k_e^2 t\right), \label{comparePsiC}
\end{eqnarray}
где $\mathbf{k}_e=\dot a\boldsymbol{\xi}$. Таким образом, пространственное распределение
пилотной волновой функции $\Psi(\boldsymbol{\xi},t)$ при больших $t$ просто пропорционально
импульсному распределению амплитуды ионизации. Это, впрочем, очевидно
и из классических соображений: за время $t$ электрон улетает от центра на расстояние
$r={k}_e t$. При этом точка, фиксированная в сопутствующих координатах, в обычных координатах тоже равномерно удаляется от начала координат, так что в сопутствующих координатах улетающий электрон будет стремиться к состоянию неподвижности.

Заметим, что преобразование (\ref{psiphi}) удаляет из волновой функции квадратичную фазу, быстро растущую с увеличением времени. Как следует из (\ref{comparePsiC}),
при наличии кулоновского потенциала зависимость фазы пилотной функции от времени сохраняется, но поскольку эта зависимость логарифмическая, существенных проблем для численных расчетов это не вызывает \cite{Kaschiev2003}.

Представляет интерес скорость сходимости квадрата модуля волновой функции к квадрату модуля амплитуды ионизации (который пропорционален дифференциальному сечению).
При $\xi \gg R_{b}/a(t)$, где $R_{b}$ --- типичный радиус связанных состояний системы,
\begin{equation}
|C({\mathbf k}_e)|^2=\frac{1}{\dot a_{\infty}^{3N_e}}~|\Psi({\mathbf
k}_e/\dot a_{\infty},t)|^2+O(Z\ln t/t)+O(R_{i}/t),
\label{spectrphi}
\end{equation}
где $R_{i}$ --- радиус волнового пакета до расширения.

\subsection{Применение метода сопуствующих координат к расчету ионизации атома гелия}
Рассмотрим применение этого подхода для двухкратной ионизации атома гелия одним фотоном.
 В этом случае, т.е. для двух электронов в поле неподвижного ядра, ур.(\ref{PREq})  имеет вид
\begin{equation}
i\frac{\partial
\Psi(\boldsymbol{\xi}_1,\boldsymbol{\xi}_2,t)}{\partial t}=
\left\{\hat{h}_1(t)+\hat{h}_2(t)+\frac{1}{a(t)|\boldsymbol{\xi}_2-\boldsymbol{\xi}_1|}\right\}\Psi(\boldsymbol{\xi}_1,\boldsymbol{\xi}_2,t).\label{BasicEq0}
\end{equation}
Здесь $\hat{h}_{1,2}(t)$ --- одноэлектронные гамильтонианы в сопутствующих координатах,
\begin{equation}
\hat{h}_{\alpha}(t)=-\frac{1}{2a^2(t)}\nabla^2_{\boldsymbol{\xi}_{\alpha}}
-\frac{Z}{a(t)\xi_{\alpha}}+\frac {a(t) \ddot
a(t)}{2}\xi_{\alpha}^{2}.\label{H_soputstv}
\end{equation}

Легко показать \cite{Kasansky2003}, что если решить уравнение (\ref{BasicEq0}) с начальным условием
\begin{equation}
 \Psi(\mathbf{r}_1,\mathbf{r}_2,0)=
(\mathbf{e}\cdot\mathbf{r}_1+\mathbf{e}\cdot\mathbf{r}_2)
 \varphi_0(\mathbf{r}_1,\mathbf{r}_2),\label{InitCondPhotoIon}
\end{equation}
где $\varphi_0(\mathbf{r}_1,\mathbf{r}_2)$ --- волновая функция начального состояния,
 $\mathbf{e}$ --- направление поляризации падающего излучения, то коэффициент разложения
 $C(\mathbf{k})$ в (\ref{int}) будет совпадать с амплитудой однофотонной ионизации
 в координатной калибровке.

Для двухкратной ионизации фотоном с энергией $\omega=E+I_{\text{DI}}$
($I_{\text{DI}}$ --- потенциал
двойной ионизации, $E$ --- полная энергия двух вылетевших электронов) сечение дается
формулой
\[
 \sigma_{\omega}^{(3)}(\Omega_1,E_2,\Omega_2)=\frac{4\pi^2 \omega}{c} k_1 k_2
 \dot{a}_{\infty} ^{-6}
 \lim_{t\to\infty} \left|
 \Psi({\mathbf k}_1/\dot a_{\infty},{\mathbf k}_2/\dot a_{\infty},t)
 \right|^2.
\]
Сечение однократной фотоионизации
\[
 \sigma_{\omega}^{(1)}(\Omega_2)=\frac{4\pi^2 \omega}{c}
k_2
 \dot{a}_{\infty}^{-3}
 \lim_{t\to\infty} a^{3/2}(t)\left|
 \langle\varphi_{nlm}(a(t)\boldsymbol{\xi}_1)|\Psi(\boldsymbol{\xi}_1,{\mathbf k}_2/\dot
 a_{\infty},t)\rangle \right|^2,\]
где $\varphi_{nlm}(\mathbf{r})$ --- волновая функция остаточного иона.

В работах \cite{Serov2007,Serov2008,SerovSergeeva2010} использовался
масштабный множитель в виде
\begin{eqnarray}
 a(t)=\left\{
 \begin{array}{ll}
 1,& t\leq t_{int}; \\
 \left[ 1 + \gamma^2 (t-t_{int})^2\right]^{1/2},& t>t_{int},
 \end{array} \right.,
\end{eqnarray}
где $t_{int}$ --- время окончания внешнего воздействия на атом.
При таком выборе $\dot a_{\infty}=\gamma$, а переход к расширению происходит гладко.
Преимуществом данного конкретного выбора $a(t)$ (хотя и малосущественным)
является то, что $a(t)/\gamma$ совпадает с зависимостью от времени радиуса
свободного гауссового волнового пакета с начальным значением $1/\gamma$.

Основным достоинством метода сопутствующих координат в приложении к вычислению
однофотонной фотоионизации является возможность получения многократного
дифференциального сечения ионизации для всех значений энергии падающего фотона
 за один прогон программы. В результате получается непрерывная зависимость
 дифференциального сечения от энергии, тогда как другие широко распространенные
 {\it ab initio} методы подразумевают отдельное вычисление для каждого значения энергии.
 Детали численной схемы для решения двухэлектронного временного уравнения Шредингера
 см. в Приложении \ref{sectChangFano}.

\begin{figure}[t]
\begin{center}
\includegraphics[width=0.5\textwidth]{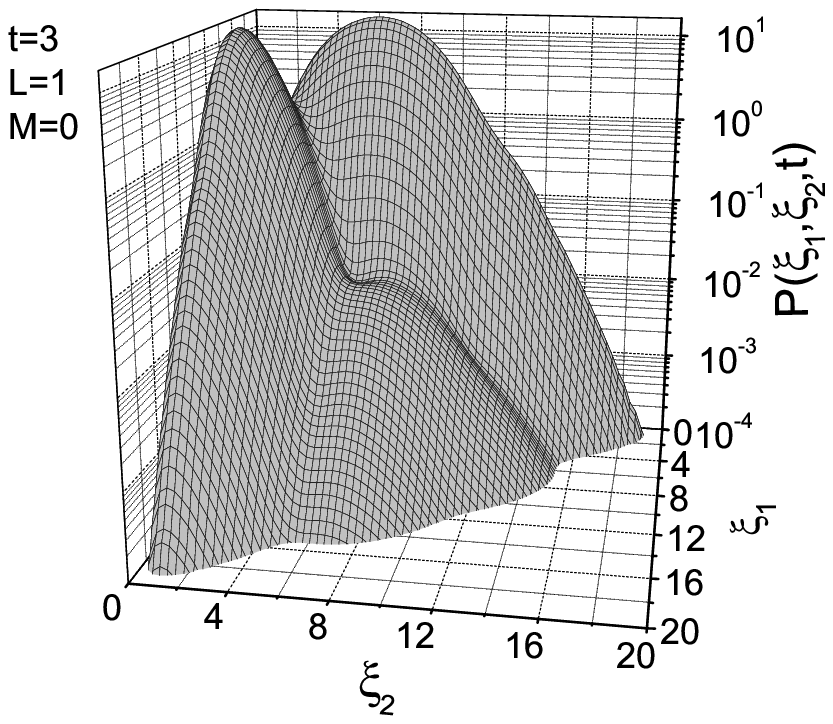}\\[-0.5cm]{\textit{а}}\\[0.5cm]
\includegraphics[width=0.5\textwidth]{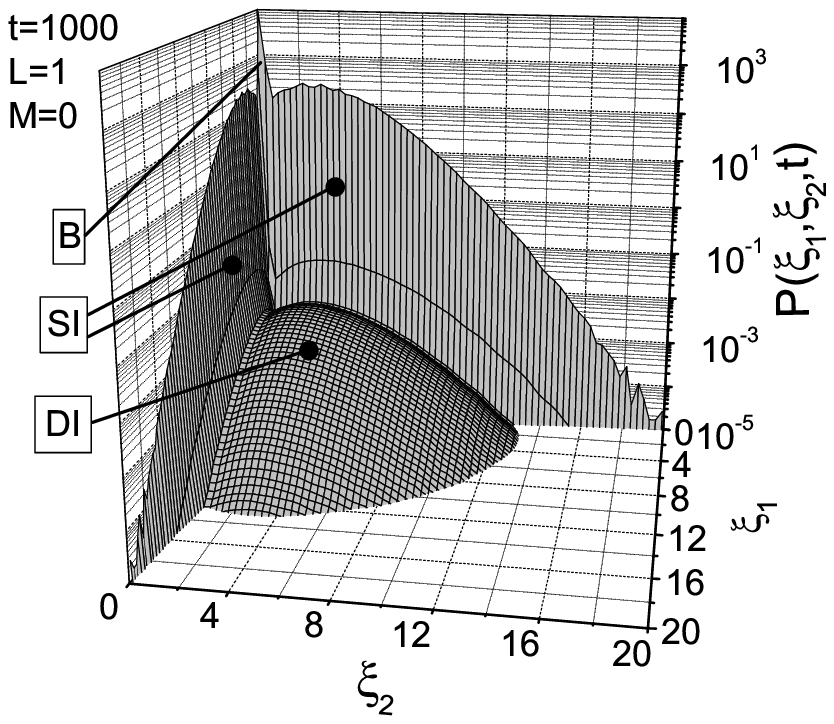}\\[-0.5cm]{\textit{б}}
\end{center}
 \caption{Распределение плотности вероятности обнаружения электронов в атоме гелия после удара быстрым электроном, проинтегрированное по угловым переменным в зависимости от радиальных сопутствующих координат $\xi_1$, $\xi_2$: \textit{а}) после удара, но до начала расширения координатной сетки ($t=3$ а.е.); \textit{б}) после расширения ($t=1000$).}\label{AbsPsi10}
\end{figure}

 Существует, однако, вычислительная трудность, связанная с
использованием сопутствующих координат: в этих координатах размер
связанных и однократно ионизированных (связанных по одной
координате) состояний убывает со временем, так что радиальное
конечно-разностное приближение (см. Приложение \ref{sectChangFano})
в конечном счете становится плохим. Для любых практически
достижимых параметров сетки характерный размер основного состояния становится меньше
шага сетки задолго до того, как достигается сходимость сечения.
В качестве иллюстрации к сказанному на рис. \ref{AbsPsi10}, \textit{а} показана плотность вероятности, проинтегрированная по
угловым переменным, в зависимости от радиальных координат двух
электронов при $t=t_{int}$. На рис. \ref{AbsPsi10}, \textit{б} то же самое
распределение показано при большом $t$. ``Боковые стенки''
соответствуют однократно ионизированным  состояниям, а пик вблизи
центра --- связанным состояниям. Остальная гладкая часть
распределения отвечает двукратно ионизированному состоянию.
Важно отметить, что для кулоновских потенциалов эта проблема несущественна, так
как связанные состояния коллапсируют в узел, ближайший к ядру. Это
не влияет на сечение двойной ионизации, основной вклад в которое
дают удаленные узлы.

Больше вычислительных проблем возникает из-за двукратно
возбужденных (автоионизационных) состояний. Численная ошибка,
растущая с $t$, сдвигает действительную часть энергии таких
состояний. В результате появляются нефизические осцилляции в
зависимости сечения однократной ионизации от энергии вылетающего
электрона. Для устранения
этого артефакта в работе \cite{Serov2007} мы подавляли эти состояния до начала
расширения координат с помощью фильтрации исходной волновой
функции:
 \[\Psi^{LM}_{flt}(t_{int})=
 \exp(-\hat H_0 t_{flt})\prod_{n=1}^{N_{flt}}(\hat H_0-E_{n})\Psi^{LM}(t_{int}),\]
где $E_{n}$ - энергии двукратно возбужденных состояний, $N_{flt}$
--- число двукратно возбужденных состояний, заметно заселяемых во
время удара. Умножение на $(\hat H_0-E_{n})$ существенно уширяет
волновой пакет в энергетическом пространстве. Для коррекции этого
можно подавить высокоэнергетическую часть спектра
с помощью экспоненциального множителя, что легко реализуется путем
распространения в мнимом времени в течении $t_{flt}$. Во всех расчетах в \cite{Serov2007}
использовалось $N_{flt}=1$ и $E_{1}=-0.625$
с $t_{flt}=0.2$, чего было достаточно, чтобы сделать нефизические осцилляции пренебрежимо малыми.

Поскольку длина волны нефизических осцилляций, возникающих из-за автоионизационных состояний,
 всегда того же порядка, что и шаг радиальной сетки $h$, в работе \cite{SerovSergeeva2010}
  использовался другой метод избавления от них. Вместо фильтрации до расширения
  применялась фильтрация волновой функции после завершения эволюции путем исключения компонент
  с длинами волн меньше, чем $4h$.

В работе \cite{Serov2007} метод сопутствующих координат (TDS) (time-dependent scaling)
в комбинации с PA1B был использован для расчета двойной ионизации атома гелия ударом
быстрого электрона, см. раздел \ref{sectionPA} и рис. \ref{e3e}.

\subsection{Использование рассчитанного 3ДС для анализа угловой корреляции электронов в ходе двойной фотоионизации}

Энерго-угловая зависимость 3ДС, полученная с помощью нашего метода, позволяет полно охарактеризовать процесс ДФИ. Природа последнего такова, что он возможен лишь благодаря наличию межэлектронных корреляций, так что анализ рассчитанного 3ДС даёт возможность изучения этого явления. Общепринятым приближённым аналитическим подходом к рассмотрению влияния межэлектронных корреляций на двойную ионизацию является теория Ванье. Мы использовали полученное в численных расчётах 3ДС для проверки следствий этой теории.

Более 50 лет назад Ванье \cite{Vanier} показал, что если два медленных электрона
покидают в результате двойной ионизации положительно заряженный остов,
то полное сечение $\sigma$ зависит от полной энергии электронов $E$
по закону $\sigma \propto E^\alpha$, где $\alpha>1$ зависит только от заряда остаточного иона.
Это выражение было получено путем разделения окружающего ион пространства на три области:
зона реакции, из которой стартуют электроны, кулоновская зона,
 в которой энергия кулоновского взаимодействия электронов и положительного остова
 намного больше $E$, и зона свободного движения, где, напротив, кинетическая энергия электронов намного больше потенциальной. При этом в кулоновской зоне движение электронов
 описывается классическим образом. Из теории Ванье также следует,
 что наиболее вероятен вылет электронов в противоположных направлениях.
 До недавнего времени экспериментальному измерению были доступны лишь
 полные сечения двойной ионизации, пока не появилась техника совпадений \cite{Becker2000},
 позволяющая  измерять многократные дифференциальные сечения, зависящие от того,
 как энергия делится между вылетающими электронами и каковы углы их вылета.
 Эти данные несут информацию о корреляции между электронами.

Одним из простейших процессов, определяемых корреляцией электронов,
является двойная ионизация  гелия одним фотоном \cite{Briggs1}.
Было показано \cite{Huetz1991}, что если падающее излучение линейно поляризовано в
направлении оси $z$, то трехкратное дифференциальное сечение (3ДС)
двойной фотоионизации атома одним фотоном можно выразить через сумму
четной и нечетной амплитуд
\begin{eqnarray}
\frac{d^{3}\sigma}{dE_1d\Omega_1d\Omega_2}=
|a_g(E_1,E_2,\theta_{12})(\cos\theta_{1}+\cos\theta_{2})
 +
 a_u(E_1,E_2,\theta_{12})(\cos\theta_{1}-\cos\theta_{2})|^2,\label{agau}
\end{eqnarray}

где $E_{1,2}$  --- энергии вылетевших электронов (их полная энергия $E=E_1+E_2$), $\theta_{1,2}$ --- углы вылета электронов относительно направления поляризации падающего излучения, $\theta_{12}$ --- относительный угол между направлениями их разлета, $a_g$ и $a_u$ --- четная и нечетная амплитуды соответственно (первую обычно называют корреляционным параметром). Сомножители $\cos\theta_1\pm \cos\theta_2$ представляют собой кинематическую компоненту 3ДС, явным образом отделённую в этой формуле от динамической. Кроме того, при $E_1=E_2$ $a_u=0$.
Четную амплитуду $a_g$ обычно называют корреляционным
параметром. Следуя теории Ванье \cite{Vanier}, потенциал
межэлектронного взаимодействия можно аппроксимировать квадратичным
членом разложения Тейлора по степеням величины
$\left(\theta_{12}-\pi\right)$ в седловой точке
$\left(\theta_{12}=\pi,\,\,\,r_1=r_2\right)$, то есть потенциалом
гармонического осциллятора с частотой, зависящей от гиперрадиуса
$R=\sqrt{r_1^2+r_2^2}$. Рау \cite{4} предположил, что угловое
поведение волновой функции вблизи седловой точки совпадает с
волновой функцией основного состояния гармонического осциллятора для
каждого значения $R$ в пределах кулоновской зоны, а в свободной зоне
эта зависимость ``замораживается'' и остается такой же, как на
границе между ними. Это приводит к
гауссовой форме корреляционного параметра \cite{5}
\begin{eqnarray}
a_g(E_1, E_2, \theta_{12})\simeq A\exp\left[-2\ln
2\frac{(\theta_{12}-\pi)^2}{\gamma^2}\right] ,\label{GaussianAppr}
\end{eqnarray}
с гауссовой шириной
\begin{equation}
\gamma=\gamma_0 E^{1/4},\label{gammalaw}
\end{equation}
где $A$ --- постоянная, а приведенная
 гауссова ширина $\gamma_0$ зависит от выбора гиперрадиуса границы между кулоновской
 и свободной зонами.  Гауссова ширина характеризует
  угловое распределение корреляционного параметра: большая величина $\gamma$ означает слабую корреляцию и наоборот. По этой причине данный параметр часто используют для оценки
   степени электрон-электронной корреляции. По аналогии с пороговым законом Ванье для полного сечения, равенство (\ref{gammalaw}) обычно также называют законом Ванье,
  хотя сам Ванье не имеет к нему прямого отношения.

 Несмотря на то, что область энергий, в которой справедливо выражение (\ref{gammalaw}),
 до сих пор не определена, и экспериментаторы, и теоретики часто используют его при интерпретации
 данных, пытаясь найти приведенную гауссову ширину $\gamma_0$. Множество формул для $\gamma_0$
 предложено различными авторами \cite{6}. Между тем, Казанский и Островский \cite{7}
 путем использования замены переменных, обеспечивающей сведение задачи к эволюции волнового пакета
  в потенциале гармонического осциллятора переменной частоты, показали, что ширина пакета
  не может адиабатически следовать за изменением частоты осциллятора из-за замедления
  электронов полем ядра.  А это противоречит допущениям, на основании которых  в \cite {5} была выведена формула (\ref{gammalaw}). При учете неадиабатичности поведение
  гауссовой ширины вблизи порога сильно изменяется. Предположение о том, что при малых энергиях из реакционной зоны может выйти волновой пакет, содержащий только основную осцилляторную моду, также неверно, как было показано теми же авторами. Следовательно, согласно \cite{7}, $a_g\left(\theta_{12}\right)$ зависит от деталей протекания процесса внутри реакционной зоны и может иметь негауссов вид даже при $E\rightarrow 0$.

Первоначальной мотивацией детального анализа \cite{SerovSergeeva2010} хорошо известной и, на первый взгляд, понятной задачи об энергетической зависимости гауссовой ширины корреляционного параметра для атома гелия и гелиеподобных ионов при низких энергиях  послужило обнаруженное в \cite{Serov2008} явление: для отрицательного иона водорода (протон с двумя электронами) в области низких энергий $\gamma$ начинает расти по мере убывания энергии, что находится в явном противоречии с пороговым законом Ванье.
В расчетах \cite{SerovSergeeva2010} использован метод сопутствующих координат для решения временного
уравнения Шредингера \cite{Serov2008,Serov2007}, изложенный в разделе \ref{expgridTDSE}.

Результаты расчета 3ДС затем использовались для определения гауссовой ширины $\gamma$.
 Квадрат модуля корреляционного параметра $|a_g (E_1,E_1,\theta_{12})|^2$ можно извлечь
 из 3ДС с помощью уравнения (\ref{agau}) с последующей аппроксимацией выражением (\ref{GaussianAppr})
 по методу наименьших квадратов \cite{11}. Альтернативный подход \cite{12}
 основан на аппроксимации двукратного дифференциального сечения (2ДС)
  $\sigma^{(2)}(E_1,E_2,\theta_{12})=\frac{d^2 \sigma}{dE_1 d\theta_{12}}$ выражением
\begin{equation}
\label{4}
\sigma^{(2)}(E_1,E_2=E_1,\theta_{12})\simeq \frac{32 \pi^2}{3}|A|^2 \exp\left[-\frac{4 \ln 2(\pi-\theta_{12})^2}{\gamma^2}\right]\cos^2\frac{\theta_{12}}{2},
\end{equation}
которое получается из (\ref{agau}) посредством интегрирования по всем углам,
кроме $\theta_{12}$. Поскольку корреляционный параметр $a_g(\theta_{12})$
 может значительно отклоняться от гауссовой формы, величины $\gamma$, вычисленные двумя способами,
 различаются. Так, в методе наименьших квадратов значения 3ДС вблизи $\theta_{12}=\pi$
вносят вклад в сумму квадратов с меньшим весом при подгонке $\sigma^{(2)}(\theta_{12})$,
чем при подгонке $|a_g(\theta_{12})|^2$.  Гауссова ширина, полученная подгонкой $|a_g(\theta_{12})|^2$,
 будет ниже обозначаться как $\gamma(|a_g|^2)$, а полученная подгонкой
 $\sigma^{(2)} (\theta_{12})$ --- как $\gamma(\sigma^{(2)})$.

\subsubsection{Фотоионизация гелия в основном состоянии}

\begin{figure}[t]
\vspace{-0.5cm}
\begin{center}
\includegraphics[angle=-90,width=0.5\textwidth]{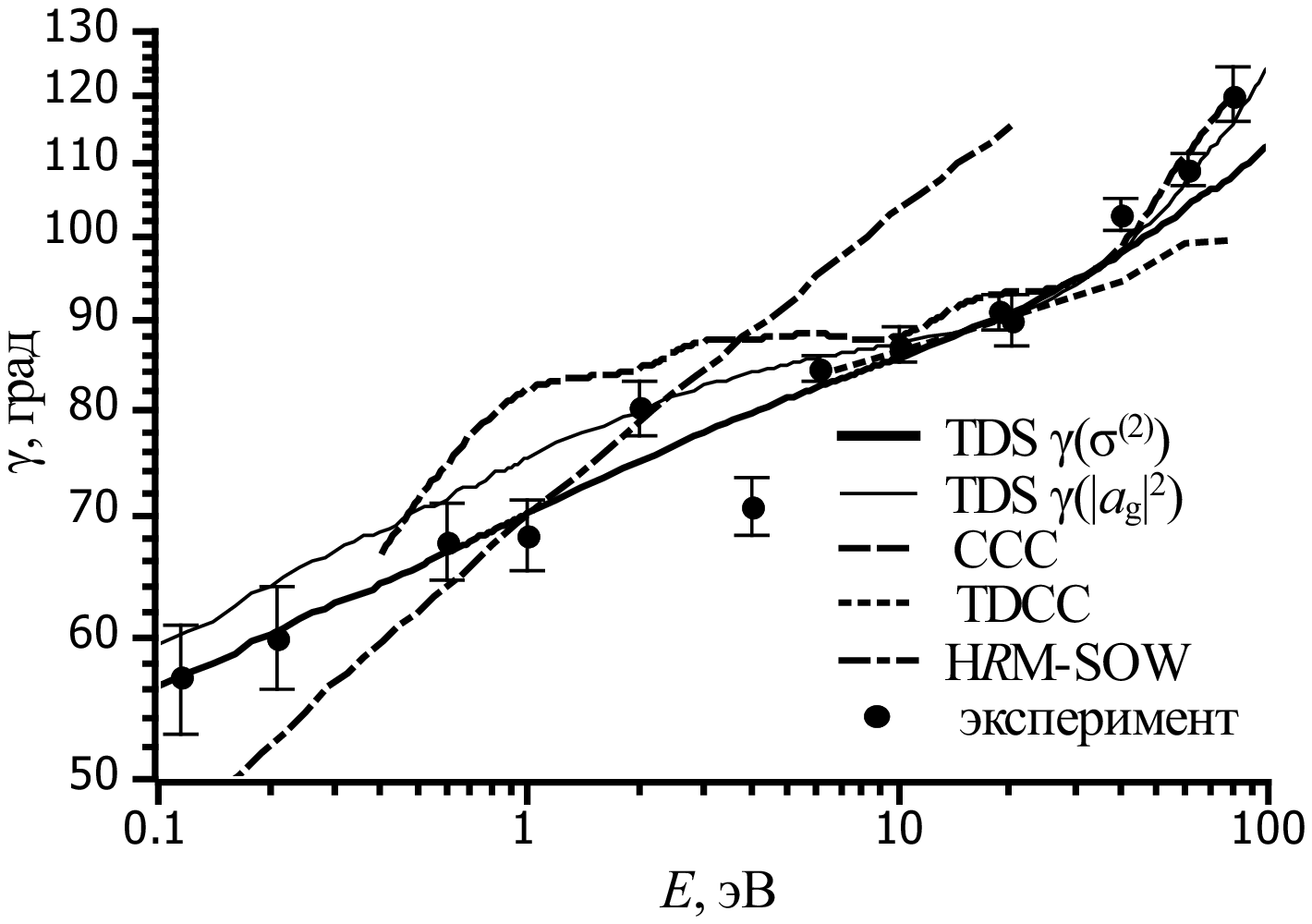}
\end{center}
\vspace{-0.5cm}
\caption{Гауссова ширина $\gamma$ как функция полной энергии испущенных электронов $E$: результаты, полученные НК-аппроксимацией
$\sigma^{(2)}(\theta_{12})$ (толстая сплошная кривая) и
$|a_g(\theta_{12})|^{2}$ (тонкая сплошная кривая), вычисленных с помощью TDS, а также результаты CCC \cite{10,11} (штриховая кривая), TDCC \cite{13} (пуниктирная линия), H$R$M-SOW \cite{14} (штрих--пунктирная кривая), и экспериментальные данные \cite{12,15,Dorner,17,18,19} (кружки).}\label{FIGgammaHe}
\end{figure}

На рисунке \ref{FIGgammaHe} приведена зависимость гауссовой ширины $\gamma $ от полной
энергии вылетевших электронов $E$ для двойной фотоионизации гелия из основного состояния
при условии, что энергия делится между электронами поровну.
Результаты метода сопутствующих координат (TDS) демонстрируют отличное согласие с экспериментом
во всем диапазоне энергий от 0.1 эВ до 100 эВ, кроме точки при 4 эВ из работы \cite{17}.
Особо отметим точное совпадение кривой TDS $\gamma(\sigma^{(2)})$ с экспериментальными
точками при 0.116 и 0.209 эВ \cite{12}.  Кривая $\gamma(|a_g|^2)$ хуже согласуется с этими точками, и это легко объясняется тем, что
в эксперименте \cite{12} определялась именно $\gamma(\sigma^{(2)})$.
При высоких энергиях экспериментальная точка на 80 эВ \cite{19} была получена
подгонкой $|a_g(\theta_{12})|^2$, так что неудивительно, что эта точка находится гораздо
ближе к кривой TDS $\gamma(|a_g|^2)$, чем к TDS $\gamma(\sigma^{(2)})$.
Вообще говоря, различие между $\gamma(|a_g|^2)$ и $\gamma(\sigma^{(2)})$ может быть проявлением
степени отклонения $a_g(\theta_{12})$ от гауссовой формы, хотя совпадение этих кривых не означает,
что функция $a_g(\theta_{12})$ чисто гауссова. На рис. \ref{FIGgammaHe}
показаны также результаты других расчетов из первых принципов. Согласие результатов ССС \cite{10,11}
 с результатами TDS хорошее для $E>10$ эВ и удовлетворительное при более низких энергиях.
 Результаты TDCC \cite{13} там, где они имеются, близки к TDS, а результаты HRM-SOW \cite{14}
 сильно отличаются от всех остальных кривых.
 
Графики на рис. \ref{FIGgammaHe} построены в логарифмическом масштабе по обеим осям,
 в котором степенные зависимости типа закона Ванье (\ref{gammalaw}) выглядят как наклонные
 прямые линии. И действительно, TDS графики близки к прямым линиям, когда $E$ не превышает
 нескольких электронвольт. Однако показатель степени вовсе не равен 1/4.
 В интервале энергий от 0.1 до 2 эВ аппроксимация кривой $\gamma(\sigma^{(2)})$ степенным законом
 общего вида
\begin{equation}
\label{gengamma}\gamma=\tilde{\gamma_0}E^s
\end{equation}
с использованием метода наименьших квадратов дает показатель
$s=1/10$  и коэффициент пропорциональности
$\tilde{\gamma}_0=70^{\circ}$ эВ$^{-s}$. Столь значительное
отклонение от порогового закона Ванье, часто используемого при
интерпретации экспериментальных и теоретических данных, выглядит
обескураживающе.

\begin{figure}[t]
\begin{center}
\includegraphics[angle=-90,width=0.5\textwidth]{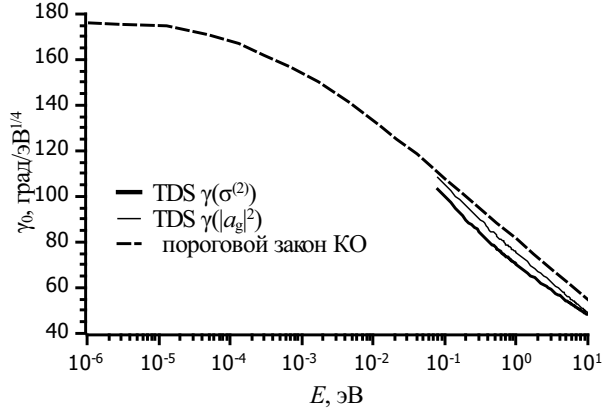}
\caption{Приведенная гауссова ширина $\gamma_0$ для He как функция $E$: результаты, полученные НК-аппроксимацией вычисленных $\sigma^{(2)}(\theta_{12})$ (толстая линия) и $|a_g(\theta_{12})|^{2}$ (тонкая кривая), вычисленных с помощью TDS, и  пороговый закон Казанского--Островского \cite{7} (штриховая кривая).}\label{FIGgamma0He}
\end{center}
\end{figure}

На рис. \ref{FIGgamma0He} показана приведенная гауссова ширина
$\gamma_0(E)=\gamma(E)/E^{1/4}$ в сравнении с кривой, полученной в
\cite{7} путем использования квадратичной аппроксимации потенциала
взаимодействия между электронами вблизи $(\theta_{12}-\pi)$ и
полуклассического приближения для радиального движения с учетом
неадиабатической зависимости волновой функции от гиперрадиуса, а также
предположения о том, что только наинизшая мода по $\theta_{12}$ заселена
на границе реакционной зоны, гиперрадиус которой полагался равным
$R=4$. Последнюю закономерность ниже мы будем называть пороговым законом
Казанского-Островского (КО). Видно что кривые TDS близки к
пороговому закону КО вплоть до минимальной энергии $E=0.1$ эВ,
достигнутой в расчетах. На рис. \ref{FIGgamma0He} пороговый закон
Ванье (\ref{gammalaw}) выглядел бы как горизонтальная прямая линия.
При очень низких энергиях порядка $10^{-5}$ эВ кривая КО кажется
выходящей на горизонталь, однако  это происходит лишь потому, что
энергии ниже $10^{-6}$ эВ не показаны на рисунке. Согласно \cite{7},
зависимость $\gamma_0(E)$  при убывании энергии имеет осциллирующий
характер с периодом, который постоянен в логарифмической шкале (на
рис. \ref{FIGgamma0He} показана половина этого периода), и никогда
не переходит в пороговый закон Ванье.

\subsubsection{Фотоионизация мишеней с сильно асимметричной конфигурацией начального состояния}
\begin{figure}[t]
\begin{center}
\parbox{0.5\textwidth}{\includegraphics[angle=-90,width=0.5\textwidth]{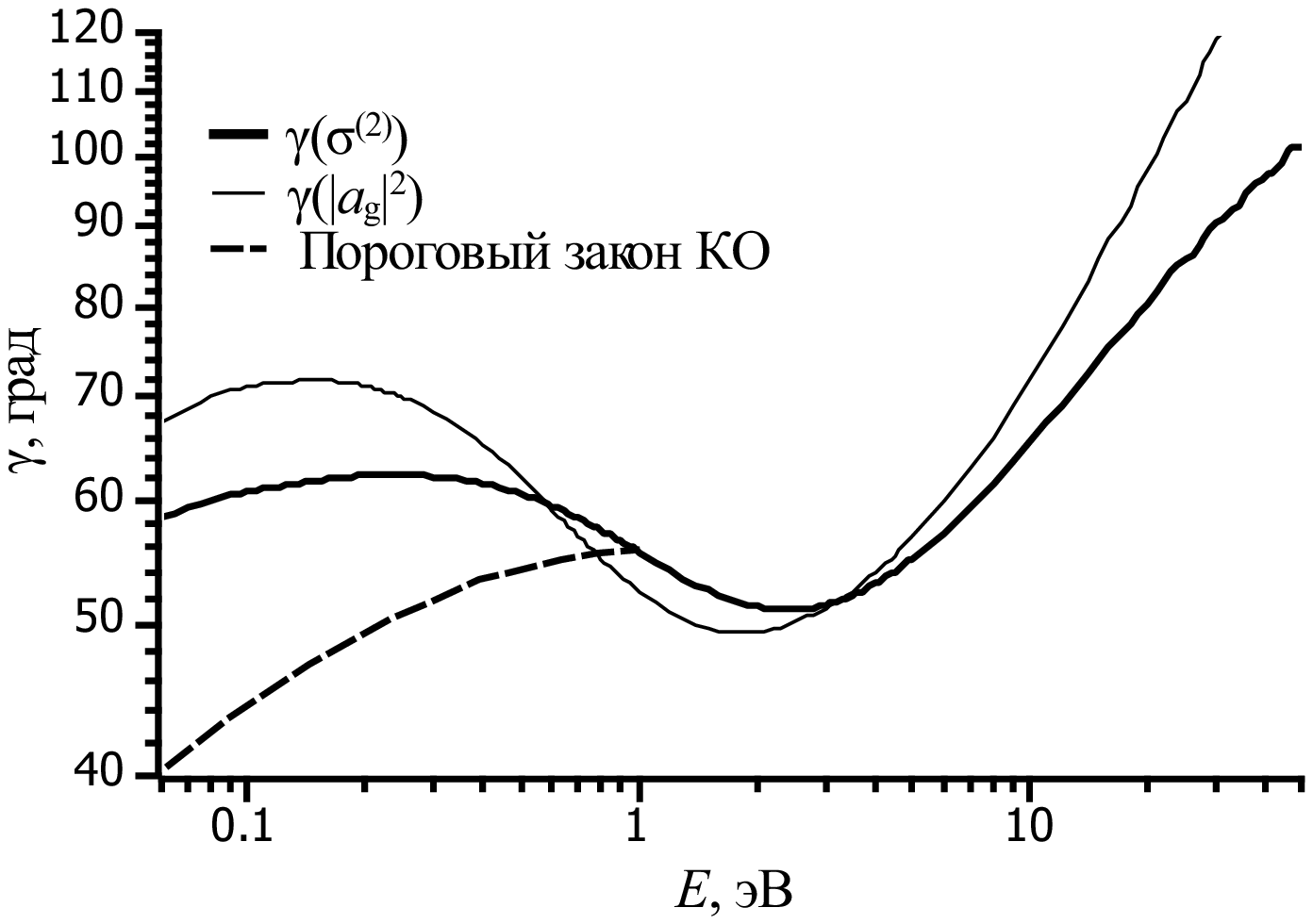}\\ \textit{а}}
\parbox{0.5\textwidth}{\includegraphics[angle=-90,width=0.5\textwidth]{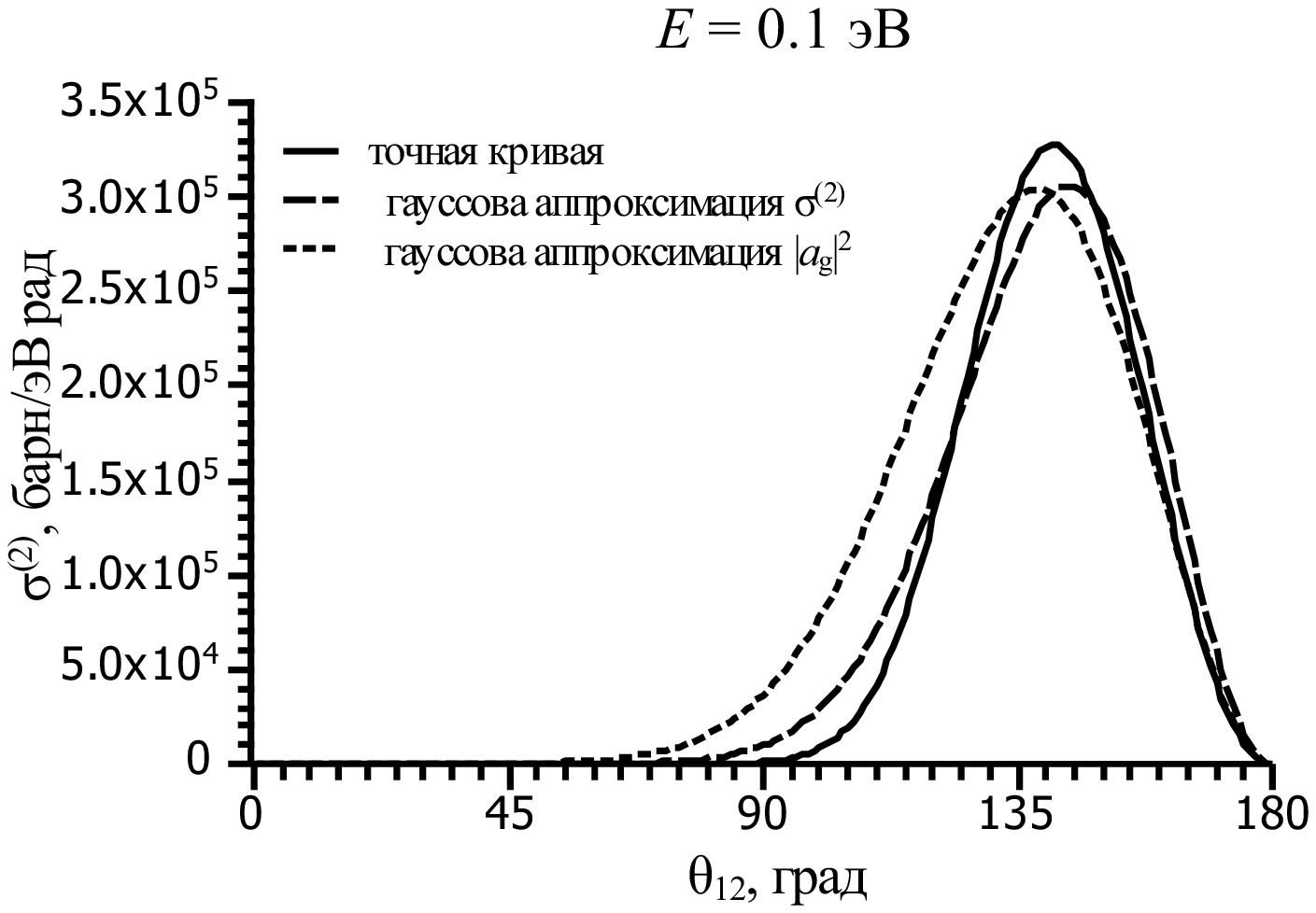}\\ \textit{б}}
\end{center}
\caption{Двукратная ионизация отрицательного иона водорода. \textit{а}) Гауссова ширина $\gamma$ как функция $E$: результаты TDS (сплошные кривая) и пороговый закон Казанского--Островского \cite{7} (штриховая кривая). \textit{б}) Двукратное дифференциальное сечение как функция межэлектронного угла $\theta_{12}$ для $E$=0.1 эВ (сплошная линия), ее прямая аппроксимация гауссовой кривой (штриховая кривая) и через гауссову аппроксимацию $|a_g(\theta_{12})|^{2}$ (пунктирная кривая).}\label{FIGgammaHminus}
\end{figure}

В работе \cite{Serov2008} сравнивались зависимости $\gamma(E)$ для различных гелиеподобных ионов.
Было обнаружено, что для отрицательного иона $\mathrm{H}^-$ эта функция начинает расти
 при энергиях ниже 2.5 эВ. Это резко отличается от поведения кривых, полученных для гелия и положительных гелиеподобных ионов. В \cite{SerovSergeeva2010} эти расчеты были продолжены до меньшей энергии 0.06 эВ; результаты представлены на рис. \ref{FIGgammaHminus}.
Из рис. \ref{FIGgammaHminus}, \textit{а} видно, что $\gamma(\sigma^{(2)})$ растет
с уменьшением энергии от 2.6 эВ до 0.23 эВ. На рис. \ref{FIGgammaHminus}, \textit{б}
показана зависимость 2ДС от межэлектронного угла $\theta_{12}$ и результаты гауссовой
аппроксимации этой зависимости для $E$ ниже значения, которое соответствует максимуму функции на \ref{FIGgammaHminus}, \textit{а}. При энергиях ниже 0.09 эВ функция $\gamma(\sigma^{(2)})$
оказывается степенной функцией энергии с показателем степени $s=0.083$
и $\tilde{\gamma_0}=74^{\circ}$ эВ$^{-s}$, однако, диапазон энергий слишком узок,
чтобы рассматривать это наблюдение как строгий вывод.
Как видно из рис. \ref{FIGgammaHminus}, \textit{а}, для отрицательного иона водорода
результаты TDS существенно отличаются от порогового закона КО, тогда как для атома
гелия (рис. \ref{FIGgamma0He}) имело место хорошее согласие.
Также следует отметить, что зависимость $\gamma(E)$, полученная Казанским и Островским,
монотонна (рис. \ref{FIGgammaHminus}, \textit{а}), хотя $\gamma_0(E)$ осциллирует.

Основная гипотеза, выдвигаемая в \cite{SerovSergeeva2010}, чтобы объяснить отличие результатов
для $\mathrm{H}^-$ от таковых для гелия и гелиеподобных ионов \cite{Serov2008},
заключается в сильном различии конфигураций начального состояния мишеней.
Действительно, при $r_{1,2} \rightarrow \infty$ волновая функция связанного
состояния  $\mathrm{H}^-$ имеет асимптотический вид
\begin{equation}
\varphi_0(\mathbf{r}_1,\mathbf{r}_2 )\sim e^{-r_1}\frac{e^{-0.235r_2}}{r_2}+e^{-r_2}\frac{e^{-0.235r_1}}{r_1}.\label{hydrowave}
\end{equation}
То есть $\mathrm{H}^-$ --- дейтроноподобная слабосвязанная система, состоящая из атома
водорода и электрона, проводящего большую часть времени вне области действия потенциала притяжения.
Для проверки этой гипотезы в \cite{SerovSergeeva2010} были проведены расчеты для других мишеней
с сильно асимметричной конфигурацией исходного связанного состояния, а именно для атома гелия в возбужденных состояниях $\mathrm{2s^1S}$ и $\mathrm{3s^1S}$.

\begin{figure}[t]
\begin{center}
\parbox{0.48\textwidth}{\includegraphics[angle=-90,width=0.45\textwidth]{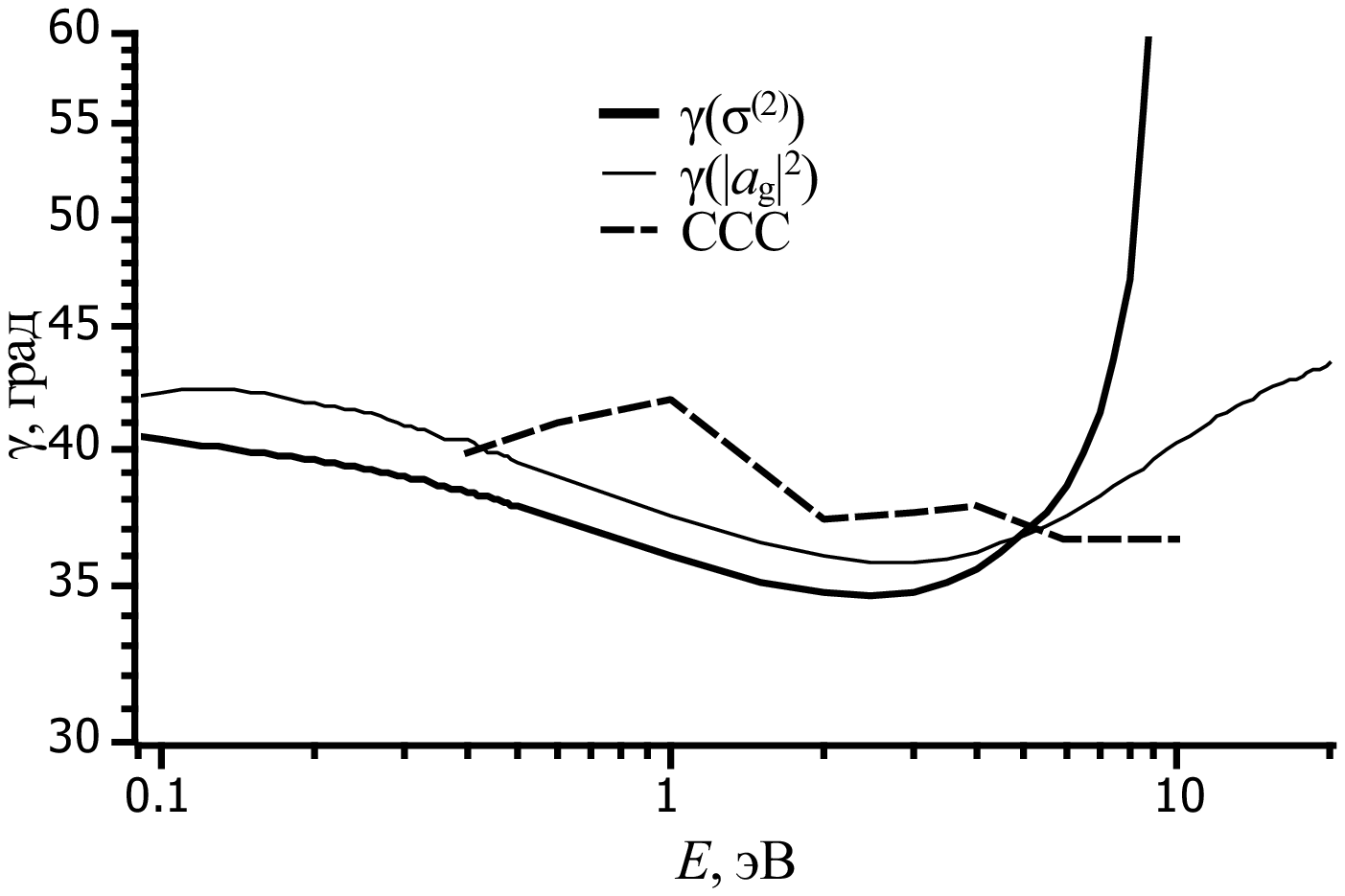}\\ \textit{а}}
\parbox{0.48\textwidth}{\includegraphics[angle=-90,width=0.5\textwidth]{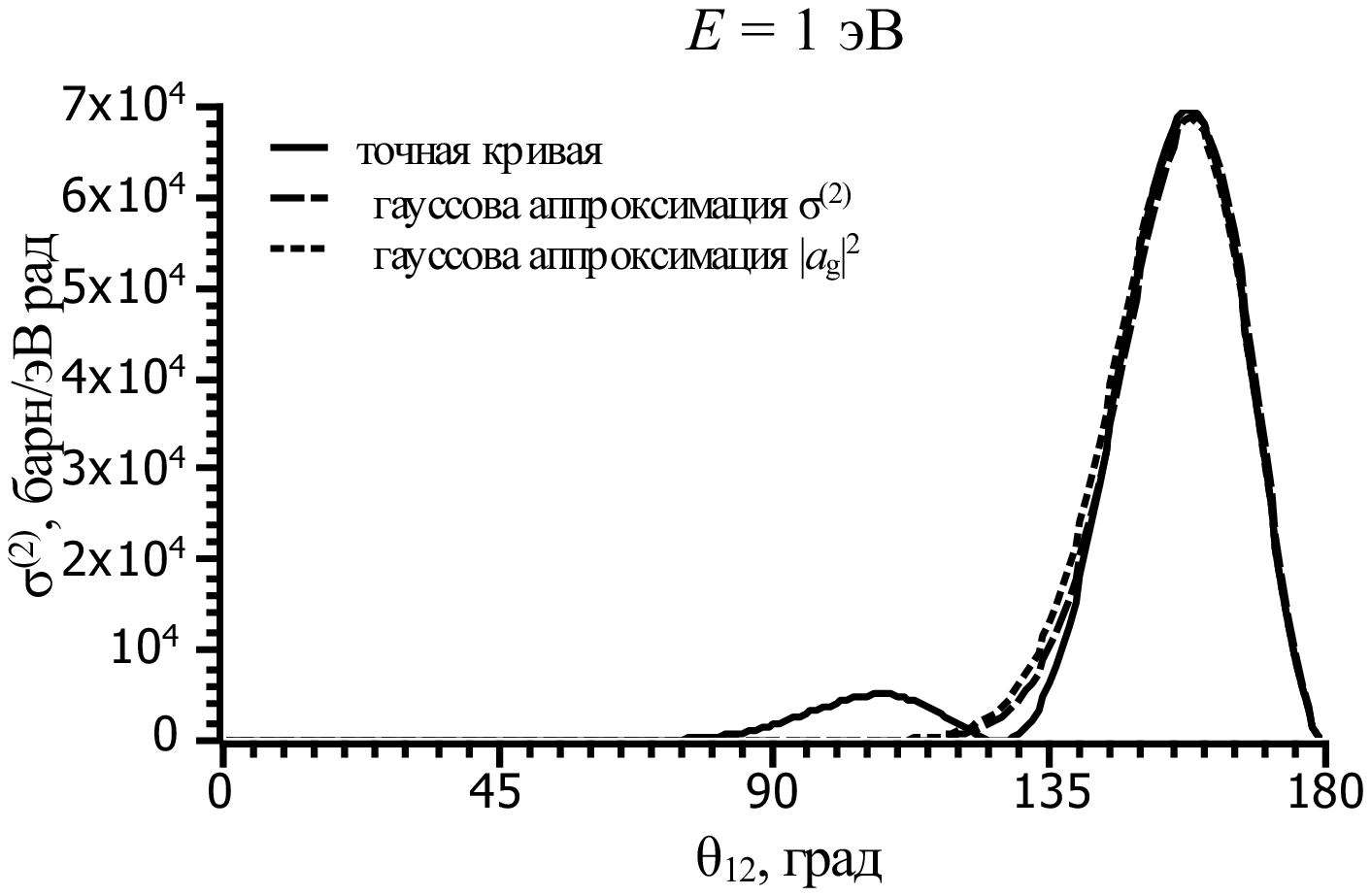}\\ \textit{б}}\\
\vspace{1cm}
\parbox{0.48\textwidth}{\includegraphics[angle=-90,width=0.5\textwidth]{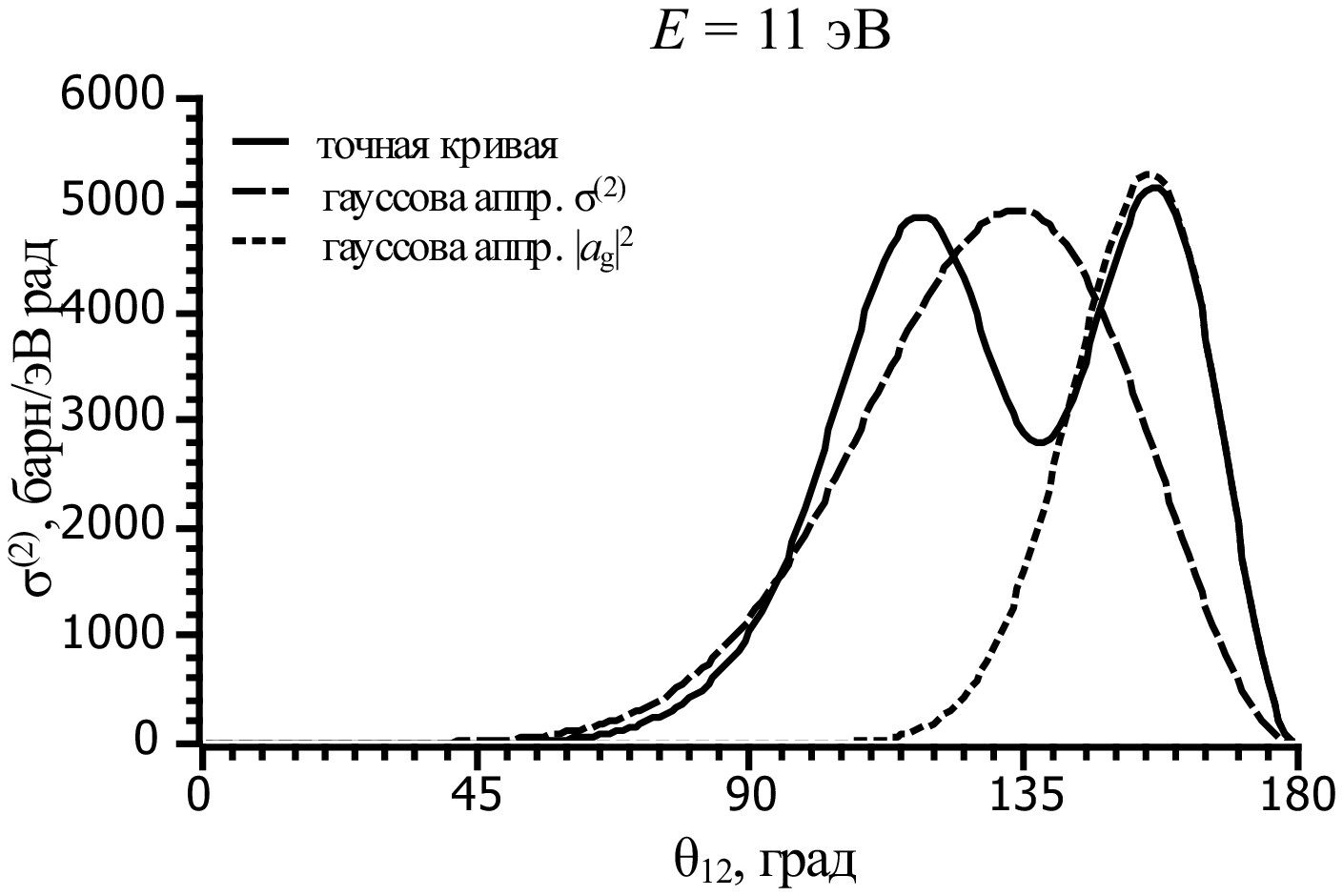}\\ \textit{в}}
\parbox{0.48\textwidth}{\includegraphics[angle=-90,width=0.5\textwidth]{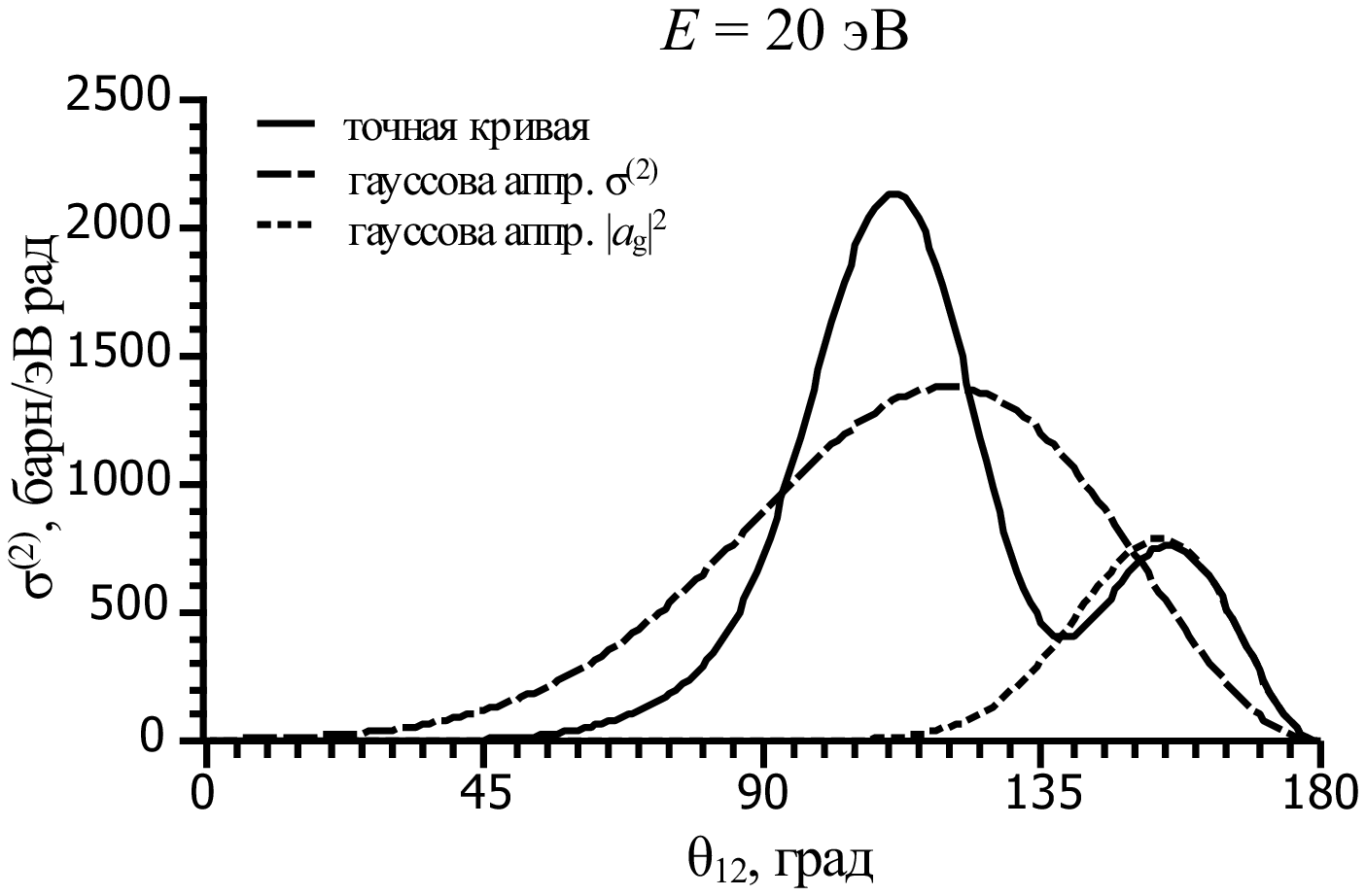}\\ \textit{г}}
\end{center}
\caption{Фотоионизация гелия из метастабильного состояния $\mathrm{2s^1S}$. \textit{а}) Гауссова ширина $\gamma$ как функция $E$: результаты TDS
(сплошные линии) и CCC \cite{10} (штриховая кривая).
2ДС как функция межъэлектронного угла $\theta_{12}$ для \textit{б}) $E$=1 эВ; \textit{в}) $E$=11 эВ; \textit{г}) $E$=20 эВ: точная (сплошная кривая), ее прямая аппроксимация гауссовой кривой (штриховая кривая) и через гауссову аппроксимацию $|a_g(\theta_{12})|^{2}$ (пунктирная кривая).}\label{FIGgammaHe2s}
\end{figure}

На рис. \ref{FIGgammaHe2s}, \textit{а} показана $\gamma$ как функция $E$ для фотоионизации гелия
из метастабильного состояния  $\mathrm{2s^1S}$. Наши результаты мало отличаются
от ССС-расчетов \cite{10} по величине, но сильно отличаются отсутствием осцилляций.
При понижении энергии $\gamma$, как и для H$^-$, вначале убывает, а затем начинает расти
(для $\gamma(\sigma^{(2)})$ это происходит при $E=2.5$
 эВ). К нашему удивлению, при $E>5$
эВ  разность между  $\gamma(|a_g|^2)$ и $\gamma(\sigma^{(2)})$
становится громадной. Причина этого становится ясной из рис.
\ref{FIGgammaHe2s}, \textit{б}-\textit{г}. Коэффициент корреляции
имеет форму, сильно отличающуюся от гауссовой, с двумя пиками даже
при низкой энергии $E=1$ эВ. При убывании энергии побочный пик
уменьшаетя и распределение превращается в гауссово, как следует из
теории Ванье. Когда энергия растет, побочный пик также увеличивается,
становясь больше основного при $E=11.3$  эВ, как видно из рис.
\ref{FIGgammaHe2s}, \textit{в} и \ref{FIGgammaHe2s}, \textit{г}. В
этом случае аппроксимация гауссовой функцией, естественно,
неприменима.

\begin{figure}[t]
\begin{center}
\parbox{0.48\textwidth}{\includegraphics[angle=-90,width=0.45\textwidth]{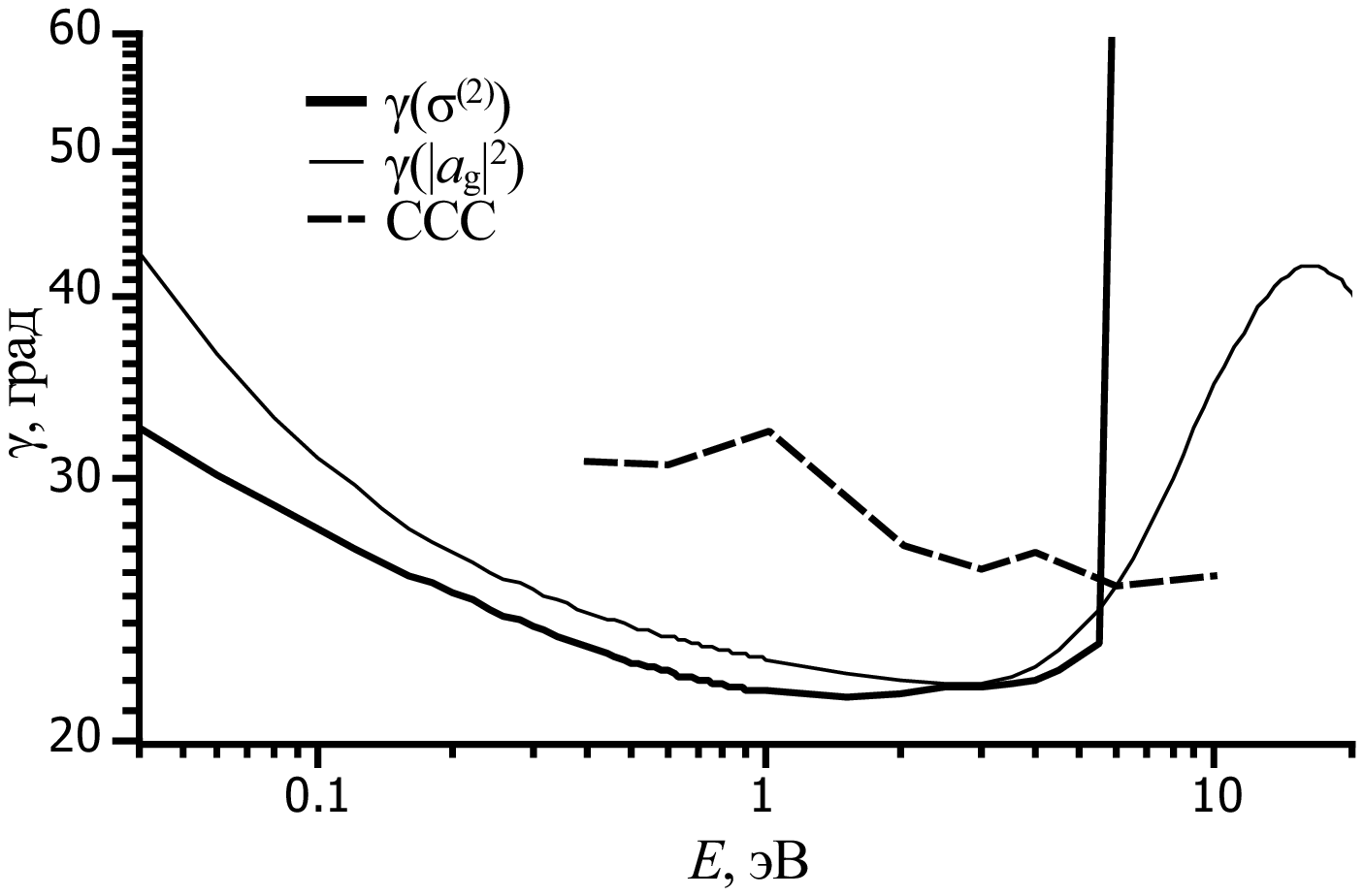}\\
\textit{а}}
\parbox{0.48\textwidth}{\includegraphics[angle=-90,width=0.5\textwidth]{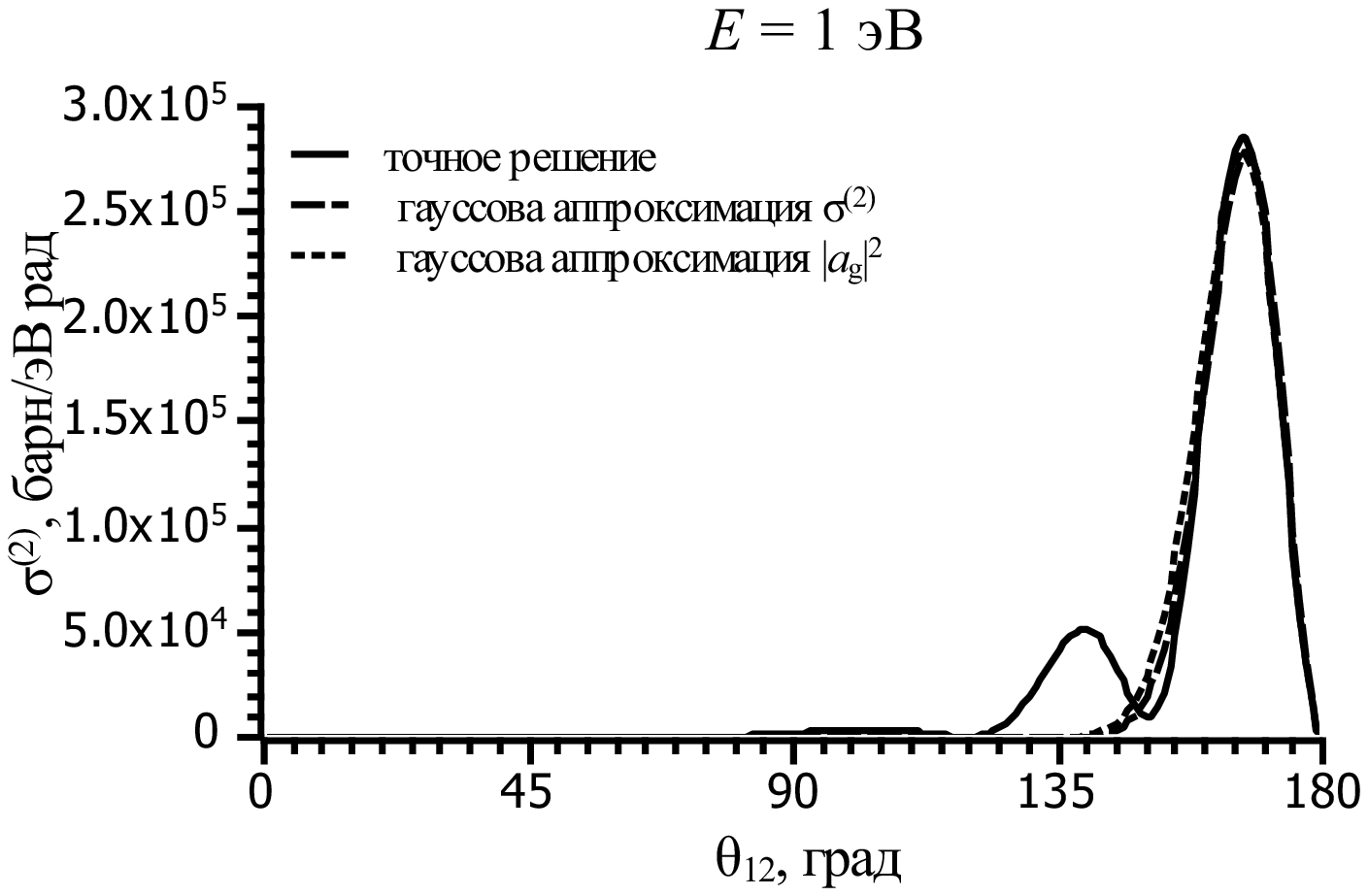}\\
\textit{б}}\\
\vspace{1cm}
\parbox{0.48\textwidth}{\includegraphics[angle=-90,width=0.5\textwidth]{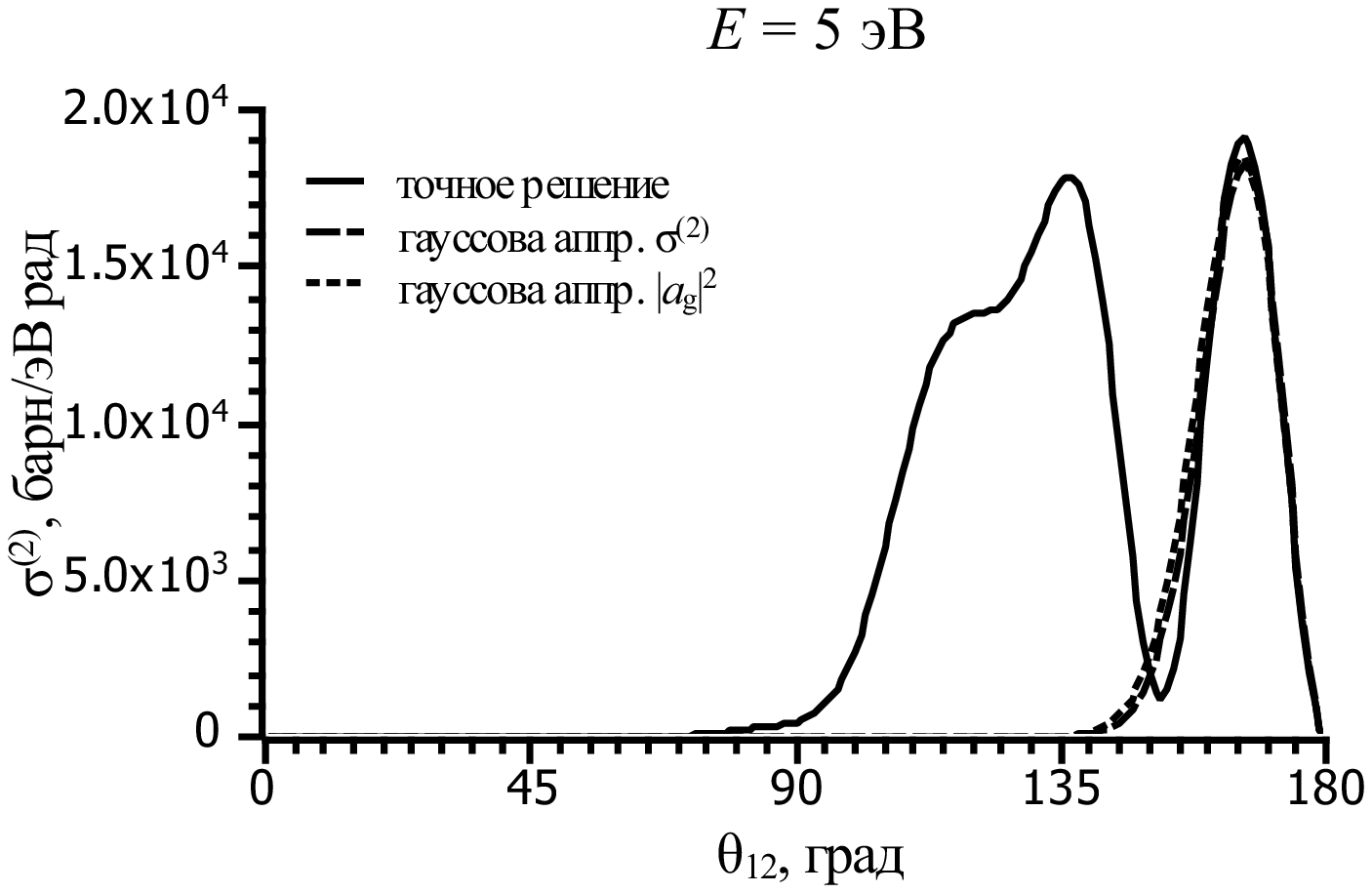}\\
\textit{в}}
\parbox{0.48\textwidth}{\includegraphics[angle=-90,width=0.5\textwidth]{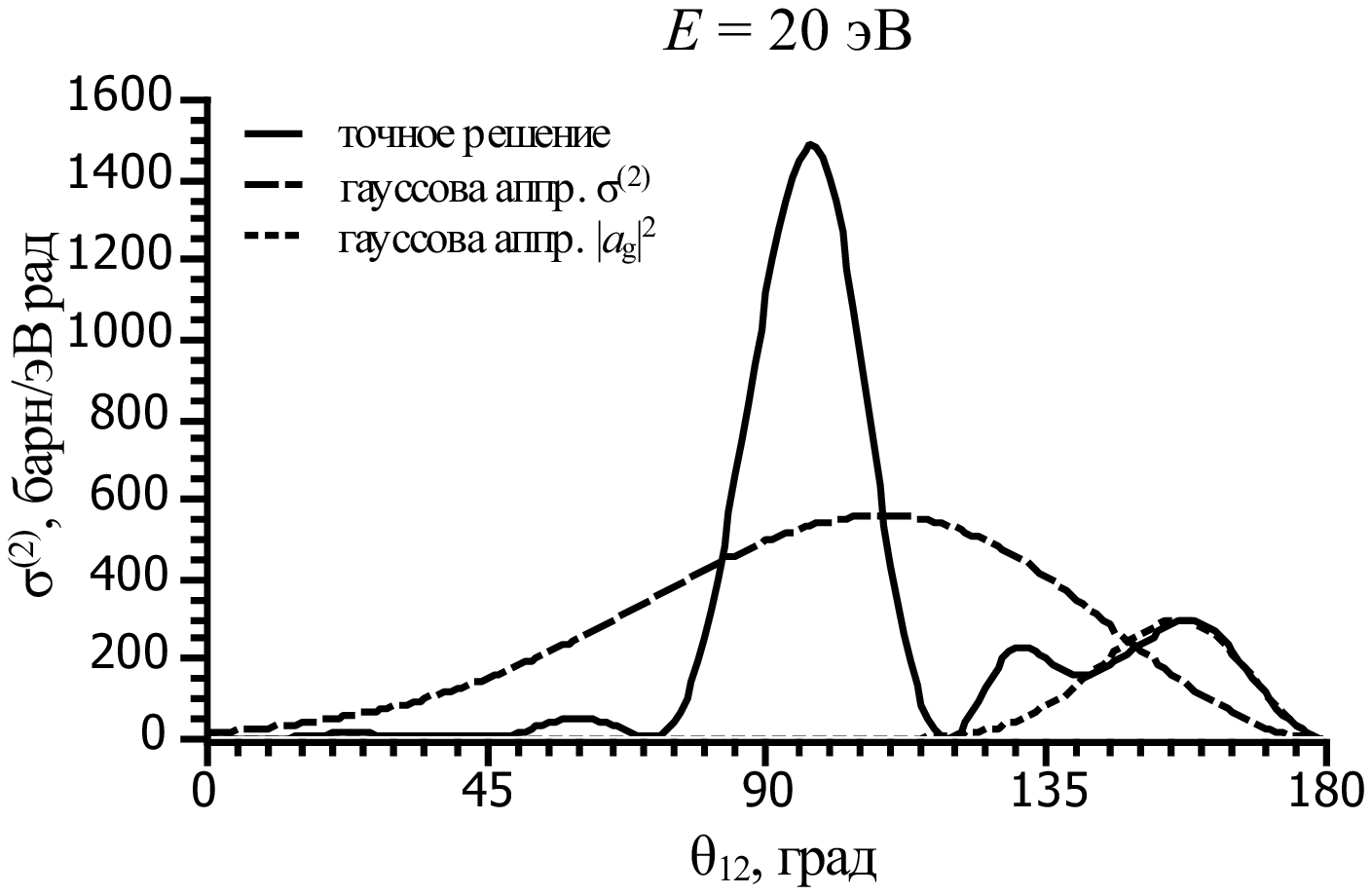}\\
\textit{г}}
\end{center}
\caption{Фотоионизация гелия из возбужденного состояния  $\mathrm{3s^1S}$. \textit{а}) Гауссова ширина $\gamma$ как функция $E$: результаты TDS (сплошные кривые) и  CCC \cite{10} (штриховая кривая).
2ДС как функция межъэлектронного угла $\theta_{12}$ для \textit{б}) $E$=1 эВ; \textit{в}) $E$=5 эВ; \textit{г}) $E$=20 эВ: точная (сплошная кривая), ее прямая аппроксимация гауссовой кривой (штриховая кривая) и расчет через гауссову аппроксимацию $|a_g(\theta_{12})|^{2}$ (пунктирная кривая).}\label{FIGgammaHe3s}
\end{figure}

На рис. \ref{FIGgammaHe3s}, \textit{а} показана зависимость $\gamma$ от $E$ для фотоионизации атома
гелия из состояния  $\mathrm{3s^1S}$. Наши результаты сильно отличаются от таковых,
полученных методом ССС \cite{10} как по величине, так и по характеру поведения.
Общий вид кривой аналогичен случаю ионизации из состояния $\mathrm{2s^1S}$,
однако $\gamma(\sigma^{(2)})$ достигает локального максимума при $E=1.5$
эВ. Из рис. \ref{FIGgammaHe3s}, \textit{б}--\textit{г} видно, что энергетическая зависимость
корреляционного параметра сильно негауссова, как и для  $\mathrm{2s^1S}$, но имеет не два,
а три пика при низких энергиях. С ростом энергии два меньших пика становятся выше и сливаются, так что при больших значениях $E$ форма зависимости $\sigma^{(2)}(\theta_{12})$ становится весьма сложной.

Можно заметить, что число пиков  $\sigma^{(2)}(\theta_{12})$ при низких энергиях равно числу пиков радиального
распределения плотности ``внешнего'' электрона в начальном состоянии мишени. Это число равно единице
для $\mathrm{H}^-$ (рис. \ref{FIGgammaHminus}, \textit{б}), двум для $\mathrm{He(2s^1S)}$
(рис. \ref{FIGgammaHe2s}, \textit{б}) и трем для $\mathrm{He(3s^1S)}$ (рис. \ref{FIGgammaHe3s},
\textit{б}). Причины возникновения дополнительных пиков в угловом
распределении были позже объяснены в работе \cite{Kheifets2011} на
примере $\mathrm{He(2s^1S)}$. Данный эффект вызван тем, что в
$\mathrm{2s^1S}$-состоянии у возбужденного электрона имеется два
хорошо изолированных облака вероятности, для которых характерны существенно
различающиеся уровни угловых корреляций с невозбужденным электроном.
При двойной ионизации каждое из этих облаков дает гауссово
распределение вида (\ref{GaussianAppr}) с шириной, соответствующей
уровню начальной угловой корреляции, а конечная симметричная
амплитуда является суммой этих двух гауссовых распределений с
противоположными знаками (из-за того, что знаки двух облаков в
начальной волновой функции противоположны).
В результате в зависимости $a_g(\theta)$ возникает узел и два пика.

Очевидно, что аналогичные рассуждения верны и для $\mathrm{3s^1S}$- состояния, с той разницей, что конечная амплитуда --- сумма трёх распределений (\ref{GaussianAppr}) с разными значениями гауссовой ширины. Это даёт три пика и два узла, что и наблюдается на рис. \ref{FIGgammaHe3s},
\textit{б}.

\section{Заключение}
В настоящем обзоре мы рассмотрели численные методы вычисления трехчастичных кулоновских волновых функций
непрерывного спектра и сечений физических процессов на их основе.
К таким процессам относятся ионизация ударом быстрого электрона и двойная фотоионизация.
 Изучение этих процессов важно для понимания явления межэлектронной корреляции, которое,
 в свою очередь, играет важнейшую роль в химии и физике твердого тела.

По причине обширности темы в данной работе мы сосредоточились в
основном на методах, которые применяли в своих собственных
работах. В обзоре было рассмотрено применение внешнего
комплексного скейлинга для обеспечения граничных условий уходящей
волны. Продемонстрировано, как расчет однократной и двукратной
ионизации одним фотоном или быстрым электроном можно свести к
решению шестимерного уравнения Шредингера с правой частью и
граничными условиями уходящей волны. Путем сопоставления результатов
разных авторов показана предпочтительность применения
сфероидальных координат для расчета ионизации двухатомных молекулы
по сравнению со сферическими. Рассмотрено использование
параксиального приближения волноводной оптики для расчета
ионизации ударом быстрой частицы и его пределы применимости.
Показано удобство применения метода расширяющейся координатной
сетки для расчета многократного сечения двухкратной ионизации по
сравнению с другими методами.

Нам представляется, что наиболее существенные физические результаты, которые получены авторами настоящего обзора с помощью изложенных в нем методов, заключаются в том, что:
  \begin{enumerate}
  \item Продемонстрирована некорректность закона Ванье для углового
  распределения двойной ионизации даже при очень малых энергиях.
  \item Показана прямая связь количества пиков в симметричной амлитуде
  двойной ионизации с количеством узлов в волновой функции начального
  состояния.
  \item Показано, что при ионизации атомов и двухатомных молекул ударом электрона
  промежуточной энергии отклонения углового распределения вылетивших
  электронов от результатов первого борновского приближения для
  рассеянных электронов в основном вызваны взаимодействием испущенного
  электрона с рассеянным после испускания.
  \item Показано, что, вопреки ожиданиям,
   бидипольный второй борновский член, описывающий последовательную
 двойную ионизацию ударом электрона промежуточных энергий, не вносит
  существенного вклада в многократное дифференциальное сечение.
  \item Для молекулы водорода с неравновесным расстоянием между атомами
продемонстрировано, что интерференция электронов, испущенных разными
центрами при однократной ионизации существенно влияет на вероятность двухкратной
ионизации и угловое распределение вылетевших при двухкратной ионизации электронов.
Это открывает возможность экспериментального наблюдения зависимости
двухцентровой интерференции от расстояния между ядрами, так как
двойная ионизация сопровождается диссоциацией, после которой по
энергии разлетевшихся протонов можно определить начальное расстояние
между ядрами. 
 \end{enumerate}

Работа поддержана грантом Президента РФ для молодых кандидатов наук MK-2344.2010.2
 и грантом РФФИ 11-01-00523-a  "Математическое моделирование
воздействия быстрых частиц,
лазерных импульсов и магнитных полей на атомы, молекулы и
полупроводниковые наноструктуры". Авторы благодарны проф. Б. Жулакяну (B. Joulakian)
за многолетнее сотрудничество и коллективу Лаборатории молекулярной физики и столкновений
Университета г. Мец (Франция)
(Laboratoire de Physique Mol\'eculaire et Collisions, Universit\'e Paul Verlaine--Metz, France)
 за поддержку исследований в рамках совместного проекта.
В заключение авторы благодарят И.В. Пузынина, В.В. Пупышева и участников семинара
по малочастичным системам Лаборатории Теоретической Физики
Объединенного Института Ядерных Исследований за полезные обсуждения
представленного  обзора.

\appendix

\section{Численные методы}

\subsection{Численная схема на основе представления дискретной переменной для решения
пятимерного параксиального уравнения}\label{sect5Dsolution}

Численная схема для расчета эволюции по времени основана на методе
расщепления \cite{Marchuk1982}. Для этого перпендикулярный (трансверсальный)
гамильтониан в уравнении (\ref{paraxialeq})
\begin{eqnarray}
\hat{H}_\perp=-\frac{1}{2\mu}\nabla_\perp^2+V(\mathbf{r}_1,\mathbf{r}_0) +\hat{H}(\mathbf{r}_1)
\end{eqnarray}
расщеплялся на 3 слагаемых, для каждого из которых пропагация во
времени выполнялась при помощи метода Кранка-Николсона, за
исключением эффективного гамильтониана мишени (\ref{effHam}), для
которого также выполнялось расщепление. Аппроксимация
пространственных операторов выполнялась с помощью представления
дискретной переменной (DVR) \cite{SerovMetodichka}.

Поперечные переменные падающего электрона были представлены в цилиндрической системе координат.
По угловой переменной $\phi_s$ использовалось разложение по функциям
\begin{eqnarray*}
\varphi_{m}(\phi)=\left\{
\begin{array}{ll}
\frac{1}{\sqrt{2\pi}},&m=0;\\
\frac{1}{\sqrt{\pi}}\cos{m\phi},&m>0;\\
\frac{1}{\sqrt{\pi}}\sin{m\phi},&m<0.
\end{array}\right.
\end{eqnarray*}
Для радиальной переменной $\rho_s$ использовался метод конечных элементов
на квадратурах Гаусса--Лобатто (гибрид ПДП и метода конечных элементов, МКЭ-ПДП
\cite{Rescigno2000, SerovMetodichka}), для обеспечения корректных граничных условий
при $\rho=0$ в первом конечном элементе использовалась квадратура Гаусса--Радау.
После выполнения шага с оператором поперечной части кинетической энергии падающего электрона осуществлялся (при помощи дискретного преобразования Фурье) переход к ПДП по угловой переменной $\phi_s$ с узлами квадратуры
\begin{eqnarray*}
\varphi_j=\frac{2\pi}{N_{\varphi}}(j-1);\;\;\; j=1,\ldots,N_{\varphi}
\end{eqnarray*}
($N_\phi$ - число узлов сетки по $\phi$), так что оператор потенциала становился диагональным.
А чтобы избежать сингулярности в точке $\mathbf{r}_1=\mathbf{r}_0$, вместо $1/|\mathbf{r}_1-\mathbf{r}_0|$ использовалось соотвествующее разложение Неймана. Для  обеспечения второго порядка точности чередование шагов метода расщепления менялось, то есть шаги выполнялись в порядке:
\begin{itemize}
\item действие оператора $-\frac{1}{2\mu}\nabla_\perp^2$,
\item обратное преобразование Фурье по переменной $\phi_s$,
\item действие оператора $U_i(\mathbf{r}_0)+\frac{1}{|\mathbf{r}_1-\mathbf{r}_0|}$,
\item действие оператора $2\hat{H}_i(\mathbf{r}_1)$,
\item действие оператора $U_i(\mathbf{r}_0)+\frac{1}{|\mathbf{r}_1-\mathbf{r}_0|}$,
\item прямое преобразование Фурье,
\item действие оператора $-\frac{1}{2\mu}\nabla_\perp^2$.
\end{itemize}
Поскольку шаг по времени полагался равным $\tau$, после прохождения всех слоев расщепления в прямом и обратном порядке получалась волновая функция в момент времени $t+2\tau$.

Для выполнения шага расщепления с гамильтонианом мишени в случае
одноцентровых мишеней использовался метод \cite{Melezhik1996} с тем отличием, что для радиальной переменной $r$
использовался МКЭ-ПДП \cite{SerovMetodichka}, а для подавления
нефизического отражения от границ сетки по $r$ применялся метод
внешнего комплексного скейлинга (см. раздел \ref{sectECS}).
Поскольку для двухцентровых мишеней с большими зарядами ядер
волновая функция сходится с увеличением числа базисных функций чрезвычайно медленно, для них использовались
сфероидальные (эллиптические) координаты и разложение функции по
базису
\begin{eqnarray*}
\Phi_{ijm}(\xi,\eta,\phi)=\sqrt{\frac{8}{R^3(\xi_i^2-\eta_j^2)}}f_{mi}(\xi)\varsigma_{mj}(\eta)\varphi_m(\phi).
\end{eqnarray*}
Здесь $R$ --- расстояние между ядрами (ориентация молекулярной оси относительно направления
падения $\mathbf{k}_i$ могла выбиратся произвольной),
\begin{eqnarray*}
f_{mi}(\xi)=\left\{\begin{array}{ll}
f_i(\xi),& \text{ нечетное } m;\\
\frac{\xi_i}{\sqrt{\xi_i^2-1}}\frac{\sqrt{\xi^2-1}}{\xi}f_i(\xi),& \text{ четное } m;
\end{array}\right.
\end{eqnarray*}
где $f_i(\xi)$ --- базисные функции МКЭ-ПДП, составленные из кусков полиномов Лагранжа,
удовлетворяющие условию $f_i(\xi_{i'})=\delta_{ii'}/\sqrt{w_i}$, $\xi_i$ и $w_i$ --- узлы и веса квадратуры, составленной из квадратуры Гаусса--Радау для первого конечного элемента и Гаусса--Лобатто для остальных, при $\rho=\rho_{max}$ ставилось граничное условие Неймана. Введение $f_{mi}(\xi)$ обеспечивает корректную асимпототику при $\xi=1$ для нечетных $m$ при сохранении основного свойства функции Лежандра $f_{mi}(\xi_{i'})=\delta_{ii'}/\sqrt{w_i}$. В качестве угловых базисных функций использовались функции Лежандра
\begin{eqnarray*}
\varsigma_{mj}(\eta)=\sqrt{\varpi_j}\sum_{l=0}^{l_{max}}\bar{P}^m_l(\eta_j)\bar{P}^m_l(\eta),
\end{eqnarray*}
где $\eta_j,\varpi_j, j=1,\ldots,l_{max}+1$ --- узлы и веса квадратуры Гаусса--Лежандра на отрезке $\eta=[-1,1]$, $\bar{P}^m_l(\eta)$ --- присоединенные полиномы Лежандра, ортонормированные на квадратуре Гаусса--Лежандра \cite{Melezhik1996}. Таким образом, в матрице гамильтониана мишени выделяются два слагаемых: квазирадиальный гамильтониан
\begin{eqnarray*}
H^{mj}_{\xi ii'}=\frac{2}{R^2\sqrt{(\xi_i^2-\eta_j^2)(\xi_{i'}^2-\eta_{j}^2)}}\left[\int_{0}^{\xi_{max}}f_{mi}'(\xi)(\xi^2-1)f_{mi'}'(\xi)d\xi+R(Z_1+Z_2)\xi_i\delta_{ii'}\right],
\end{eqnarray*}
являющийся матрицей с полушириной ленты $s$  ($s$ --- порядок конечных элементов)
и рангом $N_{\text{nodes}}=sN_{r}$  ($N_{r}$ --- число конечных элементов), и квазиугловой гамильтониан
\begin{eqnarray*}
\!H^{mi}_{\eta
jj'}\!=\!\frac{2}{R^2\sqrt{(\xi_i^2\!-\!\eta_j^2)(\xi_i^2\!-\!\eta_{j'}^2)}}
\left[\sqrt{\varpi_j\varpi_{j'}}\sum_{l=|m|}^{l_{max}+|m|}\overline{P}^m_l(\eta_j)l(l+1)
\overline{P}^m_l(\eta_{j'})\!+\!R(Z_1-Z_2)\eta_j\delta_{jj'}\right]
,
\end{eqnarray*}
являющийся полностью заполненной квадратной матрицей ранга $l_{max}+1$. Таким образом,
гамильтониан мишени расщеплялся на четыре слагаемых:
$\hat{H}_{\xi}$, $\hat{H}_{\eta}$, $U_{sh}(\mathbf{r})=U_i(\mathbf{r})+Z_1/|\mathbf{r}_1-\mathbf{R}/2|+Z_2/|\mathbf{r}_1+\mathbf{R}/2|$ и $\hat{X}_i$. После выполнения шагов для $\hat{H}_{\xi}$ и $\hat{H}_{\eta}$ выполнялся переход к ПДП для $\phi$ с помощью Фурье-преобразования, благодаря чему матрица потенциала $U_{sh}(\mathbf{r})$ (содержащего средний потенциал неактивных оболочек и всех ядер, кроме первых двух) становилась диагональной. Шаг с приближенным обменным оператором $\hat{X}_i$ выполнялся таким образом: волновая функция раскладывалась по волновым функциям оболочек $\varphi_n(\mathbf{r})$, после чего выделялась часть волновой функции, ортогональная им всем:
\begin{eqnarray*}
C_n(t)&=&\int\varphi_n^*(\mathbf{r})\psi(\mathbf{r},t)d\mathbf{r};\\
\psi_{rest}(\mathbf{r},t)&=&\psi(\mathbf{r},t)-\sum_{n=1}^{N}C_n(t)\varphi_k(\mathbf{r}).
\end{eqnarray*}
Шаг по времени для коэффициентов $C_n$ выполнялся по схеме Кранка--Николсона
\begin{eqnarray*}
\mathbf{C}(t+\tau)&=&\left[\mathbf{I}+\frac{i\tau}{2}\mathbf{X}\right]^{-1}\left[\mathbf{I}-\frac{i\tau}{2}\mathbf{X}\right]\mathbf{C}(t),
\end{eqnarray*}
где $I_{nk}=\delta_{nk}$ и
\begin{eqnarray*}
X_{nk}&=&\langle \varphi_n|\hat{X}_i|\varphi_k\rangle+\frac{1-\delta_{ni}}{\tau}\delta_{nk}.
\end{eqnarray*}
Ко всем диагональным элементам матрицы $\mathbf{X}$, кроме $i$-го
(который соответствует номеру активного электрона), мы добавили
большие числа $1/\tau$, чтобы подавить запрещенный принципом Паули
переход активного электрона в состояния, занятые другими
электронами. После выполнения этого шага изменившаяся за счет обмена часть
волновой функции добавлялась к остаточной части
\begin{eqnarray*}
\psi(\mathbf{r},t+\tau)&=&\psi_{rest}(\mathbf{r},t)+\sum_{n=1}^{N}C_n(t+\tau)\varphi_k(\mathbf{r}).
\end{eqnarray*}
Таким образом, конечный порядок шагов при вычислении эволюции электрона мишени: $U_{sh}(\mathbf{r})$, $\hat{X}_i$, $\hat{H}_{\xi}$, $2\hat{H}_{\eta}$, $\hat{H}_{\xi}$, $\hat{X}_i$, $U_{sh}(\mathbf{r})$.

\subsection{Решение уравнения Шредингера с правой частью в сфероидальных координатах для
двухатомных молекул с двумя активными электронами}\label{sectH2solution}

В работах \cite{Serov2009,Serov2010} уравнение (\ref{drivenSchr}) решалось для каждого
собственного значения $M$ $z$-компоненты полного углового момента двух электронов при помощи
разложения
\begin{eqnarray}
\!\!\psi^{(+)}_M(\mathbf{r}_1,\mathbf{r}_2)\!=\!
\sum_{m}\!\!\sum_{l_1=|M-m|}^{N_l-1+|M-m|}\sum_{l_2=|m|}^{N_l-1+|m|}
\!\!\!\sum_{j_1,j_2=1}^{N_{\xi}}\!\!\!\psi_{Mmj_1j_2l_1l_2}\varphi_{j_1l_1,M-m}(\xi_1,\eta_1,\phi_1)\varphi_{j_2l_2m}(\xi_2,\eta_2,\phi_2),\label{psi_expansion}
\end{eqnarray}
где
\[ \varphi_{j\,lm}(\xi,\eta,\phi)=b_{mj}(\xi)Y_{lm}(\arccos\eta,\phi). \]
В качестве ``радиального'' базиса использовались b-сплайны порядка $k$ \cite{deBoor2001},
модифицированные для согласования с квадратнокорневыми
асимптотиками при $\xi\to 1$ для $m\neq 0$
\[ b_{mj}(\xi)= \left\{ \begin{array}{ll} b^k_j(\xi),& \text{если  } m \text{  четное};  \\ \frac{\sqrt{\xi^2-1}}{\xi}b^k_j(\xi), & \text{если  } m \text{  нечетное}. \end{array} \right. \]
Здесь и далее $N_{\xi}$ --- число b-сплайнов,
$N_l$ --- число сферических гармоник на электрон. Заметим, что $l$
является квантовым числом углового момента только при
$\xi\to\infty$. Набор из $N_m$ элементов $m$ --- собственных значений
компоненты углового момента вдоль молекулярной оси --- выбирается
так, чтобы не нарушать симметрию волновой функции.

При вычислении матричных элементов потенциала электрон-электронного взаимодействия в вытянутых сфероидальных координатах, чтобы избежать сингулярностей, удобно использовать разложение Неймана \cite{Neumann}
\begin{eqnarray}
\frac{1}{|\mathbf{r}_1-\mathbf{r}_2|}=\sum_{\mu=-\infty}^{\infty}\sum_{\lambda=|\mu|}^{\infty}U_{\lambda\mu}(\xi_1,\xi_2)Y_{\lambda\mu}(\arccos\eta_1,\phi_1)Y_{\lambda,-\mu}(\arccos\eta_2,\phi_2),
\label{Neuman}
\end{eqnarray}
где
\begin{eqnarray}
U_{\lambda\mu}(\xi_1,\xi_2)=\frac{8\pi}{R}\left[\frac{(\lambda-\mu)!}{(\lambda+\mu)!}\right]P_\lambda^\mu(\xi_<)Q_\lambda^\mu(\xi_>),
\end{eqnarray}
Здесь $P_l^m(x)$ и $Q_l^m(x)$ ---
регулярная и нерегулярная функция Лежандра. Так
как разложение угловой волновой функции (\ref{psi_expansion})
ограничено величинами  $N_l$ и $N_m$, мы взяли пределы
разложения Неймана (\ref{Neuman}) такими, что
$|\mu|\leq\mu_{max}\leq N_m-1$,
$\lambda\in[\mu_{max},\mu_{max}+\lambda_{max}]$,
$\lambda_{max}\leq N_l-1$.
 В результате мы получили линейную систему с разреженной матрицей
\begin{eqnarray}
(\mathbf{H}_1\otimes\mathbf{S}_2 +
 \mathbf{S}_1\otimes\mathbf{H}_2 + \mathbf{U}_{12}
 -E\;\mathbf{S}_1\otimes\mathbf{S}_2)\cdot\boldsymbol{\psi}=\mathbf{f}.
 \label{linsys}
\end{eqnarray}
Здесь $\mathbf{S}_\alpha$ --- одночастичные матрицы перекрытия с
элементами
\begin{eqnarray}
 S_{jlm,j'l'm'}&=&\langle
 jlm|j'l'm'\rangle=\frac{R^3}{8}\int\varphi_{j\,lm}^*\varphi_{j'\,l'm'}\,(\xi^2-\eta^2)d\xi\,d\eta\,d\phi\nonumber\\
 &=&\frac{R^3}{8}\delta_{mm'}\left[\delta_{ll'}\int_1^{\xi_{max}}b_{mj}\xi^2b_{mj'}d\xi-\langle
 lm|\eta^2|l'm\rangle\int_1^{\xi_{max}}b_{mj}b_{mj'}d\xi\right] ,
\end{eqnarray}
а $\mathbf{H}_\alpha$ --- матрицы одночастичного гамильтониана с
элементами
\begin{eqnarray}
 H_{jlm,j'l'm'}&=&\frac{R}{4}\delta_{mm'}\delta_{ll'}\nonumber\\
 &\times&\left\{\int_1^{\xi_{max}}b_{mj}'\,(\xi^{2}-1)b_{mj'}'d\xi+\int_1^{\xi_{max}}b_{mj}\left[\frac{m^2}{\xi^{2}-1}+l(l+1)-\frac{RZ_+}{2}\xi\right]b_{mj'}d\xi\right\}
 \nonumber
\end{eqnarray}
($Z_+$ --- полный заряд ядер, полагавшийся для H$_2$ равным 2).

Система линейных уравнений (\ref{linsys}) решается методом
сопряженных градиентов (СГ) с прекондиционером \cite{Kalitkin1978}
\begin{eqnarray*}
\mathbf{\tilde{A}}=\mathbf{H}_1\otimes\mathbf{S}^D_2 +
 \mathbf{S}^D_1\otimes\mathbf{H}_2
 -E\;\mathbf{S}^D_1\otimes\mathbf{S}^D_2 ,
\end{eqnarray*}
где $\mathbf{S}^D$ представляет диагональную часть $\mathbf{S}$
относительно индекса $l$,
$S^D_{jlm,j'l'm}=S_{jlm,j'lm}\delta_{ll'}$, а $\mathbf{S}$
трехдиагональна по $l$.

Для разбиения оператора возмущения первого борновского порядка (\ref{FBexp})
на парциальные компоненты применялось разложение (\ref{PWE}).
Пределы этого разложения выбирались в соответствии с модулем вектора передачи импульса $K$.
В расчетах, касающихся простой (e,2e) ионизации, мы брали парциальные волны
до $L_{max}=4$, $M_{max}=3$ в случае $\theta_s=1^{\circ}$ ($K=0.3233$) и $L_{max}=5$,
$M_{max}=4$ для $\theta_s=3^{\circ}$ ($K=0.9087$). В расчетах (e,3e), когда $K=0.6682$  а.е.,
мы использовали $L_{max}=4$, $M_{max}=4$.

Для унификации и большей ясности определения радиальной сетки введем
вспомогательную переменную $\tilde{r}=\frac{R}{2}(\xi-1)$. В
расчетах мы использовали b-сплайны четвертого порядка на равномерной
сетке с шагом  $\Delta\tilde{r}=0.5$ и $\tilde{r}_{max}=50$, что
делает число сплайнов равным $N_{\xi}=104$. Параметры углового
базиса выбирались равными $N_l=6$, $N_m^{\Sigma}=7$ для четных $M$ и
$N_m^{\Pi}=8$ для нечетных $M$. Чтобы гарантировать выполнение
граничных условий уходящей волны, мы применяли метод внешнего
комплексного скейлинга \cite{Simon1979,ReviewMcCurdy2004}, в соответствии с которым осуществлялся поворот контура в точке $\tilde{r}_{ECS}=40$ в комплексную плоскость на
угол $\theta_{ECS}=45^{\circ}$. Разложение Неймана (\ref{Neuman})
было ограничено $\lambda_{max}=N_l-1$, $\mu_{max}=N_m^{\Pi}/2$. Для
таких параметров метод сопряженных  сходился к относительной ошибке
$10^{-6}$ после примерно  20--30 итераций.

В расчетах (e,2e) мы использовали меньший угловой базис: $N_m=3$ для $M=0$ и $N_m=2$
для других значений $M$, $N_l=6$, $\lambda_{max}=3$ и
$\mu_{max}=1$. Наблюдение показывает, что для достижения той же
относительной точности в методе СГ требуется большее число
итераций (около 60--70).  Это, очевидно, объясняется тем фактом,
что конечная энергия $E<0$ лежит ниже порога двойной ионизации.
Сравнение сечений, рассчитанных при различных размерах базиса (см.
таблицу \ref{DPIconv}) и других параметрах схемы  (радиус
пространственной области $\tilde{r}_{max}$, угол комплексного
скейлинга $\theta_{ECS}$, точка поворота контура $\tilde{r}_{ECS}$
и радиус выделения амплитуды $\tilde{r}_{S}$) показывает, что
численная ошибка наших расчетов не превышает  2\% для расчетом ДФИ
и  5\% в расчетах (e,3e).
\begin{table}[!ht]
\caption{Сходимость сечения двойной ионизации $H_2$ фотонами с
энергией 75 эВ в зависимости от шага сетки $\Delta\tilde{r}$ и
параметра углового базиса $N_l$. Другие параметры полягались равными $k=4$,
$\tilde{r}_{max}=50$, $\tilde{r}_{ECS}=40$,
$\theta_{ECS}=45^{\circ}$, $N_m^{\Pi}=2(N_l-2)$,
$N_m^{\Sigma}=N_m^{\Pi}-1$, $\lambda_{max}=N_l-1$,
$\mu_{max}=N_m^{\Pi}/2$ и $\tilde{r}_{S}=38$.\label{DPIconv}}
\begin{ruledtabular}
\begin{tabular}{c|ccc}
\backslashbox{$N_l$}{$\Delta\tilde{r}$}
    & 1.0     & 0.5     & 0.333333 \\
 \hline
  4 & 3165.03 & 3034.37 & 3028.05 \\
  5 & 2963.59 & 2836.08 & \\
  6 & 2899.06 & 2773.35 & \\
  7 & 2891.90 &         & \\
\end{tabular}
\end{ruledtabular}
\end{table}

Мы начали с расчета волновой функции связанного состояния H$_2$,
которая необходима как функция начального состояния
$\varphi_0$ в правой части уравнения (\ref{drivenSchr}). Будучи
собственной функцией того же гамильтониана, она ортогональна
функции конечного состояния в (\ref{trans_ampl}). Использовался
непрерывный аналог метода Ньютона (НАМН) \cite{Puzynin1976,PuzyninReview2007},
который на каждом шаге дает систему линейных уравнений
типа (\ref{linsys}). При межъядерном расстоянии  $R=1.4$ а.е.
и при вышеназванных параметрах базиса для расчета ДФИ мы получили
энергию Борна-Оппенгеймера $E_{H_2}=-1.17419$ а.е. при точном
значении  $E_{H_2}=-1.174475714220$ а.е. \cite{Sims2006}.

Для расчетов двойной фотоионизации желательно использовать b-сплайны
не ниже четвертого порядка, поскольку при проведении интегрирования
по гиперсфере возникает небольшой член перекрывания пробной функции
(\ref{DItestFunction}) двойного континуума и состояниями однократной
ионизации. Сами численно рассчитанные функции континуума  $H_2^+$,
использованные в (\ref{DItestFunction}), точно (с машинной точностью)
ортогональны функциям связанных состояний  $H_2^+$, и перекрывание
возникает только за счет интегрирования по искривленной
гиперповерхности. Мы заметили, что этот артефакт очень чувствителен
к гладкости производных волновой и пробной функций в
(\ref{f_via_flux}). Поэтому для кубических b-сплайнов, имеющих
разрыв производной третьего порядка, это перекрывание проявляется
как малые осцилляции рассчитанного сечения. Величина этих
осцилляций, вообще говоря, имеет порядок численной ошибки и обычно не
влияет на результаты. Однако в некоторых случаях, как, например, при
определении изменения параметра асимметрии $\beta$ с изменением
энергии вылетевшего электрона, эти осцилляции ясно видны. Поэтому
 для расчета ДФИ применялись b-сплайны четвертого порядка \cite{deBoor2001}, что, как мы
убедились, позволяет избавиться от указанных осцилляций.

\subsection{Численный метод для шестимерного временного уравнения Шредингера на основе
преобразования Чанга-Фано и представления дискретной переменной}\label{sectChangFano}

Рассмотрим гелиеподобный атом и используем как параксиальное приближение,
так и метод сопутствующих координат. Соответствующее двухэлектронное неоднородное уравнение
шредингеровского типа принимает вид
\begin{equation}
i\frac{\partial
\Psi(\boldsymbol{\xi}_1,\boldsymbol{\xi}_2,t)}{\partial t}=
\left\{\hat{h}_1(t)+\hat{h}_2(t)+U(|\boldsymbol{\xi}_2-\boldsymbol{\xi}_1|,t)\right\}\Psi(\boldsymbol{\xi}_1,\boldsymbol{\xi}_2,t)
+F(\boldsymbol{\xi}_1,\boldsymbol{\xi}_2,t).\label{BasicEq}
\end{equation}
Здесь $\hat{h}_{1,2}(t)$ даны выражениями (\ref{H_soputstv}),  $U(\xi_{12},t)$ --- потенциал
межэлектронного взаимодействия,
\[U(\xi_{12},t)=\frac{1}{a(t)\xi_{12}}.\]
Член $F(\boldsymbol{\xi}_1,\boldsymbol{\xi}_2,t)$ в правой части (\ref{BasicEq}) отсутствует
при расчете фотоионизации \cite{Serov2007,SerovSergeeva2010}, но появляется при использовании PA1B приближения для двухкратной ионизации
ударом быстрого электрона \cite{Serov2007}.

После окончания внешнего воздействия полный угловой момент и его проекция сохраняются,
так что решение можно представить как
\begin{equation}
\Psi(\boldsymbol{\xi}_1,\boldsymbol{\xi}_2,t)=\sum_{LM}\Psi^{LM}(\boldsymbol{\xi}_1,\boldsymbol{\xi}_2,t),
\end{equation}
где $L$ and $M$ - квантовые числа полного углового момента и его
проекции, соответственно. Разложим $\Psi^{LM}$ по бисферическим
гармоникам \cite{Varshalovich1989}
\begin{equation}
\Psi^{LM}(\boldsymbol{\xi}_1,\boldsymbol{\xi}_2,t)= \frac{1}{\xi_1
\xi_2} \sum_{l_1,l_2}
 Y^{LM}_{l_1 l_2}(\theta_1,\varphi_1,\theta_2,\varphi_2) \psi^{LM}_{l_1
 l_2}(\xi_1,\xi_2,t).\label{BSBasis}
\end{equation}
Коэффициенты аналогичного разложения источника
$F(\boldsymbol{\xi}_1,\boldsymbol{\xi}_2,t)$ обозначим
\begin{equation*}
f^{LM}_{l_1 l_2}(\xi_1,\xi_2,t)=\xi_1 \xi_2 <LM l_1
l_2|F(\boldsymbol{\xi}_1,\boldsymbol{\xi}_2,t)>.
\end{equation*}
Для решения получившейся системы уравнений мы используем метод
двуциклического расщепления \cite{Marchuk1982}, в котором каждый шаг
эволюции расщепляется на подшаги, легко реализуемые с помощью
простой схемы Кранка-Николсона \cite{CrankNicolson}. Вводя
радиальную сетку $\left\{\xi_{1i_1},\xi_{2i_2}\right\}$,
$i_{\alpha}=1..N_r$, дискретные временные точки $t_n$ и обозначение
$\psi^{LM}_{l_1 l_2 i_1 i_2 n}=\psi^{LM}_{l_1
l_2}(\xi_{1i_1},\xi_{2i_2},t_n)$, можно осуществить эволюцию от
$t_n$ к $t_{n+2}$ по следующей схеме
\begin{subequations}
\begin{eqnarray}
\psi^{LM}_{l_1 l_2 i_1 i_2 n+1/4}&=&\psi^{LM}_{l_1 l_2 i_1 i_2 n}-i\tau_n f^{LM}_{l_1 l_2 i_1 i_2 n};\\
\left[\delta_{i_1}^{j_1} + \frac{i\tau_n}{2}\mathcal{H}_{l_1
i_1}^{j_1}(t_{n+1}) \right]\psi^{LM}_{l_1 l_2 j_1 i_2 n+1/2}&=&
\left[\delta_{i_1}^{j_1} - \frac{i\tau_n}{2}\mathcal{H}_{l_1
i_1}^{j_1}(t_{n+1})\right]\psi^{LM}_{l_1 l_2 j_1 i_2 n+1/4}\\
\left[\delta_{i_2}^{j_2} + \frac{i\tau_n}{2}\mathcal{H}_{l_2
i_2}^{j_2}(t_{n+1}) \right]\psi^{LM}_{l_1 l_2 i_1 j_2 n+3/4}&=&
\left[\delta_{i_2}^{j_2} - \frac{i\tau_n}{2}\mathcal{H}_{l_2
i_2}^{j_2}(t_{n+1})\right]\psi^{LM}_{l_1 l_2 i_1 j_2 n+1/2}\\
\phi^{LM}_{k\,m_2 i_1 i_2 n+3/4}&=&\mathcal{P}^{l_2}_{k m_2}
\mathcal{C}^{L l_1}_{l_2 m_2} \psi^{LM}_{l_1 l_2 i_1 i_2 n+3/4} \label{DOTr}\\
\phi^{LM}_{k m i_1 i_2 n+5/4}&=& \frac{1 - i\tau_n U_{i_1 i_2
k}(t_{n+1})}{1 + i\tau_n U_{i_1 i_2 k}(t_{n+1})}
\phi^{LM}_{k m i_1 i_2 n+3/4}\label{phCN}\\
\psi^{LM}_{l_1 l_2 i_1 i_2 n+5/4}&=&\overline{\mathcal{C}}^{\,L
m_2}_{l_1 l_2}\overline{\mathcal{P}}^{\,k}_{l_2
m_2}\phi^{LM}_{k\,m_2 i_1 i_2 n+5/4}\label{IOTr}\\
\left[\delta_{i_2}^{j_2} + \frac{i\tau_n}{2}\mathcal{H}_{l_2
i_2}^{j_2}(t_{n+1}) \right]\psi^{LM}_{l_1 l_2 i_1 j_2 n+3/2}&=&
\left[\delta_{i_2}^{j_2} - \frac{i\tau_n}{2}\mathcal{H}_{l_2
i_2}^{j_2}(t_{n+1})\right]\psi^{LM}_{l_1 l_2 i_1 j_2 n+5/4}\\
\left[\delta_{i_1}^{j_1} + \frac{i\tau_n}{2}\mathcal{H}_{l_1
i_1}^{j_1}(t_{n+1}) \right]\psi^{LM}_{l_1 l_2 j_1 i_2 n+7/4}&=&
\left[\delta_{i_1}^{j_1} - \frac{i\tau_n}{2}\mathcal{H}_{l_1
i_1}^{j_1}(t_{n+1})\right]\psi^{LM}_{l_1 l_2 j_1 i_2 n+3/2}\\
\psi^{LM}_{l_1 l_2 i_1 i_2 n+2}&=&\psi^{LM}_{l_1 l_2 i_1 i_2
n+7/4}-i\tau_n f^{LM}_{l_1 l_2 i_1 i_2 n+2}.
\end{eqnarray}
\end{subequations}
Для упрощения записи уравнений использовалось правило суммирования Эйнштейна.
На каждом временном шаге $\tau_n=(t_{n+2}-t_n)/2$  среднее время полагалось равным
$t_{n+1}=(t_{n+2}+t_n)/2$. Матрицы $\mathcal{H}_{l_{\alpha}
i_{\alpha}}^{j_{\alpha}}(t)$ являются конечно-разностными
аппроксимациями одномерных радиальных гамильтонианов
\[\hat{H}_{\alpha l_{\alpha}}(t)=-\frac{1}{2a^2(t)}\left(\frac{\partial^2}{\partial
\xi_{\alpha}^2}-\frac{l_{\alpha}(l_{\alpha}+1)}{\xi_{\alpha}^2}\right)-\frac{Z}{a
(t)\xi_{\alpha}} +\frac {a(t) \ddot a(t)}{2}\xi_{\alpha}^{2}. \]

Чтобы диагонализовать потенциал электрон-электронного взаимодействия,  используем
два ортогональных преобразования. Сначала применим
преобразование Чанга-Фано \cite{Fano}, обеспечивающее переход
от бисферического базиса  (\ref{BSBasis}) к $D$-базису
\cite{Varshalovich1989}
\begin{equation}
\Psi^{LM}(\boldsymbol{\xi}_1,\boldsymbol{\xi}_2,t)= \frac{1}{\xi_1
\xi_2} \sqrt{\frac{2L+1}{4\pi}} \sum_{l_2,m_2} D^{L*}_{M
m_2}(\varphi_1,\theta_1,\varphi_{12})
  \; Y_{l_2\,m_2}(\theta_{12},0)\;\chi^{LM}_{l_2
  m_2}(\xi_1,\xi_2,t),
\end{equation}
где $D^l_{m_1 m_2}$ --- функция Вигнера. Матрицы размерности
$(2\min(L,l_2)+1)\times (2\min(L,l_2)+1)$, осуществляющие прямое и
обратное преобразование Чанга-Фано, имеют вид
\begin{subequations}
\begin{eqnarray}
\mathcal{C}^{L l_1}_{l_2 m_2}&=&(-1)^{l_2+m_2}
C^{l_1\,0}_{l_2\,-m_2\,L\,m_2};\\
\overline{\mathcal{C}}^{\,L m_2}_{l_1 l_2}&=&(-1)^{l_2+m_2}
C^{l_1\,0}_{l_2\,-m_2\,L\,m_2},
\end{eqnarray}
\end{subequations}
где $C_{l_1 m_1\, l_2 m_2}^{l_3 m_3}$ - коэффициенты
Клебша-Гордана \cite{Varshalovich1989}, $l_2=0,\ldots, l_{2max}$,
$l_1=|L-l_2|,\ldots, L+l_2$,
$m_2=-\min(L,l_2),\ldots,\min(L,l_2)$. А для финальной
диагонализации  используем представление дискретной переменной
\cite{Melezhik1999}, основанное на преобразовании
\begin{eqnarray}
\phi^{LM}_{km}(\xi_1,\xi_2,t)&=&\sum_{l=0}^{N_m-1}\chi^{LM}_{lm}(\xi_1,\xi_2,t)P_{l}^{m}(\eta_k^{m}),
\end{eqnarray}
где $\phi^{LM}_{km}(\xi_1,\xi_2,t)$ --- значение волновой функции в
 $k$-м узле угловой решетки, $\eta=\cos\theta_{12}$,  $\eta_k^{m_2}$
 и
$\lambda_k^{m_2}$ --- узлы и веса соответствующих квадратур
Гаусса-Лежандра, $k=1,\dots, N_{m_2}$, $P_{l}^{m}(\eta)$ ---
нормированные присоединенные полиномы Лежандра, $N_{m_2}$ --- число
 полиномов в разложении для данного  $m_2$. Матрицы
размерности $N_{m_2}\times N_{m_2}$ прямого и обратного
преобразования имеют вид
\begin{subequations}
\begin{eqnarray}
\mathcal{P}^{l}_{k m}&=&P_{l}^{m}(\eta_k^{m});\\
\overline{\mathcal{P}}^{\,k}_{l
m}&=&P_{l}^{m}(\eta_k^{m})\lambda_k.
\end{eqnarray}
\end{subequations}

Прямые преобразования выполняются на шаге (\ref{DOTr}) алгоритма.
Благодаря им потенциал электрон-электронного взаимодействия
приобретает диагональный вид. В результате применение схемы
Кранка-Николсона на шаге (\ref{phCN}) сводится к умножению на
фазовый множитель. После этого выполняются обратные преобразования
(\ref{IOTr}), чтобы вернуться в бисферический базис, удобный для
вычислений, включающих одноэлектронные гамильтонианы.

Для представления радиальных дифференциальных операторов в
работах \cite{Serov2007,Serov2008,SerovSergeeva2010} использовалась
пятиточечная конечно-разностная аппроксимация четвертого
порядка на сетке с узлами
\[ \xi_{\alpha i}=hi; \;\;\; i=1..N_{r};\;\;\; h= \frac{\xi_{\mathrm{max}}}{N_{r}}, \]
где $\xi_{\mathrm{max}}$ --- размер сетки в сопутствующих
координатах, $N_{r}$ --- число радиальных узлов. Поскольку нас в
основном интересовала ионизация с испусканием электронов
относительно малых энергий, мы помимо перехода к сопутствующим
координатам использовали внешний комплексный скейлинг (см. раздел
\ref{sectECS}), обеспечивающий поглощение высокоэнергичных
электронов, что позволяет уменьшить требущийся размер сетки
$\xi_{\mathrm{max}}$. Использование же неравномерных сеток, как в
работе \cite{Serov2007}, оказалось непрактичным.

Метод расщепления устойчив, если шаг по времени $\tau$ удовлетворяет условию
\begin{equation}
\tau \leq \frac{h^2}{4a^2}. \label{CondStab}
\end{equation}
Однако практические расчеты показывают, что для $L>0$ шаг $\tau$
может быть взят в несколько раз большим без какого-либо вреда. Это
связано с отсутствием трехчастичных столкновений в случае $L>0$.
Мы использовали зависящий от времени шаг $\tau_n=\tau(t_{n+1})=a(t_{n+1})\tau_0$, поскольку наименее медленно при расширении сетки убывает кулоновский потенциал, а это происходит по закону $1/a(t)$. При таком выборе шага компьютерное время, необходимое для расчета эволюции вплоть до заданного физического времени $t$, пропорционально $\log t$.

В расчетах фотоионизации гелия \cite {SerovSergeeva2010} использовались следующие параметры численной схемы: параметр углового базиса $l_{2{\mathrm{max}}}=13$, равномерная радиальная сетка с $N_r =500$ и размером $\xi_{\mathrm{max}}=25$, радиус комплексного скейлинга $\xi_{\mathrm{sc}}=22.5$, угол комплексного скейлинга $\theta_{\mathrm{sc}}=30^{\circ}$, скорость расширения сетки $\dot{a}_{\infty}=0.01$. Моделировалась эволюция вплоть до времени $t_{\mathrm{max}}=12800$. Для других мишеней использовались иные параметры радиальной сетки: $N_r=500$, $\xi_{\mathrm{max}}=50$, $\xi_{\mathrm{sc}}=45$, $\dot{a}_{\infty}=0.05$ для $\mathrm{H}^-$; $N_r=1000$, $\xi_{\mathrm{max}}=50$, $\xi_{\mathrm{sc}}=40$, $\dot{a}_{\infty}=0.05$ для $\mathrm{He}$ в возбужденном состоянии $\mathrm{1s2s^1 S}$ и $N_r=1400$, $\xi_{\mathrm{max}}=70$, $\xi_{\mathrm{sc}}=60$, $\dot{a}_{\infty}=1/30$ для $\mathrm{He}$ в возбужденном состоянии $\mathrm{1s3s^1 S}$.

Начальная волновая функция мишени $\varphi_0(\mathbf{r}_1,\mathbf{r}_2)$, необходимая в  (\ref{InitCondPhotoIon}) и члене-источнике в PA1B, расчитывалась методом эволюции в мнимом времени
\cite{Pindzola2001}.

\section{Процедура интегрирования при вычислении второго борновского оператора возбуждения}\label{appendixSB}

В лабораторной системе координат с  $Oz||\mathbf{k}_i$ можно записать
\begin{eqnarray}
 \mathcal{I}(\mathbf{k}_i,\mathbf{K};\mathbf{r})&=&\int_0^{\infty} \frac{dq}{q}\oint  d\Omega_{q}\frac{\frac{1}{|\mathbf{K}-\mathbf{q}|^2}\exp(i\mathbf{q}\mathbf{r})}{\cos\theta_{q}-\frac{q^2/2+E_{t}-E_0}{k_iq}+i\epsilon}. \label{2BIntegral}
\end{eqnarray}
Мы заменили переменную интегрирования $\mathbf{k}$ на $\mathbf{q}$,
потому что при этом угловая зависимость в $1/|\mathbf{q}|^2$
пропадает. Сначала сведем этот интеграл к сумме более простых
интегралов с помощью двух хорошо известных разложений
\cite{Varshalovich1989}
\begin{eqnarray}
\exp(i\mathbf{q}\mathbf{r})\simeq \sum_{\ell=0}^{\ell_{max}} i^\ell j_\ell(qr) (2\ell+1) P_\ell(\cos\hat{\mathbf{q}\mathbf{r}}), \label{expqr}
\end{eqnarray}
и
\begin{eqnarray}
\frac{1}{|\mathbf{K}-\mathbf{q}|^2}\simeq
\frac{1}{2Kq}\sum_{l=0}^{l_{max}}Q_l\left(\frac{K^2+q^2}{2Kq}\right)P_l(\cos\hat{\mathbf{K}\mathbf{q}}),
\label{onedivq1s}
\end{eqnarray}
где
\begin{eqnarray}
P_l(\cos\hat{\mathbf{n}\mathbf{n}'})=\frac{4\pi}{2l+1}\sum_{m=-l}^{l}Y_{lm}^*(\theta,\phi)Y_{lm}(\theta',\phi').
\end{eqnarray}
Интегрирование по азимутальному углу $\phi_q$ тривиально.
Интегрирование по  $\theta_q$ выполняется с применением тождества
для произведений сферических гармоник \cite{Varshalovich1989}, что
приводит к
\begin{eqnarray}
\int_{-1}^{1}\frac{P_{l}(x')}{x'-x+i\epsilon}dx=\left\{
 \begin{array}{ll}
 -2Q_l(x)-i\pi P_l(x),& x\in [-1,1];\\
 -2Q_l(x),& x>1,
 \end{array}
 \right.
\end{eqnarray}
где $P_l(x)$ --- полином Лежандра, $Q_l(x)$ --- функция Лежандра
второго рода, определенная на интервале  $x\in [-1,\infty)$ и
удовлетворяющая соотношению
$Q_0(x)=\frac{1}{2}\ln\left|\frac{x+1}{x-1}\right|$,
$Q_1(x)=xQ_0(x)-1$,
$Q_{l}(x)=\frac{2l-1}{l}xQ_{l-1}(x)-\frac{l-1}{l}Q_{l-2}(x)$.
Интегрирование по  $q$ затем производится численно, с помощью
квадратурных формул, учитывающих наличие логарифмических
сингулярностей подынтегрального выражения в точках
$q_{a,b}=k_i\mp\sqrt{k_i^2-2(E_t-E_i)}$. Для $q>q_b$ подынтегральное
выражение быстро стремится к нулю. Это позволяет использовать
конечный верхний предел интегрирования $q_{max}=2k_i$.

Прямое применение разложения (\ref{onedivq1s}) представляет
дополнительную трудность, так как не гарантирует сходимости
внеоболочечной части интеграла $\mathcal{I}$  с ростом $l_{max}$,
поскольку выражение (\ref{onedivq1s}) имеет особенности на сфере
$q=K$ для любого конечного числа $l_{max}$. Чтобы избежать эту трудность, мы использовали разложение
\begin{eqnarray*}
\frac{1}{|\mathbf{K}-\mathbf{q}|^2}=\frac{1}{K^2+q^2-2(\mathbf{K}\cdot\mathbf{q})}\simeq
 \frac{1}{K^2+q^2}\sum_{s=0}^{l_{max}}\left[\frac{2(\mathbf{K}\cdot\mathbf{q})}{K^2+q^2}\right]^s,
\end{eqnarray*}
которое не имеет особенностей для $K>0$. Это эквивалентно замене
 $Q_l(x)$ в (\ref{onedivq1s}) его разложением в конечный ряд
\begin{eqnarray}
Q_{l}^{[l_{max}]}(x)=\frac{1}{x}\sum_{n=0}^{\lfloor(l_{max}-l)/2\rfloor}
 \frac{(2n+l)!}{(2n)!!(2n+2l+1)!!}\frac{1}{x^{2n+l}}.
\end{eqnarray}
При таком подходе  рассчитанное значение  (\ref{2BIntegral})
монотонно сходится с ростом $l_{max}$. Во всех расчетах в \cite{Serov2010}
использовались параметры $l_{max}=99$ и $\ell_{max}=3$.

\end{document}